\documentclass[12pt]{article}
\usepackage{dsfont}

\usepackage{amsmath,amssymb,graphicx}
\usepackage{diagbox}
\usepackage{hyperref}
\usepackage[nosort]{cite}

\usepackage{tikz}
\usetikzlibrary{calc} 
\usetikzlibrary{patterns} 
\usetikzlibrary{decorations.pathreplacing} 
\usetikzlibrary{decorations.markings} 
\usetikzlibrary{decorations.pathmorphing} 
\usetikzlibrary{positioning}
\usetikzlibrary{arrows.meta}

\usepackage{color,xcolor}
\definecolor{darkred}{rgb}{0.65,0.15,0}
\definecolor{newgreen}{rgb}{0.2,0.62,0.14}
\hypersetup{pdfborder={0 0 0},colorlinks=true,urlcolor=blue,citecolor=blue,linkcolor=darkred,linktocpage=true}

\numberwithin{equation}{section}

\def\nn{\nonumber}

\def\spa#1.#2{\left\langle#1\,#2\right\rangle}
\def\spb#1.#2{\left[#1\,#2\right]}

\def\ep{\epsilon}

\def\ap{\alpha'}

\def\beq{\begin{equation}}
\def\eeq{\end{equation}}
\let\Re\relax
\let\Im\relax
\DeclareMathOperator{\Re}{Re}
\DeclareMathOperator{\Im}{Im}

\newcommand{\eq}{\begin{equation}}
\newcommand{\eqe}{\end{equation}}
\newcommand{\eqa}{\begin{eqnarray}}
\newcommand{\eqae}{\end{eqnarray}}

\newcommand{\p}{\partial}

\newcommand{\bea}{\begin{eqnarray}}
\newcommand{\eea}{\end{eqnarray}}
\newcommand{\dd}{\mathrm{d}}

\newcommand{\newf}{\tilde f}
\newcommand{\newg}{\widetilde G}

\newcommand{\RR}{\mathbb R}

\newcommand{\CC}{\mathbb C}

\newcommand{\ZZ}{\mathbb Z}

\newcommand{\dplus}[1]{{\cal D}^+ \! \left[\begin{smallmatrix}#1\end{smallmatrix}\right]}
\newcommand{\cplus}[1]{{\cal C}^+ \! \left[\begin{smallmatrix}#1\end{smallmatrix}\right]}
\newcommand{\cminus}[1]{{\cal C}^- \! \left[\begin{smallmatrix}#1\end{smallmatrix}\right]}

\newcommand{\cplustri}[3]{\,{\cal C}^+\!\left[\protect\begin{smallmatrix}#1\protect\end{smallmatrix}\middle|\protect\begin{smallmatrix}#2\protect\end{smallmatrix}\middle|\protect\begin{smallmatrix}#3\protect\end{smallmatrix}\right]}

\def\cC{{\cal C}}
\def\cD{{\cal D}}

\def\cH{{\cal H}}

\def\a{\alpha}
\def\b{\beta}
\def\tet{\vartheta}
\def\sm{\smallskip}
\def\no{\nonumber}
\def\nabchi{\nabla}

\newbox\charbox
\newbox\slabox
\def\s#1{{      % Feynman slash
        \setbox\charbox=\hbox{$#1$}
        \setbox\slabox=\hbox{$/$}
        \dimen\charbox=\ht\slabox
        \advance\dimen\charbox by -\dp\slabox
        \advance\dimen\charbox by -\ht\charbox
        \advance\dimen\charbox by \dp\charbox
        \divide\dimen\charbox by 2
        \raise-\dimen\charbox\hbox to \wd\charbox{\hss/\hss}
        \llap{$#1$}
}}

\newcommand{\KN}{\text{KN}}
\newcommand{\ad}{\text{ad}}
\newcommand{\MKES}{\text{MKES}}

\usepackage[shadow,textwidth=2.7cm]{todonotes}
\usepackage{ifthen}
\reversemarginpar
\newcounter{todocounter}

\colorlet{edcolor}{green!40!white}

\newcommand{\edinline}[2][]{
  \ifthenelse { \equal {#1} {} }
    { \def\temp {#2} }  % if #1 == blank
    { \def\temp {#1} }   % else (not blank)
  \refstepcounter{todocounter}\todo[color=edcolor,inline,caption={\textbf{\thetodocounter. ED} \temp}]{\textbf{\thetodocounter. ED:} #2}{}}

\colorlet{oscolor}{blue!20!white}

\newcommand{\osinline}[2][]{
  \ifthenelse { \equal {#1} {} }
    { \def\temp {#2} }  % if #1 == blank
    { \def\temp {#1} }   % else (not blank)
  \refstepcounter{todocounter}\todo[color=oscolor,inline,caption={\textbf{\thetodocounter. OS} \temp}]{\textbf{\thetodocounter. OS:} #2}{}}

\colorlet{akcolor}{yellow!40!white}

\newcommand{\akinline}[2][]{
  \ifthenelse { \equal {#1} {} }
    { \def\temp {#2} }  % if #1 == blank
    { \def\temp {#1} }   % else (not blank)
  \refstepcounter{todocounter}\todo[color=akcolor,inline,caption={\textbf{\thetodocounter. AK} \temp}]{\textbf{\thetodocounter. AK:} #2}{}}

\colorlet{bvcolor}{violet!50!white}

\newcommand{\bvinline}[2][]{
  \ifthenelse { \equal {#1} {} }
    { \def\temp {#2} }  % if #1 == blank
    { \def\temp {#1} }   % else (not blank)
  \refstepcounter{todocounter}\todo[color=bvcolor,inline,caption={\textbf{\thetodocounter. BV} \temp}]{\textbf{\thetodocounter. BV:} #2}{}}

\begin{document}

 {\flushright 16 December 2020\\
revised 15 March 2021\\
 UUITP--54/20\\[5mm]}

\begin{center}

{\LARGE \bf   Elliptic modular graph forms I \\[3mm] {\Large Identities and generating series}}\\[5mm]

\vspace{6mm}
\normalsize
{\large  \bf Eric D'Hoker${}^{1}$, Axel Kleinschmidt${}^{2,3}$ and Oliver Schlotterer${}^4$}

\vspace{5mm}
${}^{1}${\it Mani L. Bhaumik Institute for Theoretical Physics}\\
 {\it Department of Physics and Astronomy }\\
{\it University of California, Los Angeles, CA 90095, USA}
\vskip 0.5 em
${}^2${\it Max-Planck-Institut f\"{u}r Gravitationsphysik (Albert-Einstein-Institut)\\
Am M\"{u}hlenberg 1, DE-14476 Potsdam, Germany}
\vskip 0.5 em
${}^3${\it International Solvay Institutes\\
ULB-Campus Plaine CP231, BE-1050 Brussels, Belgium}
\vskip 0.5 em
${}^4${\it  Department of Physics and Astronomy\\
Uppsala University, 75108 Uppsala, Sweden}

 \vskip 0.1in
 
{\tt \small dhoker@physics.ucla.edu},
{\tt \small axel.kleinschmidt@aei.mpg.de},  
{\tt \small oliver.schlotterer@physics.uu.se}

\vspace{5mm}

{\bf Abstract}

\vspace{2mm}

\begin{quote}
Elliptic modular graph functions and forms (eMGFs) are defined for arbitrary graphs as natural generalizations of modular graph functions and forms obtained by including the character of an Abelian group in their Kronecker--Eisenstein series.  The simplest examples of eMGFs are given by the Green function for a massless scalar field on the torus and the Zagier single-valued elliptic polylogarithms. More complicated eMGFs are produced by the non-separating degeneration of a higher genus surface to a genus one surface with punctures. eMGFs  may equivalently be represented by multiple integrals over the torus of combinations of coefficients of the Kronecker--Eisenstein series, and may be assembled into generating series. These relations are exploited to derive holomorphic subgraph reduction formulas, as well as algebraic and differential identities between eMGFs and their generating series. 
\end{quote}

\vspace{6mm}

\end{center}

\thispagestyle{empty}

\newpage
\setcounter{page}{1}

\setcounter{tocdepth}{2}
\tableofcontents

\newpage

%%%%%%%%%%%%%%%%%%%%%%%%%%%%%%%%%%%%%%%%%%%%%%%%%%%%%%%%%%%
%%%%%%%%%%%%%%%%%%%%%%%%%%%%%%%%%%%%%%%%%%%%%%%%%%%%%%%%%%%
\section{Introduction}
\label{sec:1}
%%%%%%%%%%%%%%%%%%%%%%%%%%%%%%%%%%%%%%%%%%%%%%%%%%%%%%%%%%%
%%%%%%%%%%%%%%%%%%%%%%%%%%%%%%%%%%%%%%%%%%%%%%%%%%%%%%%%%%%

A modular graph function (MGF) maps a decorated graph to an $SL(2,\ZZ)$ invariant function on the upper half complex plane $\cH_1$. MGFs generalize non-holomorphic Eisenstein series as well as multiple zeta values (MZVs), and may be further generalized to produce non-holomorphic modular forms instead of modular functions.\footnote{We shall also use the acronym MGF to refer to modular graph forms.} MGFs constitute the basic building blocks for the evaluation of genus-one contributions to effective low energy interactions in string theory: They arise as multiple integrals over a torus world-sheet of products of the Green function for a conformal scalar and its derivatives. Individual cases were studied in \cite{Green:1999pv} and \cite{Green:2008uj}, while their systematic investigation was initiated in \cite{DHoker:2015gmr,DHoker:2015sve, Zerbini:2015rss, DHoker:2015wxz}.

\sm

Using the procedure of holomorphic subgraph reduction \cite{DHoker:2016mwo, Gerken:2018zcy}, large families of MGFs were shown to satisfy a rich hierarchy of 
algebraic and differential identities  in \cite{DHoker:2016mwo, Basu:2016kli, DHoker:2016quv, Gerken:2020aju}. In particular, MGFs were shown to obey inhomogeneous Laplace eigenvalue equations for the two-loop case in \cite{DHoker:2015gmr}, for the Mercedes diagram in \cite{Basu:2016xrt}, and more general tetrahedral MGFs in \cite{Kleinschmidt:2017ege, Basu:2019idd}. A {\tt Mathematica} package is now available for the systematic implementation of identities among MGFs in \cite{Gerken:2020aju} and brought into wider context in the PhD thesis \cite{Gerken:review}. Their reduction to iterated Eisenstein integrals via generating series \cite{Gerken:2019cxz, Gerken:2020yii} exposes all their relations and furthermore a connection to elliptic MZVs in open-string computations \cite{Gerken:2020xfv} and may clarify the connection with Brown's construction of non-holomorphic modular forms \cite{Brown:2017qwo, Brown:2017qwo2}.

\sm

Elliptic modular graph functions and forms (eMGFs) are generalizations of modular graph functions and forms
in the same way as elliptic functions generalize the notion of modular forms. 
An eMGF maps a graph to a non-holomorphic single-valued elliptic function of one or several variables on the torus, and is invariant under the modular group transforming both on the modulus and on the torus points. 
Perhaps the simplest and earliest examples of eMGFs are the conformal scalar Green function on the torus or, for that matter, any conformal Green function on the torus, and Zagier's single-valued elliptic polylogarithms \cite{Ramakrish}. 
More sophisticated examples of eMGFs  have emerged recently as a result of systematic investigations into the non-separating degenerations of higher-genus MGFs in \cite{DHoker:2017pvk}. 

\sm

For example, in the non-separating degeneration limit of a genus-two MGF, the genus-two surface degenerates to a torus with two punctures, and the genus-two MGF degenerates to a modular function on the torus that depends on the locations of the two punctures \cite{DHoker:2018mys}. The resulting limit of the genus-two MGF is therefore an eMGF on the torus. Actually, it is  not just the non-separating limit but rather the full Laurent expansion of the genus-two MGF near the non-separating node that systematically produces eMGFs as coefficients of the Laurent expansion. Higher genus MGFs will reduce to a torus with multiple punctures upon taking multiple non-separating degenerations simultaneously, thereby giving rise to eMGFs which depend on multiple points on the torus.  In each case, the effect of the punctures may be encoded in terms of a  {\sl group character}, and it is this point of view that we shall adopt to define eMGFs in all generality in this paper.  

\sm

Elliptic modular graph functions inherit the implications of identities satisfied by their MGF ancestors. 
An example of this phenomenon was uncovered in \cite{DHoker:2020tcq} where an algebraic identity between genus-two MGFs was shown to imply a highly non-trivial identity between its genus-one eMGF descendants.  The latter was proven shortly thereafter via the direct use of genus-one methods in \cite{Basu:2020pey}. Further relations among genus-one eMGFs involving examples built from up to five Green functions were recently studied in \cite{Basu:2020iok}.

\sm

In the present paper, we shall present a general definition of eMGFs in various different but equivalent formulations. 
The first represents the eMGF associated with an arbitrary (decorated) graph in terms of a multiple Kronecker--Eisenstein sum in which the dependence on the points on the torus are introduced through the character of an Abelian group. The co-moving coordinates which are used to represent the points of the torus may be viewed as characteristics, in complete analogy with Jacobi $\tet$-functions with characteristics where arbitrary real characteristics may be traded for a point on the torus. The second equivalent representation of an eMGF is in terms of multiple integrals over the torus of products of non-holomorphic modular forms $\cD^+$ which are equivalent to Zagier's single-valued elliptic polylogarithms. This formulation is an immediate generalization of the manner in which MGFs arise in string theory  as multiple (configuration space) integrals over the torus. The third equivalent formulation to be detailed below is through the use of a generating series for entire families of eMGFs in terms of Kronecker--Eisenstein series. The relation with iterated modular integrals gives a fourth formulation, whose study will be relegated to a separate forthcoming paper \cite{toappsoon}.

\sm

Following the definition of eMGFs in these various formulations, we proceed to deriving algebraic and differential relations for the characters and for the eMGFs, in close parallel to the corresponding derivations in the case of MGFs. In particular, we shall prove the generalization of the holomorphic subgraph reduction procedure to the case of eMGFs, using the integral formulation of eMGFs and the Fay identities between the coefficient functions of the Kronecker--Eisenstein series. We shall also show the validity of Laplace eigenvalue equations for all two-loop eMGFs, again in close parallel to the case of MGFs studied in \cite{DHoker:2015gmr}. Finally, we shall provide examples of algebraic and differential  identities between eMGFs of low weight and various loop orders.

\sm

A complementary perspective on eMGFs in one variable and their differential properties is given on the basis of generating series of Koba--Nielsen integrals with eMGFs in their expansion coefficients. These generating series can be obtained from those of genus-one integrals in closed-string amplitudes \cite{Gerken:2019cxz, Gerken:2020yii} by leaving two rather than one of the punctures unintegrated (one of the unintegrated punctures can always be fixed at the origin). The open-string counterparts of such generating series were investigated in \cite{Broedel:2020tmd} and shown to obey Knizhnik--Zamolodchikov--Bernard(KZB)-type differential equations on a twice-punctured torus. We will spell out the analogous KZB-type equations of the generating series of eMGFs which furnish an equivalent formulation of the differential properties of eMGFs and sidestep holomorphic subgraph reduction. Moreover, the differential equations of the generating series will play a central role in the description of eMGFs in terms of iterated modular integrals \cite{toappsoon}.

\subsection*{Organization}

The remainder of this paper is organized as follows. In section \ref{sec:2} we provide the definition of eMGFs in terms of Kronecker--Eisenstein sums and characters as well as their equivalent integral formulation. In section \ref{sec:2A} we define the derivatives of eMGFs with respect to the modulus and with respect to the points on the torus, prove the holomorphic subgraph reduction procedure for eMGFs, derive the Laplace eigenvalue equations in various infinite families, and provide some examples of differential identities at low weight. In section \ref{sec:3} we derive dihedral eMGFs from generating series of Koba--Nielsen integrals, and generalize this construction in appendix \ref{app:tri} to the trihedral case and in section \ref{sec:4} to the general one-variable case.  Additional technical details and comments have been relegated to appendices~\ref{app:compdiff}--\ref{app:comm}.

\subsection*{Acknowledgments}

We are grateful to Jan Gerken, Martijn Hidding, Boris Pioline and Bram Verbeek for stimulating discussions and collaboration on related topics. The research of ED is supported in part by NSF grant PHY-19-14412. The research of OS  is supported by the European Research Council under ERC-STG-804286 UNISCAMP.

\newpage

%%%%%%%%%%%%%%%%%%%%%%%%%%%%%%%%%%%%%%%%%%%%%%%%%%%%%%%%%%%
%%%%%%%%%%%%%%%%%%%%%%%%%%%%%%%%%%%%%%%%%%%%%%%%%%%%%%%%%%%
\section{Elliptic modular graph forms as lattice sums}
\label{sec:2}
%%%%%%%%%%%%%%%%%%%%%%%%%%%%%%%%%%%%%%%%%%%%%%%%%%%%%%%%%%%
%%%%%%%%%%%%%%%%%%%%%%%%%%%%%%%%%%%%%%%%%%%%%%%%%%%%%%%%%%%

In this section we shall introduce elliptic modular graph forms (eMGFs) as multiple Kronecker--Eisenstein sums (MKES), generalizing the corresponding sums for modular graph forms (MGFs) by including a character of an Abelian group in the summand. These characters may equivalently be parametrized by points on the torus. Following the definition of eMGFs, we shall  in the next section derive their basic properties, obtain the integral and differential equations they satisfy, extend the procedure of holomorphic subgraph reduction developed for MGFs to the case of eMGFs, and provide examples.

%%%%%%%%%%%%%%%%%%%%%%%%%%%%%%%%%%%%%%%%%%%%%%%%%%%%%%%%%%%
\subsection{Basics}
\label{sec:2.1}
%%%%%%%%%%%%%%%%%%%%%%%%%%%%%%%%%%%%%%%%%%%%%%%%%%%%%%%%%%%

\begin{figure}[htb]
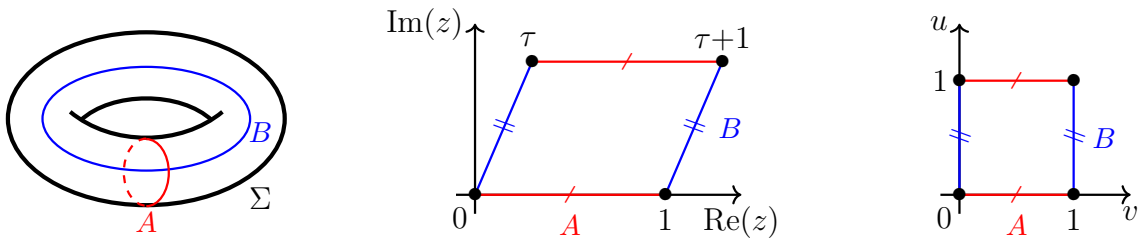

\begin{center}
\tikzpicture[scale=0.46, line width=0.30mm]
\draw[ultra thick] (0,0) ellipse  (4cm and 2.5cm);
\draw[ultra thick] (-2.2,0.2) .. controls (-1,-0.8) and (1,-0.8) .. (2.2,0.2);
\draw[ultra thick] (-1.9,-0.05) .. controls (-1,0.8) and (1,0.8) .. (1.9,-0.05);
\draw[blue](0,0) ellipse  (3cm and 1.5cm);
\draw[red] (0,-2.53) arc (-90:90:0.65cm and 0.97cm);
\draw[red,dashed] (0,-0.575) arc (90:270:0.65cm and 0.97cm);
\draw[blue] (3.3,-0.4) node{\small $B$};
\draw[red] (0,-3) node{\small $A$};
\draw (3.3,-2.3) node{$\Sigma$};
\scope[xshift=9.5cm,yshift=-2.2cm,scale=1.1]
\draw[->](-0.5,0) -- (7,0) node[below]{${\rm Re}(z)$};
\draw[->](0,-0.5) -- (0,4.5) node[left]{${\rm Im}(z)$};
\draw(-0.4,-0.6)node{$0$};
\draw[blue](0,0) -- (1.5,3.5);
\draw[blue] (0.75,1.75)node[rotate=-20]{$=$};
\draw(1.4,4.1)node{$\tau$};
\draw[red](0,0) -- (5,0);
\draw[red] (2.5,0)node[rotate=60]{$-$};
\draw (5,-0.65)node{$1$};
\draw[red](1.5,3.5) -- (6.5,3.5);
\draw[red] (4,3.5)node[rotate=60]{$-$};
\draw[blue](5,0) -- (6.5,3.5);
\draw[blue] (5.75,1.75)node[rotate=-20]{$=$};
\draw(6.5,4.1)node{$\tau{+}1$};
\draw[red](2.5,-0.8)node{\small $A$};
\draw[blue](6.7,1.75)node{\small $B$};
\draw (5,0)node{$\bullet$};
\draw(0,0)node{$\bullet$};
\draw (1.5,3.5)node{$\bullet$} ;
\draw(6.5,3.5)node{$\bullet$};
\endscope
\scope[xshift=23.5cm,yshift=-2.2cm,scale=1.1]
\draw[->](-0.5,0) -- (4.5,0) node[below]{$v$};
\draw[->](0,-0.5) -- (0,4.5) node[left]{$u$};
\draw(-0.4,-0.6)node{$0$};
\draw[blue](0,0) -- (0,3);
\draw[blue](0,1.5)node[rotate=-20]{$=$};
\draw[red](0,0) -- (3 ,0);
\draw[red] (1.5,0)node[rotate=60]{$-$};
\draw (3,-0.65)node{$1$};
\draw[red](0,3) -- (3,3);
\draw[red] (1.5,3)node[rotate=60]{$-$};
\draw[blue](3,0) -- (3,3);
\draw[blue] (3,1.5)node[rotate=-20]{$=$};
\draw[red](1.5,-0.8)node{\small $A$};
\draw[blue](3.8,1.5)node{\small $B$};
\draw (3,0)node{$\bullet$};
\draw(0,0)node{$\bullet$};
\draw (0,3)node{$\bullet$}node[left]{$1$} ;
\draw(3,3)node{$\bullet$};
\endscope
\endtikzpicture
\caption{\textit{The torus $\Sigma = \CC/\Lambda$ with a choice of canonical homology cycles $A$ and $B$ (left figure), is represented in the plane by a parallelogram with complex coordinates $z,\bar z$ (middle figure) or a square with real coordinates $0 \leq u,v \leq 1$ (right figure) related by $z=u\tau + v$, and opposite sides pairwise identified.}
\label{figtorus}}
\end{center}
\end{figure}

A torus $\Sigma$ of modulus $\tau \in \cH_1 = \{ \tau \in \CC, \, \Im(\tau)>0 \}$ is a compact Riemann surface of genus one without boundary and may be given as the quotient of $\CC$  by a lattice $\Lambda$, 
\bea
\Sigma = \CC / \Lambda 
\hskip 1in 
\Lambda =  \ZZ \tau  +  \ZZ 
\label{basic.1}
\eea
The torus $\Sigma$ may be represented in $\CC$ as a parallelogram parametrized by local complex coordinates $z,\bar z$ subject to the identifications $z \sim z+1$ and $z \sim z+\tau$, or as a square in $\RR^2$ parametrized by real coordinates $u,v$ subject to the identifications $u\sim u+1$ and $v \sim  v+1$, as shown in Figure \ref{figtorus}. The relation between these representations is given by
\bea
\label{basic.0}
z=u \tau +v \hskip 1in u,v \in [0,1]
\eea 
The coordinate $z$ has the advantage of being complex, while  $(u,v)$ has the advantage of being {\sl co-moving coordinates} whose range is independent of $\tau$. The trade-offs are familiar in the context of Jacobi $\tet$-functions with real characteristics $(u,v)$ which may converted into a point $z \in \Sigma$ using (\ref{basic.0}). Integrations over the torus $\Sigma$ are normalized according to,
\begin{align}
\label{basic:int}
\int_\Sigma \frac{\dd^2z}{\Im \tau} = \int_0^1 \dd u \int_0^1 \dd v = 1
\hspace{15mm} \frac{\dd^2 z}{\Im\tau} = \frac{i \dd z\wedge \dd\bar{z} }{2\Im\tau}  = \dd v \wedge \dd u
\end{align}

The points in the lattice $\Lambda$ correspond to the allowed momenta on the torus and may be parametrized as follows,
\bea
\label{basic.2}
p = m\tau + n \in \Lambda \hskip 0.6in m,n \in \ZZ
\eea
A modular transformation performs a change of canonical homology basis of $A$ and $B$ cycles which permutes the individual points in $\Lambda$ but leaves the lattice $\Lambda$ invariant. The action of modular transformations  on the complex data $\tau, z, p$, 
\bea
\label{modPSL}
 \tau' =  {\a \tau+\b \over \gamma \tau+\delta }
 \hskip 0.5in
 z' = { z \over \gamma \tau +\delta } 
 \hskip 0.5in
  p' = { p \over \gamma \tau +\delta } 
\hskip 0.5in 
 \left ( \begin{matrix} \a & \b \cr \gamma  & \delta \end{matrix} \right ) \in SL(2,\ZZ)
\eea
induces the corresponding transformations on the real data $(u,v)$ and $ (m,n)$, related to $z$ and $p$ by (\ref{basic.0}) and (\ref{basic.2}) respectively,  
\bea
\label{modPSL1}
\left ( \begin{matrix} v' \cr - u' \cr \end{matrix} \right ) 
=  \left ( \begin{matrix} \a & \b \cr \gamma  & \delta \end{matrix} \right ) \left ( \begin{matrix} v \cr - u \cr \end{matrix} \right ) 
\hskip 1in
\left ( \begin{matrix} n' \cr - m' \cr \end{matrix} \right ) 
=  \left ( \begin{matrix} \a & \b \cr \gamma  & \delta \end{matrix} \right ) \left ( \begin{matrix} n \cr - m \cr \end{matrix} \right ) 
\eea
The ranges of $(u,v) $ and $(u',v')$ generally differ from one another, but are such that the area of the fundamental region for $\Sigma$ is 1. Note that the modular group used here is $SL(2,\ZZ)$ rather than $PSL(2,\ZZ)$ because the element $-I$ acts non-trivially on $z, p, (u,v) $ and $(m,n)$ even if it acts trivially on $\tau$.

\subsection{The Eisenstein series and the scalar Green function}

The prototype of an MGF is the non-holomorphic Eisenstein series $E_k(\tau)$ defined by the following Kronecker--Eisenstein sum,
\bea
\label{Ek}
E_k (\tau) = \bigg( \frac{ \Im \tau }{\pi} \bigg)^{\! k} \, \sum _{p \in \Lambda '} { 1 \over   |p|^{2k}} 
= \bigg( \frac{ \Im \tau }{\pi} \bigg)^{\! k} \sum _{{(m,n) \in \ZZ^2 \atop (m,n) \not= (0,0)}} { 1 \over  |m \tau + n |^{2k}} 
\eea
where $\Lambda' = \Lambda \setminus \{0 \}$. 
In fact, $E_k$ is invariant under modular transformations~\eqref{modPSL1} and thus an example of a modular graph \textit{function}.
The series is absolutely convergent for $\Re(k) >1$ and may be analytically continued in $k$ to the full complex plane with a simple pole at $k=1$.

\sm

The prototype of an eMGF is the non-holomorphic elliptic function $g_k(z|\tau)$, defined using the relation $z=u\tau+v$ with $u,v \in \RR$,
\bea 
\label{gk}
g_k(z|\tau) =  \bigg( \frac{ \Im \tau }{\pi} \bigg)^{\! k} \sum _{{(m,n) \in \ZZ^2 \atop (m,n) \not= (0,0)}} {e^{2\pi i (nu-mv)} \over   |m \tau + n |^{2k}} 
\eea
For $\Re(k) >1$, the series is absolutely convergent, while for  $\Re(k)=1$ its convergence is conditional and defined by the Eisenstein summation convention in which the sum over $n$ is carried out first.  The scalar Green function $g(z|\tau)$ on the torus of modulus $\tau$ corresponds to $g(z|\tau) = g_1(z|\tau)$ which  may alternatively be expressed in terms of the Jacobi $\tet$-function and the Dedekind $\eta$-function  by,
\bea
g(z|\tau) = - \log \left| \frac{ \tet_1(z|\tau) }{\eta(\tau)} \right|^2 + \frac{ 2\pi (\Im z)^2 }{\Im \tau}
\eea
where $\eta (\tau) = q^{{1 \over 24}} \prod _{n=1}^\infty (1-q^n)$ with $q=e^{2\pi i \tau}$ and the $\tet$-function is normalized as follows,
\bea
\tet_1(z|\tau) = q^{1/8} (e^{i\pi z} - e^{-i\pi z} ) \prod_{n=1}^\infty (1-q^n)(1-e^{2\pi i z} q^n)(1-e^{-2\pi i z} q^n)
\eea 
For $\Re(k) >1$ we have $g_k(0|\tau)=E_k(\tau)$ and for integer $k>1$  the functions $g_k$ may be obtained recursively from the scalar Green function $g$  by convolution with $g$,
\bea
g_{k+1}(z|\tau) = \int _\Sigma {\dd^2w \over \Im \tau} \, g_k(z-w|\tau) g(w|\tau)
\eea
The functions $g_k(z|\tau)$ are modular invariant
and thus elliptic modular graph {\it functions}, 
\bea
g_k(z'  | \tau' ) = g_k(z|\tau) 
\eea
where the transformation law for $z$ and $\tau$ is given in (\ref{modPSL}).

\subsection{Characters and characteristics}

The distinction between the summands in the Kronecker--Eisenstein sums for $E_k$ and $g_k$ lies entirely in the exponential factor, which may be viewed as a {\sl  character } for the Abelian group corresponding to the lattice $\Lambda$ in a representation labelled by the characteristics $(u,v) \in \RR^2 /\ZZ^2$. It may also be viewed as a character for the Abelian group corresponding to the torus $\Sigma$ in a representation labelled by the pair $(m,n) \in \ZZ^2$. 

\sm

Points in $\Sigma$  may equivalently be labelled by $z$ or by $(u,v)$ related to one another by (\ref{basic.0}) while points in $\Lambda$ may equivalently be labelled by $p$ or $(m,n)$ related to one another by (\ref{basic.2}). We shall choose to label the characters by complex variables $z$ and $p$ rather than by the pairs $(u,v)$ and $(m,n)$ because this notation is more compact and because holomorphicity will be paramount to us.\footnote{For a given value of $\tau$, we shall indiscriminately refer to $z$ or to $(u,v)$ as {\sl characteristics}, by a slight abuse of nomenclature. Our conventions for the sign of the exponent in (\ref{basic.3}) agree with the conventions used in \cite{DHoker:2018mys,Ramakrish} but are opposite to \cite{Gerken:2018jrq, Gerken:2019cxz, Gerken:2020yii}.} Thus, we label the characters as follows,
\begin{align}
\chi_p(z|\tau) = e^{ 2\pi i(nu-mv)} =  \exp \left( \frac{2\pi i}{\tau-\bar\tau} (  \bar{p} z - p\bar{z}) \right) 
\label{basic.3}
\end{align}
The character is independent of $\tau$ when expressed in terms of $(u,v)$ and $(m,n)$ variables, but its $\tau$-dependence must be included when formulated in terms of the variables $z$ and~$p$. As a character either of a representation $\Lambda \to U(1)$ or of a representation $\Sigma \to U(1)$, $\chi$ satisfies the following group-theoretic relations,
\begin{align}
 \chi_{p_1+p_2}(z|\tau )  & =  \chi_{p_1}(z|\tau ) \,  \chi_{p_2}(z|\tau ) 
\no \\
\chi_{p}(z_1+z_2 |\tau ) & =  \chi_{p}(z_1|\tau ) \, \chi_{p}(z_2|\tau ) 
\no \\
\overline{ \chi_{p}(z|\tau )}  & =   \chi_{-p}(z|\tau ) = \chi_{p}(-z|\tau ) 
\label{basic.4}
\end{align} 
and the characters of the identity elements are given by $\chi_0(z|\tau) =\chi_p(0|\tau)=1$. The character is modular invariant when all its arguments are transformed according to (\ref{modPSL}),
\bea
\label{chimod}
\chi_{p'} (z'|\tau') = \chi_p (z|\tau)
\eea
a result that is even more transparent when the character is expressed in terms of the real pairs $(u,v)$ and $(m,n)$ due to the fact that we have  $n'u'-m'v'=nu-mv$ when these variables are related by the modular transformations given for real variables in (\ref{modPSL1}).

\sm

The functions $g_k(z|\tau)$, which were  introduced in (\ref{gk}), may be readily expressed in terms of the character $\chi$,
\bea
g_k(z|\tau) =  { (\Im \tau)^k \over \pi^k  }  \sum _{ p \in \Lambda '} {  \chi _p (z|\tau) \over  |p|^{2k} } 
\eea
They  may be generalized by assigning independent values to the exponents of $p$ and $\bar p$ in the sum over $p \in \Lambda'$, which leads us to introduce the following combinations,
\beq
\dplus{a \\ b}\!(z|\tau)
=  \frac{ (\Im \tau)^a}{\pi^b} \sum_{p \in \Lambda'} \frac{ \chi_{p}(z|\tau ) }{p^a \bar p^b}
\label{basic.13}
\eeq
For $a=b$, these functions reduce to the modular functions $g_a(z|\tau)$ defined in~\eqref{gk}. When $a\not= b$,  there is no power of $\Im \tau$ by which they can be normalized canonically. The normalization chosen here (and indicated with the $+$ subscript) is such that their modular weight $(0,b-a)$ has vanishing holomorphic part so that the forms transform as follows,  
\bea
\label{Dplus1}
\dplus{a \\ b}  \left (z' | \tau' \right ) = (\gamma \bar \tau + \delta)^{b-a} \, \dplus{a \\ b}\!(z|\tau)
\eea
where the transformations of $z$ and $\tau$ are given in (\ref{modPSL}).
They are multiples of the single-valued elliptic polylogarithms $D_{a,b}(z|\tau)$ introduced by Siegel and Zagier
\cite{Ramakrish} which are related to our normalization by,
\beq
D_{a,b}(z|\tau) = \frac{ (\tau{-}\bar \tau)^{a+b-1} }{2\pi i} 
\sum_{p \in \Lambda'} \frac{ \chi_p(z|\tau ) }{p^a \bar p^b}  = (2i)^{a+b-2}(\pi \Im \tau)^{b-1} \dplus{a \\ b}\!(z|\tau)
\label{basic.14}
\eeq
In view of this relation, it should be expected more generally that elliptic  modular graph functions and forms will be related to single-valued elliptic multiple polylogarithms.

%%%%%%%%%%%%%%%%%%%%%%%%%%%%%%%%%%%%%%%%%%%%%%%%%%%%%%%%%%%
\subsection{Kronecker--Eisenstein series and coefficients}
\label{sec:2.1.1}
%%%%%%%%%%%%%%%%%%%%%%%%%%%%%%%%%%%%%%%%%%%%%%%%%%%%%%%%%%%

Another important building block and prototype  for eMGFs will be the Kronecker--Eisenstein series, defined in terms of $\tet$-functions by,
\bea
\label{Omega}
\Omega(z,\eta|\tau) = \exp \Big( 2\pi i \eta\, \frac{ \Im z }{\Im \tau} \Big)
\frac{ \tet_1'(0|\tau) \,  \tet_1(z{+}\eta|\tau) }{\tet_1(z|\tau) \, \tet_1(\eta|\tau)}
\eea
The function $\Omega(z,\eta|\tau)$ is defined for $z, \eta \in \CC$. It is meromorphic in $\eta$ with simple poles at $\eta \in \Lambda$, and transforms with phase factors under both $\eta \to \eta +1$ and $\eta \to \eta +\tau$. By contrast, as a function of $z$ it is invariant under $z \to z+\Lambda$, but fails to be meromorphic due to its exponential prefactor. The function $\Omega(z,\eta|\tau)$ is given by the following Kronecker--Eisenstein sum over either $\Lambda$ or $\Lambda'$ in terms of the character $\chi$ in (\ref{basic.3}),
\bea
\Omega(z,\eta|\tau)  = \sum_{p \in \Lambda} \frac{ \chi_p(z|\tau ) }{\eta-p}
= \frac{1}{\eta}  + \sum_{p \in \Lambda'} \frac{ \chi_p(z|\tau ) }{\eta-p} 
\label{looprev.4}
\eea
This series is conditionally convergent and is defined here using the Eisenstein summation
convention. 

\sm

From its definition, the Kronecker--Eisenstein series has the transformation law,
\begin{align}
\label{Om:trm}
\Omega\left( \frac{z}{\gamma\tau+\delta} , \frac{\eta}{\gamma\tau+\delta} \middle| \frac{\alpha\tau+\beta}{\gamma\tau+\delta}\right) = (\gamma\tau+\delta) \Omega(z,\eta|\tau)
\end{align}
under $SL(2,\ZZ)$.
The Laurent expansion of $\Omega(z,\eta|\tau)$ in the variable $\eta$ produces 
the Kronecker--Eisenstein coefficient functions $f^{(k)}(z|\tau)$, 
\bea
\Omega(z,\eta|\tau) =  \sum_{k=0}^{\infty} \eta^{k-1} f^{(k)}(z|\tau)
\eea
where $f^{(0)} (z|\tau)=1$ while for $k \geq 1$ we have, 
\bea
f^{(k)}(z|\tau) = - \sum_{p \in \Lambda'} \frac{ \chi_p(z) }{p^k}  
\label{looprev.6}
\eea
The following equivalent expressions may be derived for $f^{(1)} (z|\tau)$,  
\bea
f^{(1)}(z|\tau) = \partial_z \log \tet_1(z|\tau) + 2\pi i \frac{ \Im z }{\Im \tau}  
=  - \p_z g(z|\tau)
\label{looprev.7}
\eea
signaling that this function is invariant under $z \to z +\Lambda$ as expected, and has a simple pole in $z$ for all $z \in \Lambda$. The function $f^{(1)}(z|\tau)$ enters string theory either as the Green function for the $(b,c)$ system of weight 0, or as the Green function for a world-sheet fermion with odd-odd spin structure. The appearance of the non-holomorphic addition results from the presence of zero modes for each of these situations. 

\sm

For $k \geq 2$, the functions $f^{(k)} (z|\tau)$ are all invariant under $z \to z +\Lambda$ as expected, without poles. For example, we have (with a prime denoting derivatives of $\tet_1$ in the first argument),
\bea
\label{f2}
f^{(2)}(z|\tau) = \frac{1}{2} \Big\{ f^{(1)} (z|\tau)^2  + \partial_z^2  \log \tet_1(z|\tau) 
- \frac{   \tet'''_1(0|\tau) }{3 \tet'_1(0|\tau)} \Big \}
= - 2\pi i \partial_\tau g(z|\tau)
\eea
where the $\tau$-derivative is performed at constant $(u,v)$ rather than at constant $z$,
see early section \ref{sec:taudervs}. From~\eqref{Om:trm} we deduce that,
\begin{align}
f^{(k)} (z'|\tau') = (\gamma\tau+\delta)^k f^{(k)}(z|\tau)
\end{align}
under the transformation~\eqref{modPSL}.

\sm

The modular forms introduced in (\ref{Dplus1}) may be expressed as convolutions of $f$ and $\bar f$,
\beq
\int_{\Sigma} \frac{ \dd^2 w}{\Im \tau} \,  f^{(a)}(z{-}w|\tau) \overline{ f^{(b)}(w|\tau)} 
=  { (-\pi)^b \over (\Im \tau)^a} \, \dplus{a \\ b}\!(z|\tau)
\label{basic.13.0}
\eeq
which will serve as a prototype for the construction of higher elliptic modular graph forms.

%%%%%%%%%%%%%%%%%%%%%%%%%%%%%%%%%%%%%%%%%%%%%%%%%%%%%%%%%%%
\subsection{Elliptic modular graph functions and forms}
\label{sec:2.1.2}
%%%%%%%%%%%%%%%%%%%%%%%%%%%%%%%%%%%%%%%%%%%%%%%%%%%%%%%%%%%

In this section we shall introduce general eMGFs.
An eMGF maps a graph $\Gamma$ to a non-holomorphic elliptic function depending on the modulus $\tau \in \cH_1$, and on a set of characteristics $(u_r,v_r)$ or equivalently  points $z_r = u_r \tau+ v_r$ on $\Sigma$. An eMGF may be represented by a multiple sum, with characters, over the lattice $\Lambda$ or $\Lambda '$ that we shall refer to as a multiple Kronecker--Eisenstein sum (MKES), or as a multiple integral over products of Kronecker--Eisenstein coefficient functions $f^{(k)}(z|\tau)$ and the functions $g_k(z|\tau)$.  

\sm

In either case, the construction of eMGFs generalizes the construction of MGFs by the inclusion of a character. The  generalization of a character to a product of $R$ copies of  $\Lambda$ and $\Sigma $ is obtained in terms of the characters on each copy by multiplication,
\bea
\Sigma ^R \to U(1) : ~ (z_1, \cdots, z_R) \to \chi_{p_1}(z_1|\tau ) \times \cdots \times \chi_{p_R} (z_R|\tau )
\label{basic.6}
\eea
where the characteristics $z_r = u_r \tau + v_r$ for $r=1,\cdots, R$ are allowed to be independent of one another.
Using the integration~\eqref{basic:int} over $\Sigma$ such product over characters satisfy the following integral formula,
\beq
\int_{\Sigma}  \frac{ \dd^2 z }{\Im \tau}  \bigg( \prod_{r=1}^R \chi_{p_r}(z_r{-}z|\tau) \bigg) = \delta\bigg( \sum_{s=1}^R p_s \bigg) \bigg( \prod_{r=1}^R \chi_{p_r}(z_r|\tau) \bigg)
\label{basic.12}
\eeq
The Kronecker $\delta$ equals 1 when the sum $\sum_s m_s= \sum_s n_s=0$ and vanishes otherwise.

\subsubsection{Definition and properties of dihedral eMGFs}

We begin by introducing {\sl dihedral} eMGFs in terms of a \MKES~over $R$ edges with momenta $p_r \in \Lambda '$ for $r=1,\cdots, R$, two vertices of valence greater or equal to 3, and an arbitrary number of bivalent vertices. The exponents $a_r$ of  the holomorphic momenta $p_r$, the exponents $b_r$ of the anti-holomorphic momenta $\bar p_r$, and the characteristics $z_r$ are arranged in $R$-dimensional arrays $A$, $B$ and $Z$, respectively, 
\beq
\begin{array}{l}
A=[a_1,a_2,\ldots,a_R]  \\
B=[b_1,\, \hspace{-0.2mm}b_2,\ldots,\, \hspace{-0.2mm}b_R] \\
Z = [z_1, z_2, \cdots, z_R] 
\end{array} \hskip 0.5in
|A| = \sum_{r=1}^R a_r
\hskip 0.5in
|B| = \sum_{r=1}^R b_r
\label{basic.15}
\eeq
where $a_r,b_r \in \mathbb Z$.
The associated dihedral eMGF is defined by the following \MKES,
\beq
\cplus{A \\ B \\ Z}\! (\tau) 
= { (\Im \tau)^{|A|} \over \pi^{ |B|}}  \sum _{p_1,\ldots ,p_R \in \Lambda '} \delta \bigg(\sum_{s=1}^R p_s \bigg) \, 
\prod _{r=1}^R { \chi_{p_r}(z_r|\tau)  \over p_r ^{a_r} \bar p_r^{b_r}}
\label{gen.66}
\eeq
which is absolutely convergent if $a_r+a_{r'}+b_r+b_{r'} >2$ for any pair $1\leq r,r' \leq R$ and defined by Eisenstein summation convention in case of conditional convergence.

\sm

Clearly, the dihedral eMGF (\ref{gen.66}) is invariant under simultaneous permutations of the $(a_r,b_r,z_r)$ in $A,B,Z$.
Its modular transformation properties follow from the modular invariance of the characters shown in (\ref{chimod}) and the customary transformation laws of $p_r$ and $\tau$ given in (\ref{modPSL}), 
\beq
\cplus{A \\ B \\ Z' }\!(\tau')  = (\gamma \bar \tau + \delta)^{|B|-|A|} \cplus{A \\ B \\ Z}\! (\tau)
\label{basic.27}
\eeq
where $Z'=[z_1', \cdots, z_R']$ with $z_r' = z_r/(\gamma \tau +\delta)$ for $r=1,\cdots, R$, as given in (\ref{modPSL}).
In view of this modular transformation law, one may view dihedral eMGFs as non-holomorphic Jacobi forms of weight $(0,|B|{-}|A|)$
and vanishing index.  For $|A|\not= |B|$ there is no canonical choice of the power of $\Im \tau$ and the normalization chosen here (indicated again with the superscript $+$) is such that the modular weight $(0, |B|{-}|A|)$ has vanishing holomorphic entry. The eMGF with conjugate normalization  (denoted with the superscript~$-$) is related as follows,
\bea
\cminus{A \\ B \\ Z }\!(\tau)  = \left ( \pi \Im \tau \right ) ^{|B|-|A|} \cplus{A \\ B \\ Z}\! (\tau) = 
\overline{\cplus{B \\ A \\ - Z}\! (\tau)} 
\label{basic.27-}
\eea
and has modular weight $(|A|-|B|,0)$. Additionally, we have a reflection symmetry,\footnote{Henceforth, when the dependence on $\tau$ is clear from the context we shall omit $\tau$.}
\bea
\cplus{A \\ B \\ -Z}  = (-1)^{|A|+|B|} \cplus{A \\ B \\ Z} 
\label{basic.22}
\eea
and momentum conservation,
\beq
\sum_{r=1}^R \cplus{A-S_r \\ B \\ Z} = \sum_{r=1}^R \cplus{A \\ B-S_r \\ Z}  = 0\, ,
\hskip 0.5in
  \cplus{A \\ B \\ Z-z S}  =  \cplus{A \\ B \\ Z} 
\label{basic.23}
\eeq
where  $S=[1,1,\cdots, 1]$, the parameter $z\in \mathbb C$ is arbitrary, and $S_r= [0, \cdots 0, \, 1, \, 0, \cdots ,0]$ has zeros in every entry except for the entry $r$ where the value is 1. The last identity expresses translation invariance  on $\Sigma$ and uses the fact that $ \prod_{j=1}^R \chi_{p_j}(z|\tau )=1$ on the support of momentum conservation.

\sm

Equivalently, we may express an eMGF as an integral over $\Sigma$ of a product of the modular forms $\dplus{a\\b}\!(z|\tau)$ defined by (\ref{basic.13}).  The equivalence with the \MKES~representation may be established by using the integral formula on characters given in (\ref{basic.12}). In the dihedral case we have,
\beq
 \cplus{A \\ B \\ Z}\! (\tau) =
  \int_\Sigma \frac{ \dd^2 z}{\Im \tau} \prod_{r=1}^R \dplus{a_r \\ b_r}\!(z_r{-}z|\tau)
 \label{basic.41}
\eeq
where $A$, $B$ and $Z$ are the arrays of $a_r$, $b_r$ and $z_r$ given in (\ref{basic.15}).
Just as for MGFs, any eMGF with $R=1$ vanishes,
\bea
\label{R=1}
\cplus{a\\b\\z}\!(\tau)=0
\eea
Making use of the special case of (\ref{basic.13}), 
\beq
\dplus{0 \\ 0}\!(z|\tau)
= \sum_{p \in \Lambda'} \chi_p(z|\tau) = (\Im \tau)  \delta^2(z,\bar z) - 1
\label{basic.42}
\eeq
the integral representation (\ref{basic.41}) implies a simple algebraic identity when a  column of exponents  in the eMGF has vanishing entries. In this case (\ref{basic.42}) implies the  identity,
\beq
 \cplus{a_1&a_2 &\ldots &a_R &0 \\ b_1&b_2 &\ldots &b_R &0 \\ z_1&z_2 &\ldots &z_R &0}
 = \bigg( \prod_{r=1}^R   \dplus{a_r\\b_r}\!(z_r)  \bigg) 
 -  \cplus{a_1&a_2 &\ldots &a_R \\ b_1&b_2 &\ldots &b_R \\ z_1&z_2 &\ldots &z_R}
 \label{basic.43}
\eeq
This relation is the generalization of the {\sl algebraic reduction} formulas  for MGFs \cite{DHoker:2016mwo}.

\subsubsection{Definition and properties of  eMGFs for arbitrary graphs}

Next, we generalize eMGFs to the case of an arbitrary graph by starting  from MGFs for arbitrary graphs and multiplying the summand in the Kronecker--Eisenstein sum by a character. The resulting general formulation is as follows. 

\sm

A decorated graph $(\Gamma, \mathcal{A},\mathcal{B},\mathcal{Z})$ with $V$ vertices and $R$ directed edges consists of a connectivity matrix with components\footnote{The components of the connectivity matrix are $\Gamma _{v \, r}=+1$ ($\Gamma _{v \, r}=-1$) if the directed edge $r$ is incoming (outgoing) for the vertex $v$ and vanish otherwise. We employ the notation $\mathcal{A}$ etc. to distinguish the case of general topology from the dihedral topology where we write $A$.} $\Gamma _{v \, r}$, for $v =1, \cdots, V$ and $r = 1,\cdots, R$ and decoration of the edges by exponents $a_r, b_r \in \mathbb Z$ and characteristics $z_r$ for $r = 1,\cdots, R$, assembled into arrays $\mathcal{A}=[a_1, \cdots, a_R]$, $\mathcal{B}=[b_1, \cdots, b_R]$ and $\mathcal{Z}=[z_1, \cdots, z_R]$. To the decorated graph {$(\Gamma, \mathcal{A},\mathcal{B},\mathcal{Z})$ we associate a function on $ \Sigma ^R \times \cH_1$ by,
\bea
\cplus{\mathcal{A}\\\mathcal{B}\\\mathcal{Z}} \!(\tau) 
= { (\Im \tau)^{|\mathcal{A}|} \over \pi ^{|\mathcal{B}|}} \sum _{p_1, \ldots, p_R \in \Lambda '} 
\left ( \prod _{r=1}^R { \chi_{p_r} (z_r|\tau)  \over (p_r)^{a_r} \, (\bar p_r )^{b_r}} \right )
\prod _{v=1}^V \delta \left ( \sum_{r=1}^R \Gamma _{v \, r} p_r \right )
\label{allemgfs}
\eea
where
\bea
|\mathcal{A}| = \sum _{r=1}^R a_r \hskip 1in |\mathcal{B}| = \sum _{r=1}^R b_r
\eea
A sufficient condition for absolute convergence of~\eqref{allemgfs} is that for any two edges $1\leq r,r'\leq R$ the sum of weights $a_r+b_r+a_{r'}+b_{r'}>2$. The normalization by powers of $(\Im \tau) $ is canonical only when $|\mathcal{A}|=|\mathcal{B}|$ and otherwise has been chosen so that the modular weight has vanishing holomorphic entry. The modular transformation law if given as follows, 
\bea
\label{emgf.mod}
\cplus{\mathcal{A}\\\mathcal{B}\\\mathcal{Z}'} \!(\tau') 
= ( \gamma \bar \tau + \delta  )^{|\mathcal{B}|-|\mathcal{A}|} \, 
\cplus{\mathcal{A}\\\mathcal{B}\\\mathcal{Z}} \!(\tau) 
\eea
where $\tau'$  and $\mathcal{Z}'=[z_1', \cdots, z_R']$ are given in (\ref{modPSL}). The momentum conservation identities take the following form, 
\bea
\sum_{r=1}^R \Gamma _{v \, r} \, \cplus{\mathcal{A}-S_r \\\mathcal{B}\\\mathcal{Z}}\! (\tau) 
= \sum_{r=1}^R \Gamma _{v \, r} \, \cplus{\mathcal{A} \\\mathcal{B} -S_r \\\mathcal{Z}}\! (\tau) 
= 0
\eea
and translation invariance generalizes to,
\beq
\cplus{\mathcal{A}\\\mathcal{B}\\\mathcal{Z}_s } \!(\tau)  = \cplus{\mathcal{A}\\\mathcal{B}\\\mathcal{Z}} \!(\tau) 
\hskip 0.8in
\mathcal{Z}_s=\mathcal{Z}- z\sum_r \Gamma_{v\, r}  S_r
\eeq
see (\ref{basic.23}) for their dihedral instances. A detailed discussion of trihedral eMGFs can
be found in appendix \ref{app:tri}.  Any convergent eMGF grows at most polynomially when $\tau\to i \infty$, see appendix C of \cite{DHoker:2018mys} for examples, similar to the case of MGFs~\cite{Zerbini:2015rss}.

\sm

According to~\eqref{emgf.mod}, the modular weight of an eMGF is given by $(0,|\mathcal{B}|-|\mathcal{A}|)$ that we shall often refer to \textit{weight} for simplicity. The graph $\Gamma$ has a definite loop order $L$ and we transfer this notion to the eMGF by saying that the eMGF has a loop order $L$. For instance, the dihedral eMGF in (\ref{gen.66}) with $R$ edges is said to have loop order $L=R{-}1$. One of the themes of this paper will be that there are algebraic relations between eMGF that relate different eMGFs of potentially different loop order up to $\tau$-independent functions. Thus, the notion of loop order of an eMGF can change when representing it differently. 

\sm

A different notion, called \textit{depth} of an eMGF, will be introduced in section~\ref{sec:sieve}. It is related to differential relations satisfied by eMGFs and leads to irreducible iterated integral representations with Kronecker--Eisenstein functions $f^{(k)}$ as integration kernels. The depth of an eMGF can differ from the loop order and we shall argue that it is a more intrinsic property of eMGFs.

%%%%%%%%%%%%%%%%%%%%%%%%%%%%%%%%%%%%%%%%%%%%%%%%%%%%%%%%%%%
\subsection{One-loop  eMGFs}
\label{sec:2.0}
%%%%%%%%%%%%%%%%%%%%%%%%%%%%%%%%%%%%%%%%%%%%%%%%%%%%%%%%%%%

In this section, we provide additional relations between the functions $\cC^+$, $\cD^+$, $f^{(k)}$ and $g_k$ in the special case of  one-loop graphs. We collect here  the following special cases of the functions $\cD^+$ which will be useful,
\bea
 \label{gen.67}
\dplus{0\\0}\! (z|\tau) & = & (\Im \tau) \delta ^2(z) -1
\no \\
\dplus{1 \\ 1}\! (z|\tau) & = & g(z|\tau)
\no \\
\dplus{k \\ k}\! (z|\tau) & = & g_k(z|\tau)
\no \\
\dplus{a \\0} \!(z|\tau) & = & - (\Im \tau)^a  \,  f^{(a)}(z|\tau) 
\no \\
\dplus{0 \\b} \!(z|\tau) & = & (-1)^{b-1}  \pi^{-b} \, \overline{ f^{(b)}(z|\tau) }
\eea
Further specialization to vanishing argument $z$ for $k\geq 2$ and $a,b\geq 3$  gives, 
\bea
 \label{gen.670}
\dplus{k \\ k}\! (0|\tau) & = & g_k(0|\tau) = E_k(\tau)
\no \\
\dplus{a \\0} \!(0|\tau) & = & - (\Im \tau)^a  \,  f^{(a)}(0|\tau) = (\Im \tau)^a \, G_a(\tau)
\no \\
\dplus{0 \\b} \!(0|\tau) & = & (-1)^{b-1}  \pi^{-b} \, \overline{ f^{(b)}(0|\tau) } = \pi^{-b} \overline{G_b(\tau)}
\eea
The holomorphic Eisenstein series $G_a(\tau)$ is a modular form of weight $(a,0)$ defined by,
\bea
G_a(\tau) = \sum _{p\in \Lambda'} { 1 \over p^a}
\eea
The series is absolutely convergent for $a\geq 3$ and vanishes for all odd $a \geq 3$. For $a=2$ the conditionally convergent series may be regularized preserving modular invariance while giving up meromorphicity, to produce the non-holomorphic but modular Eisenstein series $\widehat G_2$, which is related to $f^{(2)}$ and $\cD^+$ as follows,
\bea
\label{G2hat}
\widehat G_2 (\tau) =  (\Im \tau)^{-2} \dplus{2 \\0} \!(0|\tau) = - f^{(2)}(0|\tau)
\eea
and $f^{(2)}(0|\tau)$ is the finite limit as $z \to 0$ of the formula given in (\ref{f2}) for $f^{(2)}(z|\tau)$. 

 \sm

The simplest non-vanishing examples of dihedral eMGFs (\ref{gen.66}) have $R=2$ and thus two columns
since one-column eMGFs vanish by (\ref{R=1}). Based on (\ref{basic.22}) and (\ref{basic.23}), one can rearrange the eMGF so that the second column has vanishing entries,
\begin{align}
\cplus{a_1 &a_2 \\ b_1 &b_2 \\ z_1 &z_2}\! (\tau) 
&= (-1)^{a_2+b_2} \, \cplus{a_1+a_2 &0 \\ b_1+b_2 &0 \\ z_1-z_2 &0}\! (\tau) \label{basic.31} 
\end{align}
%

%%%%%%%%%%%%%%%%%%%%%%%%%%%%%%%%%%%%%%%%%%%%%%%%%%%%%%%%%%%
\subsection{Two-loop  eMGFs}
\label{sec:2.2loop}
%%%%%%%%%%%%%%%%%%%%%%%%%%%%%%%%%%%%%%%%%%%%%%%%%%%%%%%%%%%

The simplest generalizations of Zagier's single-valued elliptic polylogarithms are two-loop
eMGFs with $R=3$ columns in (\ref{gen.66}),
\beq
\cplus{a_1 &a_2 &a_3 \\ b_1 &b_2 &b_3 \\ z_1 &z_2 &z_3}\! (\tau) 
= { (\Im \tau)^{a} \over \pi^{ b}}  \sum _{p_1,p_2,p_3 \in \Lambda '}  {\delta(p_1{+}p_2{+}p_3) \chi_{p_1}(z_1|\tau)\chi_{p_2}(z_2|\tau)\chi_{p_3}(z_3|\tau)  \over p_1 ^{a_1} \bar p_1^{b_1} 
p_2 ^{a_2} \bar p_2^{b_2}p_3 ^{a_3} \bar p_3^{b_3}}
\label{twolp.1}
\eeq
where we denote the two-loop instances of $|A|$ and $|B|$ by
\beq 
 a=a_1+a_2+a_3\, , \ \ \ \ \ \ b=b_1 +b_2 +b_3
 \label{twolp.2}
 \eeq
Similar to (\ref{basic.31}), one can bring the two-loop eMGFs (\ref{twolp.1}) into
a canonical form with one vanishing entry among both the $a_j$ and $b_j$: 
Following the strategy in section 2.6 of \cite{DHoker:2019txf}, the vanishing
of exponents of $p_r,\bar p_r$ can be enforced by the partial-fraction decompositions
\begin{align}
\frac{1}{p_1^{a_1}p_2^{a_2} p_3^{a_3}} &= \sum_{k=1}^{a_1}  \frac{ \lambda_k(a_1,a_2)  }{p_1^k  \, p_3^{a-k}}
+\sum_{k=1}^{a_2}  \frac{ \lambda_k(a_2,a_1)  }{p_2^k  \, p_3^{a-k}}
\notag \\
\frac{1}{\bar p_1^{b_1}\bar p_2^{b_2} \bar p_3^{b_3}} &= \sum_{\ell=1}^{b_1}  \frac{ \lambda_\ell(b_1,b_3)  }{ \bar p_1^\ell  \, \bar p_2^{b-\ell}}
+\sum_{\ell=1}^{b_3}  \frac{ \lambda_\ell(b_3,b_1)  }{\bar p_3^\ell  \, \bar p_2^{b-\ell}}
 \label{twolp.3}
\end{align}
that are valid when $p_1+p_2+p_3=0$ and where
\beq
\lambda_k(a_1,a_2) = (-1)^{a_1+a_2+k} \tbinom{a_1+a_2-k-1}{a_2-1} 
 \label{twolp.4}
\eeq
Straightforward insertion of (\ref{twolp.3}) into (\ref{twolp.1}) leads to the decomposition
into two-loop eMGFs of canonical form with entries $\begin{smallmatrix} c_1 &0 &c_3 \\ 0 &d_2 &d_3 \end{smallmatrix}$ in the first two rows:
\begin{align}
\cplus{a_1 &a_2 &a_3 \\ b_1 &b_2 &b_3 \\ z_1 &z_2 &z_3}\! (\tau) = \ &
 \sum_{k=1}^{a_1} \sum_{\ell=1}^{b_1} \lambda_k(a_1,a_2) \lambda_\ell(b_1,b_3) \cplus{a-k &0 &k \\ 0 &b-\ell &\ell \\ z_3 &z_2 &z_1}\! (\tau) \notag \\
 &+ \sum_{k=1}^{a_1} \sum_{\ell=1}^{b_3} \lambda_k(a_1,a_2) \lambda_\ell(b_3,b_1) \cplus{k &0 &a-k \\ 0 &b-\ell &\ell \\ z_1 &z_2 &z_3}\! (\tau)
 \notag \\
 &+ \sum_{k=1}^{a_2} \sum_{\ell=1}^{b_1} \lambda_k(a_2,a_1) \lambda_\ell(b_1,b_3) \cplus{a-k &0 &k \\ 0 &\ell &b-\ell \\ z_3 &z_1 &z_2}\! (\tau) 
  \label{twolp.5} \\
 &+ \sum_{k=1}^{a_2} \sum_{\ell=1}^{b_3} \lambda_k(a_2,a_1) \lambda_\ell(b_3,b_1) \cplus{0 &k &a-k \\ 0 &b-\ell &\ell \\ z_1 &z_2 &z_3}\! (\tau)
\notag
\end{align}
The last line can be reduced to one-loop eMGFs (\ref{basic.31}) by the algebraic
reduction (\ref{basic.43}),\footnote{The term in the last line stems from carrying out the sum over $k$ and $\ell$  via
\[
\sum_{k=1}^{a_2} (-1)^{k} \lambda_k(a_2,a_1) = (-1)^{a_1+a_2} \frac{(a_1{+}a_2{-}1)! }{a_1!(a_2{-}1)!}
\]}
\begin{eqnarray}
&& \sum_{k=1}^{a_2} \sum_{\ell=1}^{b_3} \lambda_k(a_2,a_1) \lambda_\ell(b_3,b_1) \cplus{0 &k &a-k \\ 0 &b-\ell &\ell \\ z_1 &z_2 &z_3}\! (\tau) \notag \\
&=&\sum_{k=1}^{a_2} \sum_{\ell=1}^{b_3} \lambda_k(a_2,a_1) \lambda_\ell(b_3,b_1) 
\cplus{k &0 \\ b-\ell &0 \\ z_{21} &0}\! (\tau)
\cplus{a-k &0 \\ \ell &0 \\ z_{31} &0}\! (\tau)
  \label{twolp.6}\\
&& - (-1)^{a_3+b_1+b_3} \frac{ (a_1{+}a_2{-}1)!\,(b_1{+}b_3{-}1)!}{a_1! \, (a_2{-}1)! \, b_1! \, (b_3{-}1)!}
\cplus{a &0 \\ b &0 \\ z_{23} &0 }\! (\tau) \notag
\end{eqnarray}

\newpage

%%%%%%%%%%%%%%%%%%%%%%%%%%%%%%%%%%%%%%%%%%%%%%%%%%%%%%%%%%%
%%%%%%%%%%%%%%%%%%%%%%%%%%%%%%%%%%%%%%%%%%%%%%%%%%%%%%%%%%%
\section{Differential identities and HSR}
\label{sec:2A}
%%%%%%%%%%%%%%%%%%%%%%%%%%%%%%%%%%%%%%%%%%%%%%%%%%%%%%%%%%%
%%%%%%%%%%%%%%%%%%%%%%%%%%%%%%%%%%%%%%%%%%%%%%%%%%%%%%%%%%%

In this section, we shall show that the derivatives of eMGFs, either  in the modulus $\tau$ or in the characteristics $z_r$, produce linear combinations of eMGFs. For certain configurations of vanishing anti-holomorphic exponents, subgraphs arise which involve only holomorphic momenta. The procedure of holomorphic subgraph reduction, which is designed to simplify the Kronecker--Eisenstein sums is such cases, is generalized here from the case of dihedral MGFs introduced in \cite{DHoker:2016mwo} and generalized in \cite{Gerken:2018zcy} to  trihedral MGFs.  A simplified derivation of holomorphic subgraph reduction is presented here in terms of Kronecker--Eisenstein series and the use of the Fay identity, following \cite{Gerken:2020aju}. A major motivation for the study of eMGFs is to generalize the system of differential equations obeyed by Zagier's single-valued elliptic polylogarithms to higher depth.

\subsection{Derivatives in $\tau$}
\label{sec:taudervs}

As in the case of configuration space integrals, the most useful definition of the derivative with respect to $\tau$ (or with respect to $\bar \tau$) is obtained by keeping the co-moving coordinates $(u,v)$ of $z=u\tau+v$ and the co-moving momenta $(m,n)$ of $p=m\tau+n$ fixed.
Inspection of (\ref{basic.3}) then readily confirms that $\chi_p (z|\tau)$ is then independent of $\tau$ and $\bar \tau$ so that its derivatives in $\tau$ and $\bar \tau$ vanish,
\bea
\label{chi-tau}
\p_\tau \chi_p (z|\tau) = \p _{\bar \tau} \chi _p (z|\tau) =0 
\hskip 0.7in 
\hbox{ for } (u,v) \hbox{ and } (m,n) \hbox{ fixed}
\eea
We shall define the covariant derivative $\nabchi_\tau$, or Maass operator,  mapping
the space of modular forms of weight $(0,\mu)$
to those of weight $(0,\mu{-}2)$ while keeping all co-moving coordinates $(u,v)$ in $\Sigma$ and all co-moving momenta $(m,n)$ in $\Lambda$ fixed,  as follows,
\bea
\nabchi _\tau = 2 i (\Im \tau)^2 \p_{\tau} \hskip 0.7in \hbox{ for } (u,v) \hbox{ and } (m,n) \hbox{ fixed}
\label{crdrv}
\eea
Its complex conjugate $\overline{\nabchi}_{ \tau} = - 2 i (\Im \tau)^2 \p_{\bar \tau}$ maps
the space of modular forms of weight $(\mu, 0)$ to those of weight $(\mu{-}2,0)$ while keeping $(u,v)$ and  $(m,n)$ fixed.  In the following it is always understood that $(u,v)$ and $(m,n)$ are kept fixed in $\tau$-derivatives unless stated otherwise.

\sm

Since we have $\nabchi _\tau \chi_p(z|\tau)=0$, the relations giving the $\tau$-derivatives of MGFs \cite{DHoker:2016mwo} carry over almost verbatim to arbitrary eMGFs, and we have, 
\bea
\pi \nabchi_\tau \, \cplus{\mathcal{A} \\ \mathcal{B} \\\mathcal{Z}}\! (\tau)  & = & \sum_{r=1}^R a_r \, \cplus{\mathcal{A}+S_r \\ \mathcal{B}-S_r\\ \mathcal{Z}}\! (\tau) 
\no \\
\pi \overline{\nabchi}_{\tau} \, \cminus{\mathcal{A} \\ \mathcal{B} \\\mathcal{Z}}\! (\tau)  & = & \sum_{r=1}^R b_r \, \cminus{\mathcal{A}-S_r \\ \mathcal{B}+S_r\\ \mathcal{Z}}\! (\tau) 
\label{gen.66Ctau}
\eea
where $S_r$ was introduced below (\ref{basic.23}) and the second equation may be obtained from the first by complex conjugation and  suitable change of normalization by factors of $(\Im \tau)$. For one-loop eMGFs, the general equations (\ref{gen.66Ctau}) reduce as follows,
\bea
\pi \nabchi_\tau \dplus{a  \\ b }\! (z|\tau) & = & a  \, \dplus{a+1  \\ b-1  }\! (z|\tau) 
\no \\
(\pi \nabchi_\tau)^m \, \dplus{a  \\ b }\! (z|\tau) & = & {(a{+}m{-}1)! \over (a{-}1)!}   \dplus{a+m  \\ b-m  }\! (z|\tau) 
\no \\
(\pi \nabchi_\tau)^m \, g_{k}(z|\tau) &=&   \frac{(k{+}m{-}1)!}{(k{-}1)!} 
\dplus{k+m  \\ k-m }\! (z|\tau)  
\eea
The first equation reproduces the well-known differential equations of Zagier's single-valued elliptic polylogarithms (\ref{basic.14}), 
\bea
\nabla_\tau D_{a,b}(z|\tau) = \Im \tau \big( a D_{a+1,b-1}(z|\tau) + (b{-}1) D_{a,b}(z|\tau) \big) 
\eea
Special cases giving Kronecker--Eisenstein coefficients  are as follows,
\begin{align}
(\pi \nabchi_\tau)^b \, \dplus{a  \\ b  }\! (z|\tau)&=  
- \frac{(a{+}b{-}1)!}{(a{-}1)!} (\Im \tau)^{a+b} f^{(a+b)}(z|\tau)
\no \\
(\pi \nabchi_\tau)^k \, g_{k}(z|\tau) & = - \frac{ (2k{-}1)! }{ (k{-}1)!} (\Im \tau)^{2k} f^{(2k)} (z|\tau) 
\label{diff.6}
\end{align}
When evaluated at $z=0$, we recover the following relations for one-loop MGFs,
\bea
(\pi \nabla_\tau)^m \, \dplus{a  \\ b }\! (0|\tau) & = & {(a{+}m{-}1)! \over (a{-}1)!}   \dplus{a+m  \\ b-m  }\! (0|\tau) 
\no \\
(\pi \nabla_\tau)^b \, \dplus{a  \\ b  }\! (0|\tau)&= & 
 \frac{(a{+}b{-}1)!}{(a{-}1)!} (\Im \tau)^{a+b} G_{a+b}(\tau)
\no \\
(\pi \nabla_\tau)^m \, E_k(\tau) &=&   \frac{(k{+}m{-}1)!}{(k{-}1)!} 
\dplus{k+m  \\ k-m }\! (0|\tau) 
\label{dpt1diff}
\eea
Analogous equations may be obtained for the action of $\overline{\nabchi}_{ \tau}$.

\subsection{Derivatives in $z$}

It will be convenient to us the following  covariant derivative in $z$ and $\bar z$, at fixed $\tau$ and $p$,
\bea
\nabla_z & = &  \partial_u - \bar \tau \partial_v =  2i (\Im \tau) \, \partial_z 
\no \\
\overline{\nabla}_{ z} & = & \partial _u - \tau \partial_v =  - 2 i (\Im \tau) \, \p_{\bar z} 
\label{diff.1}
\eea
Applied to the character, we obtain,
\bea
\nabla_z \chi_p(z|\tau) &= & 2\pi i \,  \bar{p} \, \chi_p(z|\tau) 
\no \\
\overline{\nabla} _{z} \chi_p(z|\tau) & = &  2 \pi i \, p \, \chi_p(z|\tau) 
\label{basic.5}
\eea
Note that the $\tau$-derivative $\nabchi_\tau$ in (\ref{crdrv}) for $(u,v)$ and $(m,n)$ fixed only commutes
with $\nabla_z$ (as will become important for the discussion of sections \ref{sec:3.6} and \ref{sec:4.comm}) 
but not with $\overline{\nabla}_{ z}$.

\sm

The action of $\nabla _z$ on an arbitrary eMGF will be obtained by differentiating the characters that appear in the eMGF according to (\ref{basic.5}). To obtain the general formula for this derivative, we must allow for the possibility that all characters $z_r$ depend on the variable $z$ through arbitrary linear combinations compatible with the group character structure of~$\chi$. Therefore, the general formula reads as follows,
\begin{align}
  \nabla_z \, \cplus{\mathcal{A} \\ \mathcal{B} \\ \mathcal{Z}} &=  2   i  \sum_{r=1}^R { \p z_r \over \p z} \, 
\cplus{\mathcal{A} \\ \mathcal{B}-S_r \\ \mathcal{Z}}
 \label{gen.66C}
\end{align}
For the special case where $s \leq R$ of the characteristics $z_r$ coincide with $z$ while the remaining $R-s$ characteristics are independent of $z$, we obtain the following simplified and more explicit expression,
\beq
 \nabla_z \, \cplus{a_1 &\ldots &a_s &a_{s+1} &\ldots &a_R \\ b_1 &\ldots &b_s &b_{s+1} &\ldots &b_R \\ z &\ldots &z &z_{s+1} &\ldots &z_R}
=  
2   i  \sum_{r=1}^s \cplus{a_1 &\ldots &a_r &\ldots &a_s &a_{s+1} &\ldots &a_R \\ b_1 &\ldots &b_r-1 &\ldots &b_s &b_{s+1} &\ldots &b_R \\ z &\ldots &z  &\ldots &z &z_{s+1} &\ldots &z_R}
\label{diff.2}
\eeq
This formula is valid for arbitrary eMGFs.

\sm

For one-loop eMGFs, (\ref{gen.66C}) simplifies to
\bea
(  \nabla_z)^m \dplus{a  \\ b }\! (z|\tau)  & = & (2  i)^m \, \dplus{a  \\ b-m }\! (z|\tau) 
\no \\
( \nabla_z)^m g_k(z|\tau) & = &  (2  i)^m \,  \dplus{k  \\ k-m }\! (z|\tau) 
\label{diff.3}
\eea
as well as
\bea
(  \nabla_z)^b \dplus{a  \\ b }\! (z|\tau)  & = & - (2  i)^b  (\Im \tau)^a f^{(a)}(z|\tau)
\no \\
( \nabla_z)^k g_k(z|\tau) & = &-  (2  i)^k (\Im \tau)^k f^{(k)}(z|\tau)
\label{diff.3ex}
\eea
which, using  (\ref{basic.14}),  reproduces,
\begin{align}
\nabla_z D_{a,b}(z|\tau) &=  - 4\pi \Im \tau D_{a,b-1}(z|\tau)
\end{align}
 the well-known differential equations of Zagier's single-valued elliptic polylogarithms.

%%%%%%%%%%%%%%%%%%%%%%%%%%%%%%%%%%%%%%%%%%%%%%%%%%%%%%%%%%%
\subsection{Holomorphic subgraph reduction}
\label{sec:2.2.2}
%%%%%%%%%%%%%%%%%%%%%%%%%%%%%%%%%%%%%%%%%%%%%%%%%%%%%%%%%%%

As we will see, the appearance of Kronecker--Eisenstein coefficients $f^{(k)}(z|\tau)$ as in (\ref{diff.6})
 is a general feature of iterated $\tau$-derivatives of eMGFs, regardless of the graph topology. 
In order to expose factors of $f^{(k)}(z|\tau)$ in the differential equations beyond the one-loop case, we will need 
the generalization to eMGFs for the procedure of {\sl holomorphic subgraph reduction} (HSR) for MGFs \cite{DHoker:2016mwo}. 

\sm

In the dihedral case, HSR is a prescription to resolve the simultaneous vanishing of two anti-holomorphic exponents, such as in,
\bea
\cplus{a_1&a_2 & A  \\ 0&0 & B   \\ z_1&z_2 &Z}
\eea
in terms of eMGFs of lower loop number. Convergence will be guaranteed when $a_1+a_2 \geq 3$ and $z_2 \not= z_1$, which we shall assume to hold at first.  These rearrangements are necessary since the evaluation of further $\nabla_z,\nabla_\tau$ via (\ref{gen.66Ctau}) and (\ref{gen.66C}) generically leads to negative entries $b_j$ that cannot be removed via momentum conservation (\ref{basic.23}).

\sm

While HSR for MGFs was initially performed based on partial-fraction manipulations of the lattice momenta $p_j$ in nested sums \cite{DHoker:2016mwo, Gerken:2018zcy}, a reformulation in terms of Fay identities among Kronecker--Eisenstein coefficients was later on given in \cite{Gerken:2020aju}. Here, we shall follow the latter approach which is based on the Fay identity \cite{Fay,BrownLev},
\begin{align}
  \Omega(z_1, \eta_1| \tau) \Omega(z_2,\eta_2|\tau)  = 
   \Omega(z_{1}{-}z_{2},\eta_1|\tau)\Omega(z_2,\eta_1{+}\eta_2|\tau)  
   + \Omega(z_{2}{-}z_{1}, \eta_2|\tau)\Omega(z_1,\eta_1{+}\eta_2|\tau)
  \label{diff.11}
\end{align}
Upon expanding each $\Omega$ in a Laurent series in $\eta _1$ and $ \eta_2$, and identifying alike coefficients, we obtain the following identities between the Kronecker--Eisenstein coefficients \cite{Broedel:2014vla},
\begin{align}
\label{diff.12} 
f^{(a_1)}(z_1{-}z|\tau) f^{(a_2)}(z_2{-}z|\tau) &= (-1)^{a_2-1} f^{(a_1+a_2)}(z_1{-}z_2|\tau) 
\\ &\quad
+ \sum_{k=0}^{a_1} \tbinom{a_1{+}a_2{-}k{-}1}{a_1{-}k} f^{(k)}(z_1{-}z_2|\tau) f^{(a_1+a_2-k)}(z_2{-}z|\tau)
\no \\ &\quad
+ \sum_{k=0}^{a_2} \tbinom{a_1{+}a_2{-}k{-}1} {a_2{-}k} f^{(k)}(z_2{-}z_1|\tau) f^{(a_1+a_2-k)}(z_1{-}z|\tau) 
\no
\end{align}
These relations will play an important role for the generating series to be discussed in later sections as well.
Converting (\ref{diff.12}) among the $f^{(a)}$ into a relation amongst the forms $\cD^+$ using (\ref{gen.67}), and the notations $a_0=a_1+a_2$ and $z_{12}=z_1-z_2$, we find,
\bea
\label{diff.13} 
&&
\dplus{a_1 \\0}(z_1{-}z|\tau) \dplus{a_2 \\0} (z_2{-}z|\tau) = (-1)^{a_2-1} \dplus{a_0 \\0}(z_{12}|\tau) 
\\ && \hskip 0.6in 
- \tbinom{ a_0{-}1}{a_1 } \dplus{a_0 \\0} \!(z_2{-}z|\tau) + \sum_{k=1}^{a_1} \tbinom{a_0{-}k{-}1}{a_1{-}k} \dplus{k\\0}(z_{12}|\tau) \dplus{a_0-k\\0} (z_2{-}z|\tau)
\no \\ && \hskip 0.6in 
- \tbinom{ a_0{-}1}{a_2 } \dplus{a_0\\0}\!(z_1{-}z|\tau)+ \sum_{k=1}^{a_2} \tbinom{a_0{-}k{-}1} {a_2{-}k} \dplus{k\\0}(z_{21}|\tau) \dplus{a_0-k\\0}(z_1{-}z|\tau)
\qquad \no 
\eea
In obtaining (\ref{diff.13}) from (\ref{diff.12}), we have treated  the contribution $k=0$ in (\ref{diff.12}) separately as one has  $f^{(0)} (z|\tau)=1$ but $\dplus{0\\0}(z|\tau)$ would produce an unwanted $\delta$-function contribution, as may be seen from (\ref{basic.42}). The relation (\ref{diff.13}) provides a simplification for the product $\dplus{a_1\\ 0}\!(z_1{-}z|\tau)\dplus{a_2\\0}\!(z_2{-}z|\tau)$ in the integral representation (\ref{basic.41}) of dihedral eMGFs with $b_1=b_2=0$ which is equivalent to HSR, and the final formula for HSR of dihedral eMGFs is given by,
\begin{align}
 \cplus{a_1 & a_2 & A \\ 0 & 0 & B \\ z_1 & z_2 & Z}  = \ &
(-1)^{a_2} \dplus{a_0   \\ 0  }\! (z_{12}) \, \cplus{A  \\ B \\ Z}
 \notag \\ 
 & - \tbinom{a_0-1}{a_1} \, \cplus{a_0  & A \\ 0 & B \\ z_2 & Z} 
 + \sum _{k=1}^{a_1} \tbinom{a_0-1-k}{a_1-k} \,
\dplus{k  \\ 0  }\! (z_{12}) \,
\cplus{a_0-k  & A \\ 0 & B \\ z_2 & Z}
\label{diff.15}  \\ 
&- \tbinom{a_0-1}{a_2} \, \cplus{a_0  & A \\ 0 & B \\ z_1 & Z}+ 
\sum _{k=1}^{a_2} \tbinom{a_0-1-k}{a_2-k} \,
\dplus{k  \\ 0 }\! (z_{21}) \,
\cplus{a_0-k  & A \\ 0 & B \\ z_1 & Z}
\notag
\end{align}
where the one-loop lattice sums on the right-hand side may be rewritten in terms of $f^{(k)}(z_{12}|\tau)$ via (\ref{gen.67}). Similar applications of the Fay identities (\ref{diff.12}) can be used to recursively perform the
HSR for higher-point eMGFs, see \cite{Gerken:2020aju} for analogous techniques for MGFs
and \cite{DHoker:2016mwo, Gerken:2018zcy} for the earlier approach based on lattice-sum manipulations.
For trihedral eMGFs (\ref{basic.16}) with a holomorphic two-point graph, the HSR  can be found in appendix \ref{app:tri.2} which literally follows the integrand manipulations of (\ref{diff.13}). The HSR for a holomorphic three-point graph 
is spelled  out in appendix \ref{app:tri.3}. Moreover, as shown in appendix \ref{app:tri.4}, Fay 
identities also imply algebraic relations among trihedral eMGFs that are not amenable to HSR.

%%%%%%%%%%%%%%%%%%%%%%%%%%%%%%%%%%%%%%%%%%%%%%%%%%%%%%%%%%%
\subsubsection{Coincident limit of the elliptic HSR}
\label{sec:2.2.3}
%%%%%%%%%%%%%%%%%%%%%%%%%%%%%%%%%%%%%%%%%%%%%%%%%%%%%%%%%%%

In order to recover the HSR formulae for dihedral MGFs from (\ref{diff.15}), it remains to drop
the restriction $z_1\neq z_2$. Most of the terms have a straightforward limit $z_1\rightarrow  z_2$, except for the contribution of $k=1$. 
The singular behavior $\dplus{1   \\ 0  }\! (z_{12}|\tau)\rightarrow - ( \Im \tau)/  z_{12}$
gives rise to the following $z_1$-derivative in the coincident limit,
\begin{align}
\lim_{z_2\rightarrow z_1} & \dplus{1  \\ 0  }\! (z_{12}) \bigg\{
\cplus{a_0-1 & A \\ 0 & B \\ z_{2} &Z}  -  \cplus{a_0-1 & A \\ 0 & B \\ z_{1} &Z} 
\bigg\} \notag \\
&=  ( \Im \tau )  \partial_{z_1}\cplus{a_0-1 & A \\ 0 & B \\ z_{1} &Z} 
= \cplus{a_0-1 & A \\ -1 & B \\ z_{1} &Z} 
\label{diff.16}
\end{align}
which we have evaluated via (\ref{diff.2}).
Hence, the coincident limit of (\ref{diff.15}) reads,
\begin{align}
 \cplus{a_1 & a_2 & A \\ 0 & 0 & B \\ z & z & Z}   =\ &
 (-1)^{a_2} 
\dplus{a_0 \\ 0  }\! (0 )
 \, \cplus{A  \\ B \\ Z} 
 - \tbinom{a_0}{a_1} \, \cplus{a_0  & A \\ 0 & B \\ z & Z}  \notag \\
 & 
 + \sum _{k=4}^{a_1} \tbinom{a_0-1-k}{a_1-k} \,
\dplus{k  \\ 0  }\! (0 )\,
\cplus{a_0-k  & A \\ 0 & B \\ z & Z} 
\label{diff.17}  \\ 
&+ 
\sum _{k=4}^{a_2} \tbinom{a_0-1-k}{a_2-k} \,
\dplus{k  \\ 0 }\! (0 )\,
\cplus{a_0-k  & A \\ 0 & B \\ z & Z} 
\notag \\
 & +  \tbinom{a_0-2}{a_1-1} \bigg\{ 
\dplus{2\\ 0 }\! (0 )\,
\cplus{a_0-2 & A \\ 0 & B \\ z &Z} 
  +  \cplus{a_0-1 & A \\ -1 & B \\ z &Z}  \bigg\}
 \notag 
\end{align}
where the one-loop lattice sums on the right-hand side may be rewritten in terms of
$G_k(\tau)$ and $\widehat G_2(\tau)$ via (\ref{gen.670}) and (\ref{G2hat}). By additionally setting $z$ 
and all the entries of $Z$ to zero, one recovers the HSR for dihedral MGFs.

%%%%%%%%%%%%%%%%%%%%%%%%%%%%%%%%%%%%%%%%%%%%%%%%%%%%%%%%%%%
\subsubsection{Examples of the elliptic HSR}
\label{sec:2.2.9}
%%%%%%%%%%%%%%%%%%%%%%%%%%%%%%%%%%%%%%%%%%%%%%%%%%%%%%%%%%%

We will later on see that all the eMGFs in the expansion of the generating series $Y^\tau_{\vec{\eta}}$
to be introduced in sections \ref{sec:3} and \ref{sec:4} can be addressed within the following notable 
subclass of elliptic HSR-identities (\ref{diff.15}), again using the notation $a_0=a_1+a_2$,
\begin{align}
-a_1 \, &\cplus{a_1+1 & a_2 & A \\ 0 & 0 & B \\ z_1 & z_2 & Z}  
+ a_2 \, \cplus{a_1 & a_2+1 & A \\ 0 & 0 & B \\ z_1 & z_2 & Z} = 
-a_0 (-1)^{a_2} 
 \dplus{a_0+1 \\ 0}\! (z_{12}) 
\, \cplus{A  \\ B \\ Z}
 \notag \\ 
 &\ \ \ \ + \tbinom{a_0}{a_1-1} \, \cplus{a_0+1  & A \\ 0 & B \\ z_1 & Z}
- \sum _{k=3}^{a_1+1} (k{-}1) \tbinom{a_0-k}{a_2-1} \,
\dplus{k  \\ 0 }\! (z_{12})\,
\cplus{a_0+1-k  & A \\ 0 & B \\ z_2 & Z} 
\notag  \\ 
&\ \ \ \ -  \tbinom{a_0}{a_2-1} \, \cplus{a_0+1  & A \\ 0 & B \\ z_2 & Z}
+ \sum _{k=3}^{a_2+1} (k{-}1) \tbinom{a_0-k}{a_1-1} \,
\dplus{k  \\ 0  }\! (z_{21})\,
\cplus{a_0+1-k  & A \\ 0 & B \\ z_1 & Z} 
\notag \\
&\ \ \ \ + \tbinom{a_0-2}{a_1-1}  
\dplus{2  \\ 0 }\! (z_{12})
\bigg\{
\cplus{a_0-1  & A \\ 0 & B \\ z_1 & Z}\! (\tau) - \cplus{a_0-1  & A \\ 0 & B \\ z_2 & Z}
\bigg\}
\label{diff.18}
\end{align}
These combinations do not involve the singular contribution $\dplus{1   \\ 0  }\! (z_{12}|\tau)\rightarrow - ( \Im \tau)/  z_{12}$, so that the coincident limit is regular. Moreover, the coefficient of $ \dplus{2   \\ 0 }\! (z_{12}|\tau)$ in the last line   drops out in the coincident limit $z_1 \rightarrow z_2$, so that the conditionally convergent
lattice sum $\dplus{2  \\ 0}\! (0|\tau) \sim \widehat G_2(\tau)$ 
in the last line of (\ref{diff.17}) drops out as well.

\sm

An important special case is obtained by setting $a_2=1$ in (\ref{diff.18}), 
\bea
\label{HSR3}
&&
\cplus{a_1 & 2 & A \\ 0 & 0 & B \\ z_1 & z_2 & Z}
-a_1 \, \cplus{a_1+1 & 1 & A \\ 0 & 0 & B \\ z_1 & z_2 & Z}
=(a_1{+}1) \dplus{a_1+2  \\ 0 }\!(z_{12})\, \cplus{ A \\ B \\ Z} 
\\ && \hskip 1in 
-\ \cplus{ a_1+2 & A \\ 0 & B \\ z_2 & Z}
-  \sum _{k=3}^{a_1+1} (k{-}1) 
\dplus{k  \\ 0}\!( z_{12}) \,
 \cplus{a_1+2-k  & A \\ 0 & B \\ z_2 & Z}
\no \\ && \hskip 1in 
+\ \frac{a_1}{2}(a_1{+}1)\cplus{ a_1+2 & A \\ 0 & B \\ z_1 & Z} 
+  \dplus{2  \\ 0 }\!(z_{12} ) 
\bigg\{ \cplus{a_1  & A \\ 0 & B \\ z_1 & Z} - \cplus{a_1  & A \\ 0 & B \\ z_2 & Z} \bigg\}
\no
\eea
By furthermore setting $a_1=1,2,3$ in (\ref{HSR3}) for instance, we obtain
\begin{align}
\cplus{1 & 2 & A \\ 0 & 0 & B \\ z_1 & z_2 & Z}
- \cplus{2 & 1& A \\ 0 & 0 & B \\ z_1 & z_2 & Z}
&=  
 \cplus{ 3 & A \\ 0 & B \\ z_1 & Z} 
- \cplus{ 3 & A \\ 0 & B \\ z_2 & Z} 
+2\dplus{3  \\ 0 }\!(z_{12})\, \cplus{ A \\ B \\ Z}
\label{HSR3a} \\
&\ \ +  \dplus{2  \\ 0 }\!(z_{12}) \bigg\{  \cplus{1  & A \\ 0 & B \\ z_1 & Z}  -\cplus{1  & A \\ 0 & B \\ z_2 & Z} \bigg\}
\notag \\
%%%%%
\cplus{2 & 2 & A \\ 0 & 0 & B \\ z_1 & z_2 & Z}
-2 \, \cplus{3 & 1 & A \\ 0 & 0 & B \\ z_1 & z_2 & Z}
&=
3 \cplus{ 4 & A \\ 0 & B \\ z_1 & Z} 
- \cplus{ 4 & A \\ 0 & B \\ z_2 & Z}
+ 3\dplus{4 \\ 0 }\!(z_{12}) \,\cplus{ A \\ B \\ Z} 
\label{HSR3b}\\ 
& \ \ - 2  \dplus{3 \\ 0 }\!(z_{12})\, \cplus{1  & A \\ 0 & B \\ z_2 & Z}
\no \\ 
& \ \ +  \dplus{2  \\ 0 }\!(z_{12}) \bigg\{ \cplus{2  & A \\ 0 & B \\ z_1 & Z}
- \cplus{2  & A \\ 0 & B \\ z_2 & Z} \bigg\}
\notag \\
%%%%%
\cplus{3 & 2 & A \\ 0 & 0 & B \\ z_1 & z_2 & Z}
-3 \, \cplus{4 & 1 & A \\ 0 & 0 & B \\ z_1 & z_2 & Z}
&=
6 \cplus{ 5 & A \\ 0 & B \\ z_1 & Z} 
- \cplus{ 5 & A \\ 0 & B \\ z_2 & Z}
+4 \dplus{5  \\ 0 }\!(z_{12} ) \,\cplus{ A \\ B \\ Z} \label{HSR3c} \\
& \ \
- 3 \dplus{4  \\ 0 }\!(z_{12}) \,\cplus{1  & A \\ 0 & B \\ z_2 & Z}
- 2  \dplus{3 \\ 0 }\!(z_{12}) \,\cplus{2  & A \\ 0 & B \\ z_2 & Z}
\notag \\ 
&\ \ 
+  \dplus{2  \\ 0 }\!(z_{12}) \bigg\{\cplus{3  & A \\ 0 & B \\ z_1 & Z}
-  \cplus{3  & A \\ 0 & B \\ z_2 & Z} \bigg\} \notag
\end{align}

\subsubsection{Examples of two-loop eMGFs}
\label{sec:dp2}

We shall now apply some of the above examples of elliptic HSR to identify the simplest differential equations of eMGFs beyond the one-loop case (\ref{diff.6}). The elliptic generalization of the two-loop MGF $C_{1,1,1}$ was studied under the name $D_3^{(1)}$ in \cite{DHoker:2018mys}, both of which are given as follows in the present notation,
\bea
C_{1,1,1}= \cplus{1 & 1 & 1 \\ 1 & 1 & 1 \\ 0 & 0 & 0} = E_3 + \zeta_3\, ,
\hskip 0.6in
D_3^{(1)}(z) =\cplus{1 & 1 & 1 \\ 1 & 1 & 1 \\ z & 0 & 0}
\eea
and its $\tau$-derivatives following from (\ref{gen.66Ctau}) are,
\begin{align}
\label{crexpl.1} 
(\pi \nabchi_\tau)D_3^{(1)}(z)  &= \cplus{2 & 1 & 1 \\ 0 & 1 & 1 \\ z & 0 & 0}
+ 2 \cplus{2 & 1 & 1 \\ 0 & 1 & 1 \\ 0 & z & 0}  \\
(\pi \nabchi_\tau)^2 D_3^{(1)}(z) &= 4 \cplus{2 & 2 & 1 \\ 0 & 0 & 1 \\ z & 0 & 0}
+ 2  \cplus{2 & 2 & 1 \\ 0 & 0 & 1 \\ 0 & 0 & z}  
 - 4  \cplus{3 & 1 & 1 \\ 0 & 0 & 1 \\ z & 0 & 0}
 - 4  \cplus{3 & 1 & 1 \\ 0 & 0 & 1 \\ 0 & z & 0}
  - 4  \cplus{3 & 1 & 1 \\ 0 & 0 & 1 \\ 0 & 0 & z} \notag
\end{align}
Based on the HSR-identity (\ref{HSR3b}) and its coincident limit, the second derivative can be simplified to, 
\begin{align}
(\pi \nabchi_\tau)^2 D_3^{(1)} (z) &= 
8 \dplus{5  \\ 1 }\!(z) + 4 \dplus{5  \\ 1 }\!(0) + 4 \dplus{3  \\ 0 }\!(z) \dplus{2  \\ 1 }\!(z)
 \label{crexpl.2}
 \end{align} 
The entire right-hand side has been reduced to (products of) one-loop eMGFs.
Moreover, the product $\dplus{3  \\ 0 }\!(z) \dplus{2  \\ 1 }\!(z)$ with
the Kronecker--Eisenstein coefficient $\dplus{3  \\ 0 }\!(z) = -(\Im \tau)^3 f^{(3)}(z)$
is unprecedented in the differential equations (\ref{diff.6}) of one-loop eMGFs. This last term 
in (\ref{crexpl.2}) signals that $D_3^{(1)}(z)$ is what should be called a depth-two eMGF
since $f^{(3)}$ multiplies a one-loop eMGF $\sim \dplus{2  \\ 1 }\!(z)$ which in turn produces
another $f^{(3)}$ upon differentiation in $\tau$ via (\ref{diff.6}). Hence, one may view (\ref{crexpl.2})
as the simplest example of Zagier's single-valued elliptic polylogarithms at depth two. The relation between the notion of depth and the differential equations satisfied by eMGFs will be explored in more detail in section~\ref{sec:sieve}.

\sm

We also note in the limit $z\to 0$, the relation~\eqref{crexpl.2} loses its last term since $f^{(3)}(z)$ is an odd function of $z$ in agreement with the vanishing of $G_3$. Therefore, the MGF $C_{1,1,1}$ obtained from $D_3^{(1)}(z)$ in the $z\to 0$ limit is only of depth one.

\sm

As another two-loop example, one can apply the HSR (\ref{HSR3a}) to
\begin{align}
(\pi \nabchi_\tau)\cplus{1 & 1 & 1 \\ 0 & 1 & 1 \\ z & 0 & 0} &=
2 \cplus{1 & 2 & 1 \\ 0 & 0 & 1 \\ z & 0 & 0}
- 2 \cplus{2 & 1 & 1 \\ 0 & 0 & 1 \\ z & 0 & 0}
 \label{crexpl.3} \\
&= 2 \dplus{2  \\ 0 }\!(z)\dplus{2  \\ 1}\!(z)  + 2 \dplus{4  \\ 1 }\!(z)  \notag 
\end{align}
with $\dplus{2  \\ 0 }\!(z)=-(\Im \tau)^2 f^{(2)}(z)$ multiplying another one-loop eMGF 
$\dplus{2  \\ 1}\!(z)$ with $f^{(3)}$ in its $\tau$-derivative. By the same arguments as above, the eMGF $\cplus{1 & 1 & 1 \\ 0 & 1 & 1 \\ z & 0 & 0}$ is therefore of depth two.

\subsubsection{Examples of three-loop eMGFs}
\label{sec:dp3}

Similarly, the simplest examples at three-loop order include the objects
\beq
D_4^{(1)}(z) = \cplus{1 & 1 & 1 & 1 \\ 1 & 1 & 1 &1 \\ z & 0 & 0 & 0} \, , \ \ \ \ \ \
D_4^{(2)}(z) = \cplus{1 & 1 & 1 & 1 \\ 1 & 1 & 1 &1 \\ z & z & 0 & 0}
 \label{crexpl.4}
\eeq
introduced in \cite{DHoker:2018mys}. Their second $\tau$-derivatives can
again be simplified using the HSR identity (\ref{HSR3b})
\begin{align}
(\pi \nabchi_\tau)^2 D_4^{(1)}(z)  & =  18 \dplus{4 \\0}\!(z) \dplus{2 \\ 2 }\!(0)    + 18 \dplus{4 \\ 0}\!(0)  \dplus{2 \\ 2 }\!(z) +6 \dplus{3 \\ 0}\!(z)  \cplus{1&1&1 \\ 0&1&1 \\ z&0&0} \notag \\
&\ \ +36 \dplus{6 \\ 2}\!(z) 
+ 36 \dplus{6 \\ 2}\!(0)  
+ 12 \dplus{4 \\ 1}\!(z)  \dplus{2 \\ 1}\!(z)  
\notag \\
&\ \ 
- 6 \dplus{3 \\ 1 }\!(z)  ^2 
- 6 \dplus{3 \\ 1}\!(0) ^2
- 12 \dplus{3 \\ 1}\!(z)  \dplus{3 \\ 1}\!(0) 
\label{finaldd41} \\
%%%%%%%
(\pi \nabchi_\tau)^2 D_4^{(2)}(z)  & = 
 12   \dplus{4 \\0}\!(0) \dplus{2 \\ 2 }\!(0)   
+24  \dplus{4 \\0}\!(z)  \dplus{2\\ 2}\!(z)
+8  \dplus{3\\ 0}\!(z)\cplus{1&1&1 \\ 0&1&1 \\ z&0&0} 
\notag \\
& \ \ +48 \dplus{6 \\ 2}\!(z) 
+ 24 \dplus{6 \\ 2} \!(0)
+ 16 \dplus{4 \\ 1}\!(z) \dplus{2 \\ 1} \!(z)
\notag \\
&\ \ 
- 8 \dplus{3 \\ 1}\!(z) ^2 
- 16 \dplus{3 \\ 1}\!(z) \dplus{3 \\ 1} \!(0)
\label{finaldd42} 
\end{align}
where we have used the following identities in intermediate steps.
\begin{align}
\cplus{4 & 1 & 1 \\ 0 & 1 & 1  \\ 0 & 0 & 0} &= 3 \cplus{6 & 0 \\ 2 &0  \\ 0 & 0 } -  \cplus{3 & 0 \\ 1 &0  \\ 0 & 0 }^2 \notag \\
2 \cplus{4 & 1 & 1 \\ 0 & 1 & 1  \\ 0 & z & 0} + \cplus{4 & 1 & 1 \\ 0 & 1 & 1  \\ z & 0 & 0} &= 
6  \cplus{6 & 0 \\ 2 &0  \\ z & 0 }  + 3  \cplus{6 & 0 \\ 2 &0  \\ 0 & 0 }  -  \cplus{3 & 0 \\ 1 &0  \\ z & 0 }^2 
 \label{crexpl.5} \\
 & \ \ + 2  \cplus{4 & 0 \\ 1 &0  \\ z & 0 }   \cplus{2 & 0 \\ 1 &0  \\ z & 0 } 
 - 2  \cplus{3 & 0 \\ 1 &0  \\ z & 0 }   \cplus{3 & 0 \\ 1 &0  \\ 0 & 0 } 
 \notag
\end{align}
The first identity can for instance be imported from the {\tt Mathematica} notebooks
of \cite{Gerken:2020aju}, and the second one will be 
proven in section \ref{sec:sieve}.

\sm

The last terms in the first lines of (\ref{finaldd41}) and (\ref{finaldd42})
identifies $ D_4^{(1)}(z) $ and $ D_4^{(2)}(z) $ to be eMGFs at depth three:
The Kronecker--Eisenstein coefficient $ \dplus{3\\ 0}\!(z)= - (\Im \tau)^3 f^{(3)}(z)$ 
multiplies the eMGF $\cplus{1&1&1 \\ 0&1&1 \\ z&0&0} $ which was found to be
of depth two in (\ref{crexpl.3}). We shall rederive the above examples from
generating functions in section \ref{sec:3.5}.

%%%%%%%%%%%%%%%%%%%%%%%%%%%%%%%%%%%%%%%%%%%%%%%%%%%%%%%%%%%
\subsection{The sieve algorithm and the notion of depth}
\label{sec:sieve}
%%%%%%%%%%%%%%%%%%%%%%%%%%%%%%%%%%%%%%%%%%%%%%%%%%%%%%%%%%%

In this section, we formalize the notion of depth by generalizing the concepts of~\cite{DHoker:2016mwo} related to the so-called sieve algorithm. 
The basic idea is to consider powers of the Maass operators $\nabchi_\tau$ acting on an eMGF ${\cal C}^+$ with labels ${\cal A},{\cal B},{\cal Z}$ as in (\ref{allemgfs}). Since the differential operator lowers the antiholomorphic $\mathcal{B}$ labels according to~\eqref{gen.66Ctau}, this, together with momentum conservation, eventually leads to eMGFs that are amenable to HSR. One takes successive derivatives $(\nabchi_\tau)^m$ of (\ref{allemgfs}) until one can first use HSR to extract a factor of a one-loop function $\dplus{k\\0}\!(z)= - (\Im\tau)^k f^{(k)}(z)$  with $k\neq 0$ multiplying another eMGF  ${\cal C}^+_{(k)}$ in the schematic form, 
\begin{align}
(\pi \nabchi_\tau)^m {\cal C}^+= (\Im \tau)^k f^{(k)} {\cal C}^+_{(k)} + \widehat{\cal C}^+
\end{align}
where both ${\cal C}^+_{(k)}$ and $\widehat{\cal C}^+$ are $\mathbb{Q}$-linear combinations of other general eMGFs of the form (\ref{allemgfs}) and
not amenable to HSR, see for instance (\ref{crexpl.3}).
We define the depth of the original  eMGF ${\cal C}^+$ as the maximum of (the depth of ${\cal C}^+_{(k)}$ plus one) and the depth of $\widehat{\cal C}^+$.  
\sm

The eMGFs ${\cal C}^+_{(k)}$ generated by HSR and $\widehat{\cal C}^+$ therefore have to be subjected to the same procedure of taking successive derivatives $\nabchi^m_\tau$ until one can again separate out a function $(\Im \tau)^{k'}f^{(k')}(z)$. 
If the eMGF multiplying $(\Im \tau)^{k'} f^{(k')}(z)$ is constant, then ${\cal C}^+_{(k)}$
is defined to have depth one which is the base of the recursive definition. 
The recursive algorithm is guaranteed to terminate after a finite number of steps 
since the antiholomorphic indices of the eMGFs are lowered in each step, and negative entries $b_j$ are always removed by combinations of HSR and momentum conservation. Like this, the depth of an eMGF with 
antiholomorphic indices ${\cal B}$ is manifest after no more than 
$| {\cal B} |$ powers of $\nabla_{\tau}$.

\sm

Note that the functions $(\Im\tau)^k f^{(k)}(z)$ and $(\Im \tau)^{k'} f^{(k')}(z)$ generated by HSR may have 
arguments $z=0$ and thereby reduce to holomorphic Eisenstein series. Hence, the present definition of depth also applies to MGFs and coincides with the notion of depth in \cite{Broedel:2018izr, Gerken:2020yii} adapted to the iterated-Eisenstein-integral representations in the references.
Since at the one-loop order the action of both $(\nabchi_z)^b$ and $(\nabchi_\tau)^b$ on eMGFs $\dplus{a\\b}\!(z)$ produces a factor of $(\Im\tau)^k f^{(k)}(z)$
under (\ref{diff.6}) and (\ref{diff.3ex}),
one may wonder whether the definition of depth should be based on $\nabchi_\tau$ or
$\nabchi_z$. With our above definition based on $\nabchi_\tau$, the notion of depth for eMGFs reduces to that of MGFs as $z\to 0$, which is not the case for $\nabchi_z$ as it annihilates MGFs.

\subsubsection{Example of the sieve algorithm for eMGFs}
\label{sec:sieveex}

The simplifications of eMGFs by repeated $\nabchi_\tau$-derivatives and HSR not only
pinpoint an unambiguous notion of depth but also generate relations among eMGFs through
a generalization of the sieve algorithm for MGFs in~\cite{DHoker:2016mwo}. As an illustration,
we derive the relations among two-loop eMGFs in the second and third line of (\ref{crexpl.5}):
The first derivatives of the eMGFs on the left-hand side are given by
\bea
\pi \nabla_\tau \cplus{ 4&1&1 \\ 0&1&1 \\ z&0&0} & = & 
20 \cplus{ 7 & 0  \\ 1 & 0 \\ z & 0} 
- 2\cplus{ 7 & 0 \\ 1 & 0 \\ 0& 0}
- 6  \cplus{4 & 0 \\ 0 & 0 \\ z & 0} \cplus{3  & 0 \\ 1 & 0 \\ 0 & 0}
\no \\ &&
- \  2 \cplus{2 & 0 \\ 0 & 0 \\ z & 0} \cplus{5  & 0 \\ 1 & 0 \\ 0 & 0}
+2  \cplus{2 & 0 \\ 0 & 0 \\ z & 0} \cplus{5  & 0 \\ 1 & 0 \\ z & 0}
\label{sieve.1}
\eea
and
\bea
\pi \nabla_\tau \cplus{ 4&1&1 \\ 0&1&1 \\ 0&z&0} & = & 
8 \cplus{ 7 & 0  \\ 1 & 0 \\ z & 0} 
+ 10 \cplus{ 7 & 0 \\ 1 & 0 \\ 0& 0}
- 3  \cplus{4 & 0 \\ 0 & 0 \\ 0 & 0} \cplus{3  & 0 \\ 1 & 0 \\ z & 0}
- 3  \cplus{4 & 0 \\ 0 & 0 \\ z & 0} \cplus{3  & 0 \\ 1 & 0 \\ z & 0}
\no \\ &&
+\ 2\cplus{3 & 0 \\ 0 & 0 \\ z & 0} \cplus{4  & 0 \\ 1 & 0 \\ z & 0}
+ 4\cplus{5 & 0 \\ 0 & 0 \\ z & 0} \cplus{2  & 0 \\ 1 & 0 \\ z & 0}
\no \\ &&
+\ \cplus{2 & 0 \\ 0 & 0 \\ z & 0} \cplus{5  & 0 \\ 1 & 0 \\ 0 & 0}
-  \cplus{2 & 0 \\ 0 & 0 \\ z & 0} \cplus{5  & 0 \\ 1 & 0 \\ z & 0}
\label{sieve.2}
\eea
The last lines of (\ref{sieve.1}) and (\ref{sieve.2}) 
drop out from the combination in (\ref{crexpl.5}), i.e.\ 
\begin{align}
\pi \nabla_\tau \Big( 2 \cplus{4 & 1 & 1 \\ 0 & 1 & 1  \\ 0 & z & 0} + \cplus{4 & 1 & 1 \\ 0 & 1 & 1  \\ z & 0 & 0}  \Big)
= \ &36 \cplus{ 7 & 0  \\ 1 & 0 \\ z & 0}+ 18  \cplus{ 7 & 0  \\ 1 & 0 \\ 0& 0} - 6  \cplus{4 & 0 \\ 0 & 0 \\ z & 0} \cplus{3  & 0 \\ 1 & 0 \\ 0 & 0}
\notag \\
&- 6  \cplus{4 & 0 \\ 0 & 0 \\ 0 & 0} \cplus{3  & 0 \\ 1 & 0 \\ z & 0}
- 6  \cplus{4 & 0 \\ 0 & 0 \\ z & 0} \cplus{3  & 0 \\ 1 & 0 \\ z & 0}
\no \\
&
+4\cplus{3 & 0 \\ 0 & 0 \\ z & 0} \cplus{4  & 0 \\ 1 & 0 \\ z & 0}
+ 8\cplus{5 & 0 \\ 0 & 0 \\ z & 0} \cplus{2  & 0 \\ 1 & 0 \\ z & 0}
\no \\
%%%
%%%
= \ &\pi \nabla_\tau \Big(
6  \cplus{6 & 0 \\ 2 &0  \\ z & 0 }  + 3  \cplus{6 & 0 \\ 2 &0  \\ 0 & 0 }  -  \cplus{3 & 0 \\ 1 &0  \\ z & 0 }^2 
 \notag \\
 & - 2  \cplus{3 & 0 \\ 1 &0  \\ z & 0 }   \cplus{3 & 0 \\ 1 &0  \\ 0 & 0 } 
 + 2  \cplus{4 & 0 \\ 1 &0  \\ z & 0 }   \cplus{2 & 0 \\ 1 &0  \\ z & 0 } 
 \Big)
 \label{sieve.3}
\end{align}
This equality is still valid after dropping the $\nabla_\tau$ on the two sides, i.e.\ it can be
uplifted to (\ref{crexpl.5}) since the objects in the parenthesis have modular weight
$(0,-4)$ and do not admit any integration constants. First, we observe in analogy with~\cite[Lemma 1]{DHoker:2016mwo} and from the polynomial growth of eMGFs near the cusp $\tau\to i \infty$ that any integration constant can at most depend on the co-moving coordinates $(u,v)$ of the elliptic variable $z$. These transform as a vector under $SL(2,\mathbb{Z})$, see~\eqref{modPSL1}, but it is impossible to construct an $SL(2,\mathbb{Z})$ singlet using only a single (commuting) vector, therefore the integration constant must be independent of $(u,v)$ as well. As there is no constant with a non-trivial modular weight, the integration constant must vanish identically.

\subsubsection{Depth of eMGFs and iterated $\tau$-integrals}

The above definition of depth can be translated into iterated-integral representations of eMGFs.  
Similar to the MGF case~\cite{DHoker:2015wxz, Gerken:2020yii}, the differential equations \eqref{diff.6} of $\dplus{a\\b}\!(z|\tau)$
at depth one can be solved via meromorphic $\tau$ integrals (to be performed at fixed $u,v$)
\begin{align}
 \int^\tau_{i \infty}  \dd \tau_1 \, (\tau_1)^j  \, f^{(a+b)}(u\tau_1 {+}v|\tau_1)\, , \ \ \ \ \ \ 0 \leq j \leq a{+}b{-}2
  \label{sieve.99}
\end{align}
and their complex conjugates whose coefficients are rational functions of $\tau$ and $\bar \tau$.
Higher-depth eMGFs admit similar representations where (\ref{sieve.99}) generalizes to iterated
integrals with multiple integration kernels $\tau^{j_1} f^{(k_1)}$, \dots, $\tau^{j_\ell} f^{(k_\ell)}$ and
the depth of the eMGF sets the maximum value of $\ell$. The exploration of this form of eMGFs based on the generating-series considerations in later parts of the present paper will be the subject of a subsequent publication~\cite{toappsoon}.

\sm 

Alternatively, one can integrate the differential equations of eMGFs w.r.t.\ $z$ rather than $\tau$.
At depth one, this has been implemented by expressing Zagier's single-valued elliptic
polylogarithms in terms of finite linear combinations of meromorphic elliptic polylogarithms\footnote{By the tension between meromorphicity, homotopy invariance and doubly periodic integration kernels for iterated integrals on the torus,
several variants of the elliptic polylogarithms of Brown and Levin \cite{BrownLev} have appeared
in the physics literature \cite{Broedel:2014vla, Broedel:2017kkb}. The representations 
of $\dplus{a\\b}\!(z|\tau)$ in \cite{Broedel:2019tlz} are
based on the elliptic polylogarithms with meromorphic integration kernels \cite{Broedel:2017kkb}.} -- iterated 
integrals over $z$ -- and their complex conjugates \cite{Broedel:2019tlz}.

%%%%%%%%%%%%%%%%%%%%%%%%%%%%%%%%%%%%%%%%%%%%%%%%%%%%%%%%%%%
\subsection{Laplace equations}
\label{sec:2.7}
%%%%%%%%%%%%%%%%%%%%%%%%%%%%%%%%%%%%%%%%%%%%%%%%%%%%%%%%%%%

In this section, we initiate the study of Laplace equations of modular invariant eMGFs,
i.e.\ dihedral ones (\ref{gen.66}) with $|A|=|B|$ and trihedral ones (\ref{basic.44}) 
with $|A|+|C|+|E|=|B|+|D|+|F|$. The modular invariant Laplacian on functions $\phi(\tau)$ with vanishing 
modular weight is given by
\beq
\Delta \phi(\tau) = 4 (\Im \tau)^2  \partial_{\bar \tau} \partial_\tau \phi(\tau) =
 \overline{\nabchi}_\tau \Big(  (\Im \tau)^{-2} \nabchi_\tau \phi(\tau) \Big)
\label{lapl.1}
\eeq
with the complex conjugate $\overline{ \nabchi}_\tau = -2i(\Im \tau)^2 \partial_{\bar \tau}$ of
the Cauchy--Riemann derivative (\ref{crdrv}). 
We reiterate that the $\tau$-derivatives in (\ref{lapl.1}) are taken at constant
$(u_r,v_r)$ if $\phi(\tau)$ is chosen to be an eMGF depending on $z_1,z_2,\ldots,z_R$.
The Laplace action on modular invariant eMGFs
can be evaluated by combining the Cauchy--Riemann equation (\ref{gen.66Ctau}) 
with the complex-conjugation properties (\ref{basic.27-}), e.g.
\begin{align}
\Delta \dplus{k \\k} \!(z|\tau) &= \overline{ \nabchi}_\tau \bigg(  (\Im \tau)^{-2} \frac{k}{\pi} \dplus{k+1 \\k-1} \!(z|\tau) 
 \bigg) = k \pi  \overline{ \nabchi}_\tau  \overline{ \dplus{k-1 \\k+1} \!(-z|\tau)  } \notag \\
&= k(k{-}1)  \overline{ \dplus{k \\k} \!(-z|\tau)  }  = k(k{-}1)   \dplus{k \\k} \!(z|\tau)  
\label{lapl.2}
\end{align}
This is completely analogous to the derivation of the Laplace equation of non-holomorphic Eisenstein series
$E_k$ defined in (\ref{Ek}) and reproduces the well-known Laplace eigenvalue equation of the $g_k(z|\tau)$ functions (\ref{gk})
\beq
\big( \Delta  - k(k{-}1)  \big) g_k(z|\tau) = 0
\label{lapl.3}
\eeq

\subsubsection{Laplace equations of two-loop eMGFs}

For MGFs built solely from Green functions, the Laplace equations have been studied
for two-loop graph functions \cite{DHoker:2015gmr}, the Mercedes diagram \cite{Basu:2016xrt}, tetrahedral MGFs \cite{Kleinschmidt:2017ege}, certain dihedral three-loop MGFs \cite{Basu:2019idd} and generating series of Koba--Nielsen integrals \cite{Gerken:2019cxz}. 
We shall now generalize the Laplace equations
of the two-loop MGFs in \cite{DHoker:2015gmr} to the elliptic case
\beq
C_{a,b,c}(Z  ) = \cplus{a &b &c \\ a &b &c \\ z_1 &z_2 &z_3} 
= \bigg( \frac{ \Im \tau }{\pi} \bigg)^{a+b+c} \! \! \! \! \!  \! \! \! \sum_{p_1,p_2,p_3 \in \Lambda'}  \! \! \! \! \!\frac{ \delta(p_1{+}p_2{+}p_3) \chi_{p_1}(z_1)\chi_{p_2}(z_2)\chi_{p_3}(z_3) }{|p_1|^{2a} |p_2|^{2b} |p_3|^{2c} }
\label{lapl.4}
\eeq
which is the special case of the two-loop eMGFs in (\ref{twolp.1}) where the holomorphic
and antiholomorphic exponents line up.
Only two of the characteristics $Z=[z_1,\, z_2,\, z_3]$ enter as independent variables by
translation invariance (\ref{basic.23}) and the ubiquitous dependence on $\tau$ is again left implicit. 
By repeating the steps in (\ref{lapl.2}), one can straightforwardly 
evaluate the Laplace action to yield
\begin{align}
\Delta C_{a,b,c}(Z  ) &= \big(  a(a{-}1) + b(b{-}1) + c(c{-}1) \big) C_{a,b,c}(Z  )  \notag \\
&\quad+ ab \, \bigg\{ \cplus{a+1 &b-1 &c \\ a-1 &b+1 &c \\ z_1 &z_2 &z_3}
 + \cplus{a-1 &b+1 &c \\ a+1 &b-1 &c \\ z_1 &z_2 &z_3} \bigg\} \label{lapl.5} \\
&\quad + ac\, \bigg\{ \cplus{a+1 &b &c-1 \\ a-1 &b &c+1 \\ z_1 &z_2 &z_3} 
 + \cplus{a-1 &b &c+1 \\ a+1 &b &c-1 \\ z_1 &z_2 &z_3} \bigg\} \notag \\
&\quad+ bc \, \bigg\{ \cplus{a &b+1 &c-1 \\ a &b-1 &c+1 \\ z_1 &z_2 &z_3}
 + \cplus{a &b-1 &c+1 \\ a &b+1 &c-1 \\ z_1 &z_2 &z_3} \bigg\}
\notag
\end{align}
The eMGFs in the last three lines with different holomorphic and antiholomorphic entries can
be rewritten in terms of the $C_{a,b,c}(Z)$ with integer shifts in $a,b,c$ via momentum 
conservation (\ref{basic.23}), e.g.
\begin{align}
&\cplus{a+1 &b-1 &c \\ a-1 &b+1 &c \\ z_1 &z_2 &z_3}  
 + \cplus{a-1 &b+1 &c \\ a+1 &b-1 &c \\ z_1 &z_2 &z_3}  = C_{a+1,b+1,c-2}(Z)
 \label{lapl.6} \\
 & \ \ \ \  \ \  +  C_{a-1,b+1,c}(Z)+C_{a+1,b-1,c}(Z)
 -2 C_{a,b+1,c-1}(Z)-2C_{a+1,b,c-1}(Z) \notag
 \end{align}
The manipulations in (\ref{lapl.5}) and (\ref{lapl.6}) have already been used in
\cite{DHoker:2015gmr} to derive the Laplace eigenvalue equations of MGFs
$C_{a,b,c}(Z {=}[0,0,0] )$, and we find the same structure for their elliptic generalization
\begin{align}
& \quad \Big (\Delta - a(a-1)-b(b-1)-c(c-1) \Big ) C _{a,b,c} 
 \notag \\
&=
+ ab \Big ( C _{a-1,b+1,c}  + C _{a+1,b-1,c}  + C _{a+1,b+1,c-2}
 - 2 C _{a,b+1,c-1}  - 2 C _{a+1,b,c-1}   \Big )
 \label{lapl.7} \\
&\quad
+ bc \Big ( C _{a-2,b+1,c+1} + C _{a,b-1,c+1} + C _{a,b+1,c-1} 
- 2 C _{a-1,b,c+1} - 2 C _{a-1,b+1,c}  \Big )
\notag \\
&\quad
+ ca \Big ( C _{a-1,b,c+1} + C _{a+1,b-2,c+1} + C _{a+1,b,c-1} 
- 2 C _{a,b-1,c+1}  - 2 C _{a+1,b-1,c}  \Big ) \notag
\end{align}
where the common arguments of the eMGFs $(Z|\tau)$ have been suppressed.

\subsubsection{Evaluating $C _{a,b,0}$ and $C _{a,b,-1}$}

Starting out with $a,b,c\geq 1$ in (\ref{lapl.7}), its right-hand side may end up involving a lower index which vanishes or which equals $-1$. We shall now evaluate both of these functions separately. The first case can be simplified through momentum conservation~\eqref{basic.23} and the factorization property (\ref{basic.43}),
\begin{align}
C _{a,b,0} &= \cplus{a &b &0 \\ a &b &0 \\ z_1 &z_2 &z_3} =  \dplus{a\\a}(z_{13}) \dplus{b\\b}(z_{23}) - 
\cplus{a &b  \\ a &b  \\ z_{1} &z_{2} }  \notag \\
&= g_a(z_{13}) g_{b}(z_{23} ) - g_{a+b}(z_{12})
 \label{lapl.8}
 \end{align}
leading to combinations of the $g_k(z|\tau)$ functions. The second case also requires 
momentum conservation~\eqref{basic.23} and factorization~\eqref{basic.43}
\begin{align}
C _{a,b,-1} &= \cplus{a-1 &b &0 \\ a-1 &b &0 \\ z_1 &z_2 &z_3} 
+ \cplus{a &b-1 &0 \\ a &b-1 &0 \\ z_1 &z_2 &z_3} 
+ \cplus{a-1 &b &0 \\ a &b-1 &0 \\ z_1 &z_2 &z_3} 
+ \cplus{a &b-1 &0 \\ a-1 &b &0 \\ z_1 &z_2 &z_3}  \notag \\
&= g_{a-1}(z_{13}) g_{b}(z_{23} ) +g_a(z_{13}) g_{b-1}(z_{23} )   \label{lapl.9} \\
&\ \ \ \
+ \dplus{a-1\\a}(z_{13}) \dplus{b\\b-1}(z_{23})
+\dplus{a\\a-1}(z_{13}) \dplus{b-1\\b}(z_{23})
\notag
 \end{align}

\subsubsection{Examples at low weight} 

By the last line in (\ref{lapl.9}), the Laplacian of $C_{a,b,c}(Z)$ with one of $a,b,c=1$
cannot be solely expressed in terms of objects $g_k(z)$ and $C_{a,b,c}(Z)$ whose
holomorphic and antiholomorphic entries are lined up. For instance, the terms in
the second line of
\begin{align}
&\Delta C_{1,1,1}(Z) = 2 g_3(z_{12}) +  2 g_3(z_{23}) +  2 g_3(z_{31})   \label{lapl.10} \\
&\ \ \ \ + \Big\{
\dplus{1\\2}\!(z_{13}) \dplus{2\\1}\!(z_{23})
+\dplus{2\\1}\!(z_{13}) \dplus{1\\2}\!(z_{23}) + {\rm cyc}(z_1,z_2,z_3)
\Big\} \notag
\end{align}
are a signal for irreducible depth-two admixtures that can also be seen from the second Cauchy--Riemann derivative (\ref{crexpl.2}) of $C_{1,1,1}(z,0,0)= \cplus{1 & 1 & 1 \\ 1 & 1 & 1 \\ z & 0 & 0}$. We similarly obtain,
\begin{align}
(\Delta-2) C_{2,1,1}(Z) &=
 5 g_4(z_{12}) +  5 g_4(z_{13}) - g_4(z_{23})
+ g_2(z_{12}) g_2(z_{13}) 
\notag\\
& \quad+ 2 \dplus{2\\ 1}\!(z_{32}) \dplus{2\\ 3}\!(z_{12}) + 
 2 \dplus{2\\ 1}\!(z_{23}) \dplus{2\\ 3}\!(z_{13})   -  g_2(z_{13})  g_2(z_{23})\notag \\
 &\quad + 
 2 \dplus{1\\ 2}\!(z_{32}) \dplus{3\\ 2}\!(z_{12}) + 
 2 \dplus{1\\ 2}\!(z_{23}) \dplus{3\\ 2}\!(z_{13}) -    g_2(z_{12})  g_2(z_{23})
 \notag \\
 %%%
 %%%
 (\Delta-6) C_{3,1,1}(Z)&=
C_{2,2,1}(Z)+C_{2,1,2}(Z)+C_{1,2,2}(Z)
+  8 g_5(z_{12}) + 8 g_5(z_{13})
\notag\\
%%%
&\quad +  3 \dplus{1\\ 2}\!(z_{23}) \dplus{4\\ 3}\!(z_{13})  {+}  3 \dplus{2\\ 1}\!(z_{23}) \dplus{3\\ 4}\!(z_{13}) {-} 2 g_2(z_{23}) g_3(z_{13})   \notag \\
 &\quad+  3 \dplus{1\\ 2}\!(z_{32})  \dplus{4\\ 3}\!(z_{12})  {+} 3 \dplus{2\\ 1}\!(z_{32})  \dplus{3\\ 4}\!(z_{12})   {-} 2 g_2(z_{23}) g_3(z_{12})  \notag
 \\
(\Delta-4)  C_{2,2,1}(Z)   &= 
- 2 C_{2,1,2}(Z) - 2 C_{1,2,2}(Z)
+ 12 g_5(z_{12})   - 2 g_5(z_{13}) - 2 g_5(z_{23}) \notag  \\
&\quad
 + 2  g_2(z_{23}) \big( g_3(z_{12}) - g_3(z_{13}) \big)
  + 2  g_2(z_{13})\big( g_3(z_{12}) -  g_3(z_{23}) \big)  \notag \\
 %%%
&\quad  + 4 \dplus{2\\ 3}\!(z_{23}) \dplus{3\\ 2}\!(z_{13}) + 
 4 \dplus{2\\ 3}\!(z_{13}) \dplus{3\\ 2}\!(z_{23})  \label{lapl.43}
\end{align}
by combining (\ref{lapl.7}) with (\ref{lapl.8}) and (\ref{lapl.9}). Note that (\ref{lapl.10}) and  
(\ref{lapl.43}) reduce to the following Laplace equations of MGFs \cite{DHoker:2015gmr}
as $Z \rightarrow 0$:
\begin{align}
\Delta  C_{1,1,1}(Z=0) &= 6 E_3 \, , &(\Delta-2)  C_{2,1,1}(Z=0) &= 9 E_4 - E_2^2 \label{lapl.44}
\\
\Delta C_{2,2,1}(Z=0) &= 8 E_5\, , &(\Delta - 6) C_{3,1,1}(Z=0) &= 3 C_{2,2,1}(Z=0)
+ 16  E_5 - 4  E_2  E_3
\notag 
\end{align}
One may try to repeat the strategy in the MGF literature to seek linear combinations
$F$ of $C_{a,b,c}$ and $g_{a+b+c}$ such that the expression $(\Delta {-} \lambda )F$
for some eigenvalue~$\lambda$ simplifies. However, combinations of $C_{a,b,c}$ and $g_{a+b+c}$ do not suffice 
to eliminate the $\dplus{a\\b}(z_{ij})$ functions with $a\neq b$ from the Laplace equations, i.e.\ the combinations
\begin{align}
F_3(Z) &= C_{1,1,1}(Z) - \frac{1}{3}g_3(z_{12})- \frac{1}{3}g_3(z_{23})- \frac{1}{3}g_3(z_{13}) \notag \\
F_4(Z) &= -\frac{1}{2} C_{1,2,1}(Z) - \frac{1}{2} C_{1,1,2}(Z) + \frac{1}{5} g_4(z_{12}) + \frac{1}{5} g_4(z_{13})+ \frac{1}{2}g_4(z_{23}) \label{lapl.45} \\
F_5(Z) &= \frac{1}{4} C_{1,2,2}(Z) + \frac{1}{4} C_{2,1,2}(Z) + \frac{1}{4} C_{2,2,1}(Z)
 - \frac{1}{10}g_5(z_{12})- \frac{1}{10}g_5(z_{23})- \frac{1}{10}g_5(z_{13})
\notag
\end{align}
can at best be engineered to eliminate $g_{a+b+c}$ and $C_{a,b,c}$ from the right-hand sides of
\begin{align}
\Delta F_3(Z) &= \dplus{1\\2}\!(z_{13}) \dplus{2\\1}\!(z_{23})
+\dplus{2\\1}\!(z_{13}) \dplus{1\\2}\!(z_{23}) + {\rm cyc}(z_1,z_2,z_3)\label{lapl.46}  \\
\Delta F_5(Z) &= \dplus{2\\3}\!(z_{13}) \dplus{3\\2}\!(z_{23})
+\dplus{3\\2}\!(z_{13}) \dplus{2\\3}\!(z_{23}) + {\rm cyc}(z_1,z_2,z_3)
\notag
\end{align}
as well as
\begin{align}
(\Delta - 2) F_4(Z) &= g_2(z_{12}) g_2(z_{13})
+ \big\{ \dplus{2 \\ 1}(z_{12}) \dplus{2 \\ 3}(z_{23})
+\dplus{1 \\ 2}(z_{12}) \dplus{3 \\ 2}(z_{23}) \notag \\
&\ \ \    - \dplus{2 \\ 1}(z_{12}) \dplus{2 \\ 3}(z_{13})
- \dplus{1 \\ 2}(z_{12}) \dplus{3 \\ 2}(z_{13})
+(z_2\leftrightarrow z_3) \big\}
\label{lapl.47}
\end{align}
In the limit where $z_j \rightarrow 0$, the above $\dplus{a\\b}$ at odd $a+b$ vanish,
such that (\ref{lapl.46}) and (\ref{lapl.47}) reduce to well-known Laplace equations of
MGFs including \cite{DHoker:2015gmr}
\beq
(\Delta - 2 ) \bigg({-}C_{2,1,1} + \frac{9}{10} E_4 \bigg)= E_2^2
\eeq

\subsubsection{Examples beyond two-loop eMGFs}

The same interplay of Cauchy--Riemann equations, complex conjugations and
momentum conservation can be used to simplify the Laplacian of modular invariant
eMGFs associated with higher-loop dihedral graphs. For entries $A,B$ of identical weight,
we obtain
\beq
\big(\Delta +|A| \big) \cplus{A \\ B \\ Z} = \sum_{r,r'=1}^R a_r b_{r'} \cplus{A+ S_r- S_{r'} \\ B-S_r+S_{r'} \\ Z}  \, , \ \ \ \ \ \
|A|=|B|
\label{lapl.51}
\eeq
which verbatim generalizes the Laplacian of dihedral modular invariant MGFs
in \cite{DHoker:2016mwo}. When applied to the eMGFs
$D_4^{(1)}(z),D_4^{(2)}(z)$ in (\ref{crexpl.4}), the direct outcome of (\ref{lapl.51}) is
\begin{align}
\Delta D_4^{(1)}(z) & =  
3 \cplus{2&0&1&1 \\ 0&2&1&1 \\ z&0&0&0}
+3 \cplus{0&2&1&1 \\ 2&0&1&1 \\ z&0&0&0}
+6 \cplus{1&2&0&1 \\ 1&0&2&1 \\ z&0&0&0}
\notag
 \\
\Delta D_4^{(2)}(z) & =  
4 \cplus{2&0&1&1 \\ 0&2&1&1 \\ z&z&0&0}
+8 \cplus{2&1&0&1 \\ 0&1&2&1 \\ z&z&0&0}
 \label{lapl.12}
\end{align}
As detailed in \cite{Basu:2020pey}, the right-hand sides of (\ref{lapl.12}) simplify upon taking the linear combination $ D_4^{(1)}(z)  - \frac{3}{4} D_4^{(2)}(z) $ and adding products of depth-one MGFs,
\begin{align}
&\Delta\Big(  D_4^{(1)}(z)  - \frac{3}{4}  D_4^{(2)}(z) - 3 E_2 g_2(z) + \frac{3}{2} g_2(z)^2 + \frac{3}{4} E_2^2 \Big) \label{lapl.52} \\
&\ \ = 2\Big(   D_4^{(1)}(z)  - \frac{3}{4}   D_4^{(2)}(z)   \Big)  + 9 E_4 - 18 E_2 g_2(z) + 9 g_2(z)^2 + \frac{3}{2} E_2^2
\notag
\end{align}
The simplicity of this particular linear combination can also be understood from the $\tau$-derivatives 
in (\ref{finaldd41}) and (\ref{finaldd42}): The relative coefficient of $-\frac{3}{4}$ ensures
that the only depth-three term $\sim  \dplus{3 \\ 0}\!(z)  \cplus{1&1&1 \\ 0&1&1 \\ z&0&0}$ drops out from $(\pi \nabchi_\tau)^2(D_4^{(1)}(z)  - \frac{3}{4}   D_4^{(2)}(z))$. Hence, the Laplace equation (\ref{lapl.52})
eventually relates eMGFs of depth two.

In combination with the Laplace equation of the MGF $D_4=\cplus{1&1&1&1 \\ 1&1&1&1 \\ 0&0&0&0}$ \cite{DHoker:2015gmr}, the motivation for deriving (\ref{lapl.52}) in \cite{Basu:2020pey} was to prove the Laplace equation
\begin{align}
 \Delta \Big( F_4(z) - \frac{1}{2} F_2(z)^2 \Big) &= 2 F_4(z) - 3 F_2(z)^2 
\notag \\
F_2(z) &= E_2 - g_2(z) \label{lapl.53} \\
F_4(z) &= \frac{ D_4 }{12} - \frac{  D_4^{(1)}(z) }{3} + \frac{  D_4^{(2)}(z) }{4} \notag
\end{align}
which arose as a corollary of identities among higher-genus MGFs in \cite{DHoker:2020tcq}.

\newpage

%%%%%%%%%%%%%%%%%%%%%%%%%%%%%%%%%%%%%%%%%%%%%%%%%%%%%%%%%%%
%%%%%%%%%%%%%%%%%%%%%%%%%%%%%%%%%%%%%%%%%%%%%%%%%%%%%%%%%%%
\section{Dihedral eMGFs from Koba--Nielsen integrals}
\label{sec:3}
%%%%%%%%%%%%%%%%%%%%%%%%%%%%%%%%%%%%%%%%%%%%%%%%%%%%%%%%%%%
%%%%%%%%%%%%%%%%%%%%%%%%%%%%%%%%%%%%%%%%%%%%%%%%%%%%%%%%%%%

We shall now explain how eMGFs arise from world-sheet integrals involving Koba--Nielsen factors. This involves a generalization of the generating series of MGFs that were introduced in~\cite{Gerken:2019cxz,Gerken:2020yii}, and the open-string analogues of the subsequent generating functions of eMGFs have been introduced and analyzed in~\cite{Broedel:2020tmd}. This section deals with the case of two points, i.e.\ dihedral eMGFs, and the next section covers the general case.

%%%%%%%%%%%%%%%%%%%%%%%%%%%%%%%%%%%%%%%%%%%%%%%%%%%%%%%%%%%
\subsection{Two-point generating series and component integrals}
\label{sec:3.1}
%%%%%%%%%%%%%%%%%%%%%%%%%%%%%%%%%%%%%%%%%%%%%%%%%%%%%%%%%%%

Using the Kronecker--Eisenstein series~\eqref{looprev.4} we define the following $(2\times 2)$-array of generating series,
\begin{align}
\label{eq:Y2pt}
Y_{ij} (z_0,\eta,\bar\eta|\tau) &= 2i\Im\tau \int_\Sigma \frac{\dd^2 z_2}{\Im\tau} e^{s_{02} g(z_{02}|\tau) + s_{12}g(z_{12}|\tau)} 
\nn\\
&\quad \times 
\begin{pmatrix} \Omega(z_{12}, (\tau{-}\bar\tau)\eta|\tau) \overline{\Omega(z_{12},\eta|\tau)} & \Omega(z_{02}, (\tau{-}\bar\tau)\eta|\tau) \overline{\Omega(z_{12},\eta|\tau)}\\
\Omega(z_{12}, (\tau{-}\bar\tau)\eta|\tau) \overline{\Omega(z_{02},\eta|\tau)} & \Omega(z_{02}, (\tau{-}\bar\tau)\eta|\tau)\overline{\Omega(z_{02},\eta|\tau)}\end{pmatrix}_{ij}
\end{align}
where $i,j\in\{1,2\}$ and
\begin{align}
\KN_2(z_0|\tau) = e^{s_{02} g(z_{02}|\tau) + s_{12}g(z_{12}|\tau)} 
\label{KN2pt}
\end{align}
augments the usual Koba--Nielsen factor $ \exp(s_{12} g(z_{12}|\tau))$ for two points $z_1$ and $z_2$ by an additional factor depending on the extra puncture $z_0$. The latter will be the elliptic variable for the dihedral eMGFs since $z_1$ can be fixed to zero by translation invariance and $z_2$ is integrated over. There are extra propagators from the extra puncture $z_0$ only to the unfixed $z_2$ and this will generalize to higher-point cases in the next section. A factor $\exp(s_{01} g(z_{01}|\tau))$---that would be needed for a complete $(n{+}1)$-point Koba--Nielsen integral---does not depend on the integration variables and we leave it out to simplify some of the following expressions. Note that the variables $\eta$ and $\bar\eta$ do not appear symmetrically in definition~\eqref{eq:Y2pt} due to the $(\tau{-}\bar\tau)$ multiplying $\eta$ but not $\bar\eta$. As we shall see below this effects the modular transformation of the generating series.

\sm

{}From the expansion~\eqref{looprev.6} of the Kronecker--Eisenstein series we can define component integrals by Laurent expanding to a given order in the variables $\eta$ and $\bar\eta$ by
\begin{align}
Y_{ij} (z_0,\eta,\bar\eta|\tau)  = \sum_{a,b\geq 0} \eta^{a-1} \bar\eta^{b-1}  (2\pi i)^b Y_{ij}^{(a|b)}(z_0|\tau)
\label{cptexp}
\end{align}
where the factor of $(2\pi i)^b$ is a convenient convention. Henceforth, we shall suppress the arguments $z_0$ and $\tau$ on the component integrals and use the shorthand
\begin{align}
f_{ij}^{(a)} = f^{(a)}(z_{ij}|\tau)
\end{align}
to compactly represent their integrands. The explicit form of the component integrals
\begin{align}
\label{eq:Yab}
Y_{ij}^{(a|b)} = \frac{(\tau{-}\bar\tau)^a}{(2\pi i)^b} \int_\Sigma \frac{\dd^2z_2}{\Im\tau} e^{s_{02} g(z_{02}|\tau) + s_{12}g(z_{12}|\tau)} \begin{pmatrix} f_{12}^{(a)} \overline{f_{12}^{(b)}} & f_{02}^{(a)} \overline{f_{12}^{(b)}}\\
 f_{12}^{(a)} \overline{f_{02}^{(b)}} & f_{02}^{(a)} \overline{f_{02}^{(b)}} 
\end{pmatrix}_{ij}
\end{align}
manifests their modular weights $(0,b{-}a)$ under~\eqref{modPSL}
\begin{align}
Y_{ij}^{(a|b)}( z'| \tau') = (\gamma \bar\tau+\delta)^{b-a} Y_{ij}^{(a|b)}( z| \tau)
\label{cptmod}
\end{align}
The second arguments $(\tau{-}\bar \tau)\eta$ and $\bar \eta$ of the
Kronecker--Eisenstein series in the generating series (\ref{eq:Y2pt}) are engineered to attain vanishing holomorphic modular weights, at the expense of introducing factors of $\Im \tau$ into their complex-conjugation properties
\begin{align}
\overline{Y_{ij}^{(a|b)}} = (4y)^{a-b} Y_{ji}^{(b|a)} \quad\quad \text{with} \quad \quad y=\pi \Im\tau
\end{align}
The generating function~\eqref{eq:Y2pt} transforms as
\begin{align}
Y_{ij} \bigg(\frac{z_0}{\gamma\tau+\delta},(\gamma \bar \tau + \delta) \eta , \frac{ \bar \eta}{\gamma \bar \tau + \delta} \Big| \frac{ \alpha \tau + \beta}{\gamma \tau + \delta} \bigg) = Y_{ij} (z_0,\eta,\bar \eta|\tau)
\end{align}
showing the asymmetric role played by $\eta$ and $\bar\eta$.

\subsubsection{Expansion of component integrals}

Expanding the component integrals (\ref{eq:Yab}) in the Mandelstam variables $s_{02}$ and $s_{12}$ yields dihedral eMGFs, and the powers
of $\Im \tau$ in (\ref{eq:Yab}) line up with those in the normalization of the ${\cal C}^+$ in (\ref{gen.66}).
For instance, for $a,b\neq 0$ the $11$-component is
\begin{align}
\label{eq:Yabij}
Y_{11}^{(a|b)} &=  \frac{(\tau{-}\bar\tau)^a}{(2\pi i)^b}  \sum_{k,\ell\geq 0} \frac{s_{02}^k s_{12}^\ell}{k! \, \ell!}\int_\Sigma \frac{\dd^2z_2}{\Im\tau}   (g(z_{02}|\tau))^k  (g(z_{12}|\tau))^\ell  f^{(a)}_{12} \overline{f^{(b)}_{12}}\nn\\
&=(-1)^{b}  (2i)^{a-b} \sum_{k,\ell\geq 0} \frac{s_{02}^k s_{12}^\ell}{k! \, \ell!} \cplus{a & 0 & 1_k & 1_\ell\\ 0 & b & 1_k & 1_\ell \\ 0& 0 & z_k & 0_\ell} \, , \ \ \ \ \ \ (a,b) \neq (1,1)
\end{align}
Within the arrays, $z_k = (z_0,\ldots,z_0)$ denotes $k$ repeated entries and similarly for $0$ and $1$. Here and below, we rename $z_0 \rightarrow z$ within the entries of eMGFs to avoid cluttering. The case
$(a,b)=(1,1)$ is excluded from (\ref{eq:Yabij}) since the short-distance behavior 
$ f^{(1)}_{12} \overline{f^{(1)}_{12}} \sim |z_{12}|^{-2}$ and $e^{s_{12} g(z_{12}|\tau)} \sim |z_{ij}|^{-2s_{ij}}$ introduces a kinematic pole in
$s_{12}$ upon integration over $z_2$. This pole can be exposed by standard subtraction schemes
as for instance used in appendix D of \cite{Gerken:2019cxz}
\begin{align}
Y_{11}^{(1|1)} &=
 \frac{\Im \tau}{ \pi } 
\sum_{\ell\geq 0} \sum_{k\geq 1} \frac{s_{02}^k s_{12}^\ell}{k! \, \ell!}
\int_{\Sigma} \frac{ \dd^2 z_2 }{\Im \tau} 
\,(g(z_{12}|\tau))^\ell \big( g(z_{02}|\tau)^k - g(z_{01}|\tau)^k \big) f^{(1)}_{12} \overline{f^{(1)}_{12}}  \notag \\
&\ \ \ \ - \frac{1}{s_{12}} \sum_{k,\ell\geq 0} \frac{s_{02}^k s_{12}^\ell}{k! \, \ell!} g(z_{01}|\tau)^k \int_{\Sigma} \frac{ \dd^2 z_2 }{\Im \tau} \,(g(z_{12}|\tau))^\ell
\label{subpole.1}
\end{align}
see appendix \ref{app:kinsub.1} for a derivation and appendix \ref{app:kinsub.2} for the leading orders of the $\alpha'$-expansion.

\sm

For $a=0$ or $b=0$, the corresponding edges disappear from (\ref{eq:Yabij}), and we have
\begin{align}
Y_{11}^{(a|0)} &= - (2i)^{a} \sum_{k,\ell\geq 0} \frac{s_{02}^k s_{12}^\ell}{k! \, \ell!} \cplus{a & 1_k & 1_\ell\\ 0 & 1_k & 1_\ell \\ 0& z_k & 0_\ell}   &&\hspace{-2mm}\text{for $a\neq 0$}  \nn\\
Y_{11}^{(0|b)} &=(-1)^{b-1}  (2i)^{-b} \sum_{k,\ell\geq 0} \frac{s_{02}^k s_{12}^\ell}{k! \, \ell!} \cplus{0 & 1_k & 1_\ell\\ b & 1_k & 1_\ell \\ 0& z_k & 0_\ell}  &&\hspace{-2mm}\text{for $b\neq 0$}  \label{Yspecial}\\
Y_{11}^{(0|0)} &=  \sum_{k,\ell\geq 0} \frac{s_{02}^k s_{12}^\ell}{k! \, \ell!} \cplus{1_k & 1_\ell\\ 1_k & 1_\ell \\ z_k & 0_\ell}\nn
\end{align}
The other components $Y_{ij}^{(a|b)}$ only differ by the characteristics in the first
or second column of the eMGFs in the expansions (with $a,b\neq 0$ and $(a,b) \neq (1,1)$)
\begin{align}
Y_{12}^{(a|b)} &=(-1)^{b}  (2i)^{a-b} \sum_{k,\ell\geq 0} \frac{s_{02}^k s_{12}^\ell}{k! \, \ell!} \cplus{a & 0 & 1_k & 1_\ell\\ 0 & b & 1_k & 1_\ell \\ z& 0 & z_k & 0_\ell}
 \notag \\
Y_{21}^{(a|b)} &=(-1)^{b}  (2i)^{a-b} \sum_{k,\ell\geq 0} \frac{s_{02}^k s_{12}^\ell}{k! \, \ell!} \cplus{a & 0 & 1_k & 1_\ell\\ 0 & b & 1_k & 1_\ell \\ 0& z & z_k & 0_\ell}
\label{alt:Yabij} \\
Y_{22}^{(a|b)} &=(-1)^{b}  (2i)^{a-b} \sum_{k,\ell\geq 0} \frac{s_{02}^k s_{12}^\ell}{k! \, \ell!} \cplus{a & 0 & 1_k & 1_\ell\\ 0 & b & 1_k & 1_\ell \\ z& z & z_k & 0_\ell}
 \notag 
 \end{align}
and the special cases with $a=0$ or $b=0$ are similar to (\ref{Yspecial}). 
We note in particular that the pure Koba--Nielsen integrals satisfy
\begin{align}
Y_{11}^{(0|0)} = Y_{12}^{(0|0)} = Y_{21}^{(0|0)} = Y_{22}^{(0|0)}
\end{align}
The extensions of (\ref{alt:Yabij}) to $(a,b)=(1,1)$ can be reduced to
$Y_{11}^{(1|1)}$ in (\ref{subpole.1}) and
\beq
Y_{12}^{(1|1)} = - \frac{1}{s_{02}} \big( Y_{11}^{(0|0)} + s_{12} Y_{11}^{(1|1)}\big)
\label{subpole.3}
\eeq
by the following operations evident from the integral representations (\ref{eq:Yab})
\beq
Y_{22}^{(1|1)} = Y_{11}^{(1|1)} \, \big|^{z_0 \rightarrow - z_0}_{s_{12} \leftrightarrow s_{02}} \, , \ \ \ \ \ \ 
Y_{21}^{(1|1)} = \overline{ Y_{12}^{(1|1)} }
\eeq
As detailed in appendix \ref{app:kinsub.2}, the limit $z_0 \rightarrow 0$ of $Y_{12}^{(1|1)}$
does not commute with its $\alpha'$-expansion, so we define the latter through the
integration-by-parts identity (\ref{subpole.3}).

\subsubsection{Examples of component-integral expansions}

Examples of the above component integral expansions are
\begin{align}
Y_{11}^{(0|0)} &= 1 + \frac12 (s_{02}^2+ s_{12}^2)\cplus{2 & 0\\ 2 & 0 \\ 0 & 0 }+ s_{02}s_{12} \cplus{2 & 0\\ 2& 0 \\ z &0} + \frac16 (s_{02}^3 +s_{12}^3) \cplus{1&1&1\\1&1&1\\0&0&0}\nn\\
&\quad  + \frac12 s_{02} s_{12} (s_{02}+s_{12}) \cplus{1&1&1\\1&1&1\\z&0&0} + \frac{1}{24} (s_{02}^4 +s_{12}^4) \cplus{1&1&1 &1\\1&1&1&1\\0&0&0&0} \label{eq:Yabex1}\\
&\quad  +\frac{1}{6}s_{02}s_{12} (s_{02}^2+s_{12}^2) \cplus{1&1&1 &1\\1&1&1&1\\z&0&0&0}
+\frac{1}{4} s_{02}^2s_{12}^2 \cplus{1&1&1 &1\\1&1&1&1\\z&z&0&0} + \mathcal{O}(s_{ij}^5)\nn
\end{align}
as well as
\begin{align}
\label{eq:Yabex2}
Y_{11}^{(1|0)} &= 
2i s_{02} \cplus{2 &0\\1&0\\z&0} 
 +i s_{02}^2 \cplus{1&1&1\\0&1&1\\z&0&0}   +i s_{02}s_{12} \cplus{1&1&1\\0&1&1\\z&0&0}+ \mathcal{O}(s_{ij}^3) \nn\\
Y_{11}^{(2|0)} &=  4 s_{02} \cplus{3&0\\1&0\\z&0} +4 s_{12} \cplus{3&0\\1&0\\0&0} +2 s_{12}^2 \cplus{2&1&1\\0&1&1\\0&0&0}+2 s_{02}^2 \cplus{2&1&1\\0&1&1\\z&0&0}\nn\\
&\quad +4 s_{12}s_{02} \cplus{2&1&1\\0&1&1\\0&z&0} +\mathcal{O}(s_{ij}^3)\nn\\
Y_{11}^{(3|0)} &= 
- 8i s_{02} \cplus{4 &0\\1&0\\z&0} -4is_{02}^2 \cplus{3&1&1\\0&1&1\\z&0&0} + 8i s_{02} s_{12} \cplus{3&1&1\\0&1&1\\0&z&0} +\mathcal{O}(s_{ij}^3) \nn \\
%%%%%%%
Y_{12}^{(1|0)} &= -2i s_{12} \cplus{2&0\\1&0\\z&0} - i s_{02}s_{12}  \cplus{1&1&1\\0&1&1\\z&0&0}
 - i s_{12}^2 \cplus{1&1&1\\0&1&1\\z&0&0} +\mathcal{O}(s_{ij}^3)\nn\\
Y_{12}^{(2|0)} &= 4 s_{02} \cplus{3&0\\1&0\\0&0} + 4s_{12} \cplus{3&0\\1&0\\z&0} + 2 s_{02}^2 \cplus{2&1&1\\0&1&1\\0&0&0}+2s_{12}^2 \cplus{2&1&1\\0&1&1\\z&0&0} \nn\\
&\quad + 4s_{02}s_{12} \cplus{2&1&1\\0&1&1\\0&z&0} + \mathcal{O}(s_{ij}^3) \nn \\
Y_{12}^{(3|0)} &= 
 8i s_{12} \cplus{4 &0\\1&0\\z&0} +4i s_{12}^2  \cplus{3&1&1\\0&1&1\\z&0&0}  - 8i s_{02}s_{12} \cplus{3&1&1\\0&1&1\\0&z&0} +\mathcal{O}(s_{ij}^3) 
\end{align}
where we used the simplification rules from section~\ref{sec:2.1.2}, for instance translation invariance or momentum conservation to rewrite
\begin{align}
\cplus{1&1&1\\1&1&1\\z&z&0} = \cplus{1&1&1\\1&1&1\\z&0&0}
\, , \ \ \ \ \ \
\cplus{1&1&1\\0&1&1\\0&z&0} = -\frac12 \cplus{1&1&1\\0&1&1\\z&0&0}
\end{align}
In later examples, we will also take advantage of the expansions
\begin{align}
Y_{11}^{(4|0)} &= - 16 \, \Big\{
s_{02} \cplus{5&0\\1&0\\z&0}  + s_{12} \cplus{5&0\\1&0\\0&0}  
+ s_{02} s_{12} \cplus{4&1 &1\\0&1 &1\\0 &z&0}   \notag \\
& \ \ \ \ \ \ \ \
+ \frac{1}{2} s_{02}^2 \cplus{4&1 &1\\0&1 &1\\z &0&0}  
+ \frac{1}{2} s_{12}^2 \cplus{4&1 &1\\0&1 &1\\0 &0&0}   + {\cal O}(s_{ij}^3)
\Big\}  \notag\\
Y_{12}^{(4|0)} &=- 16 \, \Big\{
s_{02} \cplus{5&0\\1&0\\0&0}  + s_{12} \cplus{5&0\\1&0\\z&0}  
+ s_{02} s_{12} \cplus{4&1 &1\\0&1 &1\\0 &z&0}   \notag \\
& \ \ \ \ \ \ \ \
+ \frac{1}{2} s_{02}^2 \cplus{4&1 &1\\0&1 &1\\0 &0&0}  
+ \frac{1}{2} s_{12}^2 \cplus{4&1 &1\\0&1 &1\\z &0&0}   + {\cal O}(s_{ij}^3)
\Big\} 
\label{eq:Yabex3}
\end{align}
We note that some of the terms in the expansion are MGFs rather than eMGFs.

%%%%%%%%%%%%%%%%%%%%%%%%%%%%%%%%%%%%%%%%%%%%%%%%%%%%%%%%%%%
\subsection{$z_0$-derivatives of the generating series}
\label{sec:3.2}
%%%%%%%%%%%%%%%%%%%%%%%%%%%%%%%%%%%%%%%%%%%%%%%%%%%%%%%%%%%

The dependence of the generating series~\eqref{eq:Y2pt} on the elliptic variable $z_0$ can be determined using the differential operator $\nabla_{z_0}=2i \Im \tau \partial_{z_0}$ defined in~\eqref{diff.1}.  For this we record the following derivatives of the Kronecker--Eisenstein series~\cite{BrownLev,Gerken:2019cxz}\footnote{The contributions
from the delta function in (\ref{eq:dzOmbar}) are suppressed within Koba--Nielsen integrals since 
factors of $e^{s_{ij}g(z_{ij}|\tau)}$ scale with $|z_{ij}|^{-2s_{ij}}$ as $z_i\rightarrow z_j$. Their 
vanishing in the limit of colliding punctures stems from the analytic continuation from the region with
$\Re(s_{ij})<0$, or the ``cancelled-propagator argument'' in old string-theory lingo.}
\begin{align}
\label{eq:dzOmbar}
\partial_z \overline{  \Omega(z,\eta|\tau) } &= - \frac{2\pi i \bar \eta}{\tau{-}\bar \tau} \overline{  \Omega(z,\eta|\tau) } + \pi \delta^{(2)}(z,\bar{z})
\end{align}
and recall~\eqref{looprev.7} for $z$-derivatives of the Green function. Moreover, the following identity~\cite{Gerken:2019cxz} is a variant of the Fay identities~\eqref{diff.11}
\begin{align}
 f^{(1)}_{02}  \Omega(z_{12},\eta|\tau) &= 
f^{(1)}_{01}\Omega(z_{12},\eta|\tau) + \partial_\eta  \Omega(z_{12},\eta|\tau) 
-  \Omega(z_{01},-\eta|\tau)   \Omega(z_{02},\eta|\tau) \label{gen.9a}
\end{align}

Using these it is straight-forward to compute 
\begin{align}
\nabla_{z_0} Y_{11} &= -s_{02}\left[ (\tau{-}\bar\tau)  f^{(1)}_{01} Y_{11} + \partial_\eta Y_{11} - (\tau{-}\bar\tau)\Omega(z_{01},- (\tau{-}\bar\tau) \eta|\tau) Y_{12}\right] \notag \\
\nabla_{z_0} Y_{12} &= 2\pi i \bar\eta Y_{12} +s_{12} \left[  - (\tau{-}\bar\tau) f^{(1)}_{01} Y_{12} + \partial_\eta Y_{12} +  (\tau{-}\bar\tau) \Omega(z_{01}, (\tau{-}\bar\tau) \eta|\tau) Y_{11}\right] \notag \\
\nabla_{z_0} Y_{21} &= - 2\pi i \bar\eta Y_{21} -s_{02} \left[  (\tau{-}\bar\tau) f^{(1)}_{01} Y_{21} + \partial_\eta Y_{21} -  (\tau{-}\bar\tau) \Omega(z_{01}, -(\tau{-}\bar\tau) \eta|\tau) Y_{22}\right] \notag \\
\nabla_{z_0} Y_{22} &= s_{12} \left[  -(\tau{-}\bar\tau) f^{(1)}_{01} Y_{22} + \partial_\eta Y_{22} +  (\tau{-}\bar\tau) \Omega(z_{01}, (\tau{-}\bar\tau) \eta|\tau) Y_{21}\right]
\label{2ptzdif}
\end{align}
For deriving this equation we have used partial integration to rewrite $\partial_{z_0} \Omega(z_{02},(\tau{-}\bar\tau)\eta|\tau) = -\partial_{z_2} \Omega(z_{02},(\tau{-}\bar\tau)\eta|\tau)$ in terms of derivatives on the Koba--Nielsen factor and the $\overline{\Omega}$ term.

\sm

Upon expansion of the Kronecker--Eisenstein series we obtain the matrix form
\begin{align}
\nabla_{z_0} Y_{ij} = 2\pi i \bar\eta (j{-}i) Y_{ij} + \sum_{k\geq 0} (\tau{-}\bar\tau)^k f^{(k)}_{01}\sum_{\ell=1}^2 R_\eta(x_k)_{j\ell} Y_{i\ell}
\label{closedana}
\end{align}
where the following $2\times 2$ matrices no longer depend on $\tau$
\begin{align}
R_\eta(x_0) &= \begin{pmatrix} -s_{02} \partial_\eta & -s_{02} \eta^{-1} \\ s_{12} \eta^{-1} & s_{12} \partial_\eta\end{pmatrix} \notag \\
R_\eta(x_1) &= \begin{pmatrix} -s_{02} & s_{02} \\ s_{12} & -s_{12} \end{pmatrix} \label{eq:xops}\\
R_\eta(x_k) &= \begin{pmatrix} 0 & s_{02}(-\eta)^{k-1} \\ s_{12}\eta^{k-1}&0\end{pmatrix} \hspace{20mm}\text{for $k\geq 2$} \notag
\end{align}
The $z$-derivatives in (\ref{closedana}) generalize those of the 
generating functions of open-string Koba--Nielsen integrals studied in \cite{Broedel:2020tmd}
which were shown in the reference to obey an elliptic KZB-system \cite{KZB, EnriquezEllAss, Hain}.
The operators $R_\eta(x_k)$ furnish a $(2\times2)$-matrix representation (acting on functions of $\eta$) of abstract generators $x_k$ in the KZB equation for $z$-derivatives as discussed in~\cite[\S4]{Broedel:2020tmd} 
and agree with the operators $r_{0,2}(x_k)$ in the reference at $s_{01}=0$. However, there is
no open-string analogue of the first term
$2\pi i \bar\eta (j-i) Y_{ij} $ on the right-hand side of (\ref{closedana}).

\subsubsection{Implications for component integrals}

When evaluated for component integrals $Y_{ij}^{(a|b)}$ as defined in~\eqref{eq:Yab}, these derivatives with respect to $z_0$ become
\begin{align}
\nabla_{z_0} Y_{11}^{(a|b)} &= -s_{02}(\tau{-}\bar\tau) f^{(1)}_{01} Y_{11}^{(a|b)} - a s_{02} Y_{11}^{(a|b)} - s_{02} \sum_{k=0}^{a+1} (-1)^{k} (\tau{-}\bar\tau)^k f_{01}^{(k)} Y_{12}^{(a+1-k|b)}\nn\\
\nabla_{z_0}Y_{21}^{(a|b)} &= -s_{02} (\tau{-}\bar \tau) f^{(1)}_{01} Y_{21}^{(a|b)}
- a s_{02} Y_{21}^{(a+1|b)} - s_{02} \sum_{k=0}^{a+1} (-1)^{k} (\tau{-}\bar \tau)^k f^{(k)}_{01} Y_{22}^{(a+1-k|b)} 
- Y_{21}^{(a|b-1)}\notag \\
\nabla_{z_0}Y_{12}^{(a|b)} &= -s_{12} (\tau{-}\bar \tau) f^{(1)}_{01} Y_{12}^{(a|b)}
+ a s_{12} Y_{12}^{(a+1|b)} + s_{12} \sum_{k=0}^{a+1} (\tau{-}\bar \tau)^k f^{(k)}_{01} Y_{11}^{(a+1-k|b)} 
+ Y_{12}^{(a|b-1)} \notag \\
\nabla_{z_0}Y_{22}^{(a|b)} &= -s_{12} (\tau{-}\bar \tau) f^{(1)}_{01} Y_{22}^{(a|b)}
+ a s_{12} Y_{22}^{(a+1|b)} + s_{12} \sum_{k=0}^{a+1}  (\tau{-}\bar \tau)^k f^{(k)}_{01} Y_{21}^{(a+1-k|b)} 
\label{gen.56} 
\end{align}
see \cite{Broedel:2019gba} for similar differential equations of open-string component integrals.
By the $\ap$-expansions (\ref{eq:Yabij}) to (\ref{alt:Yabij}) in terms of eMGFs, these
differential equations are generating functions of $z$-derivatives of individual eMGFs in (\ref{gen.66C}). Moreover,
all the dihedral HSRs (\ref{diff.15}) are automatically performed on the right-hand side of (\ref{gen.56}): 
The one-loop eMGFs $(\tau{-}\bar \tau)^k f^{(k)}_{01} = -(2i)^k \dplus{k \\0} \!(z_{01}|\tau)$ generated
by HSR are already explicit, and none of the terms in the $\alpha'$-expansions (\ref{eq:Yabij}) to (\ref{alt:Yabij})
of $Y_{ij}^{(a|b)}$ are amenable to further HSR. 

\sm

Let us consider an example by using the expansions~\eqref{eq:Yabex1} and \eqref{eq:Yabex2} in terms of eMGFs. The differential equations~\eqref{gen.56} give for instance
\begin{align}
\nabla_{z_0} Y_{11}^{(0|0)} =- s_{02} Y_{12}^{(1|0)} 
\label{simp:z0}
\end{align}
which at the order of $s_{02}^2 s_{12}$ implies from~\eqref{eq:Yabex1} and \eqref{eq:Yabex2} that ($z=z_0$)
\begin{align}
\frac12 \nabla_{z} \cplus{1&1&1\\1&1&1\\z&0&0} = i   \cplus{1&1&1\\0&1&1\\z&0&0} 
\end{align}
This is consistent with the $z$-derivative of the eMGF on the left-hand side via \eqref{diff.2}. 
The $z_0$-derivatives (\ref{gen.56}) of
several further component integrals are gathered in appendix~\ref{app:compdiff.a}.

%%%%%%%%%%%%%%%%%%%%%%%%%%%%%%%%%%%%%%%%%%%%%%%%%%%%%%%%%%%
\subsection{$\tau$-derivatives of the generating series}
\label{sec:3.3}
%%%%%%%%%%%%%%%%%%%%%%%%%%%%%%%%%%%%%%%%%%%%%%%%%%%%%%%%%%%

We next determine the dependence of the generating series on the modular parameter $\tau$, using the methods of~\cite{Gerken:2019cxz,Gerken:2020yii}. Besides $\partial_\tau \overline{\Omega(z,\eta|\tau)}=0$ at fixed $u$ and $v$, an important identity for this is~\cite{BrownLev,Gerken:2019cxz},
\begin{align}
\label{eq:dtauOm}
\pi \bigg( \frac{ \nabla_\tau}{\Im \tau} + 1 + \eta \partial_\eta \bigg) \Omega(z,\eta|\tau)  &=
(\Im \tau) \partial_z \partial_\eta  \Omega(z,\eta|\tau)
\end{align}
where we recall that the $\tau$-derivative $\nabla_\tau$ defined in (\ref{crdrv}) is taken at fixed co-moving 
coordinates $(u,v)$. When the $\tau$-derivative acts on the Koba--Nielsen factor (\ref{KN2pt}) it generates
\begin{align}
\frac{ \pi \nabla_\tau}{(\Im \tau)^2} \KN_2(z_0) 
=\left(  -s_{02} f_{02}^{(2)} -s_{12} f_{12}^{(2)}\right) \KN_2(z_0)
\end{align}
and therefore we also require the following consequences of the Fay identity involving $f_{ij}^{(2)}$~\cite{Gerken:2019cxz}
\begin{align}
\label{gen.10}
\big( f^{(1)}_{02} \partial_\eta - f^{(2)}_{02}\big) \Omega(z_{12},\eta|\tau) &= 
\Big(\frac{1}{2} \partial_\eta^2 - f^{(2)}_{01}\Big) \Omega(z_{12},\eta|\tau) 
- \Omega(z_{02},\eta|\tau)  \partial_\eta \Omega(z_{01},-\eta|\tau) \\
\big( f^{(1)}_{12}\partial_\eta - f^{(2)}_{12} \big) \Omega(z_{12},\eta|\tau) &= 
\Big(\frac{1}{2} \partial_\eta^2 - \wp(\eta|\tau) \Big) \Omega(z_{12},\eta|\tau) \label{gen.9}
\end{align}
where $\wp(\eta|\tau)$ is the Weierstra\ss{} function that has the following expansion in $\eta$
\begin{align}
\wp(\eta|\tau)  = \frac{1}{\eta^2} + \sum_{k=4}^\infty (k{-}1) \eta^{k-2} G_k(\tau)
\end{align}
with $G_k(\tau)$ the holomorphic Eisenstein series~\eqref{gen.670} that vanishes for odd $k$. 

\sm

Evaluating the $\tau$-derivative of the generating series $Y_{ij}$ of~\eqref{eq:Y2pt} leads to the Cauchy--Riemann equation,
\begin{align}
-4\pi  \nabla_\tau Y_{j1} &= 2\pi i \bar \eta \partial_\eta Y_{j1}   
-  s_{02} (\tau{-}\bar \tau)^2 f^{(2)}_{01} Y_{j1} 
+ \frac{1}{2}(s_{02}{+}s_{12}) \partial_\eta^2 Y_{j1}\notag \\
& \ \ \ \  
 - s_{02} (\tau{-}\bar \tau) \big( \partial_\eta \Omega(z_{01},-(\tau{-}\bar \tau)\eta\big)  \big)Y_{j2}
 - s_{12} (\tau{-}\bar \tau)^2 \wp\big((\tau{-}\bar \tau)\eta\big)  Y_{j1} 
\notag \\
-4\pi \nabla_\tau Y_{j2} &= 2\pi i \bar \eta \partial_\eta Y_{j2} 
-  s_{12} (\tau{-}\bar \tau)^2 f^{(2)}_{01} Y_{j2} 
+ \frac{1}{2}(s_{02}{+}s_{12}) \partial_\eta^2 Y_{j2} \label{new2ptdiff} \\
& \ \ \ \
+  s_{12} (\tau{-}\bar \tau) \big( \partial_\eta \Omega\big(z_{01},(\tau{-}\bar \tau)\eta\big) \big)Y_{j1} 
 - s_{02} (\tau{-}\bar \tau)^2 \wp\big((\tau{-}\bar \tau)\eta\big)  Y_{j2}   \notag
\end{align}
By expanding both the Weierstra\ss{} function and $\partial_\eta\Omega(z_{01},\pm (\tau{-}\bar \tau)\eta)$ in $\eta$, we obtain,
\begin{align}
-4\pi \nabla_\tau Y_{ij} =\sum_{\ell=1}^2 \Big[ 
{-} R_\eta(\epsilon_0) &+ \sum_{k=2}^\infty (k{-}1) (\tau{-}\bar \tau)^k f^{(k)}_{01} R_\eta(b_k)
\label{eq:dtauY} \\
&+\sum_{k=4}^\infty (1{-}k) (\tau{-}\bar \tau)^k G_k R_\eta(\epsilon_k)
\Big]_{j\ell} Y_{i\ell} 
\notag
\end{align}
where the following $2\times 2$ matrices no longer depend on $\tau$,
\begin{align}
R_{\eta}(\epsilon_0) &= \frac{1}{\eta^2}\begin{pmatrix} s_{12} &s_{02} \\ s_{12} &s_{02} \end{pmatrix} - \frac{1}{2}(s_{02}{+}s_{12}) \partial_\eta^2 - 2\pi i \bar \eta \partial_\eta \notag \\
R_{\eta}(\epsilon_k) &= \begin{pmatrix} \eta^{k-2} s_{12} &0 \\ 0 &\eta^{k-2} s_{02} \end{pmatrix}  \ \ \ \ \ \ \ \ \ \ \ \ \ \ \ \text{for $k\geq 4$}
\label{gen.23} \\
R_{\eta}(b_2) &=\begin{pmatrix} - s_{02} &s_{02} \\ s_{12}  &-s_{12} \end{pmatrix} \notag \\
R_{\eta}(b_k) &= \begin{pmatrix} 0    &(-\eta)^{k-2}s_{02}  \\ \eta^{k-2}s_{12} &0 \end{pmatrix}  \, \ \ \ \ \ \ \ \ \ \ \text{for $k\geq 3$} 
\notag
\end{align}
The steps for this calculation are analogous to~\cite{Gerken:2019cxz,Broedel:2020tmd}.
The terms involving $f_{01}^{(k)}$ in~\eqref{eq:dtauY} are due to the elliptic variable $z_0$ and do not feature in the generating series of MGFs studied in~\cite{Gerken:2019cxz,Gerken:2020yii}, whereas they appear in an analogous open-string calculation in~\cite[\S4]{Broedel:2020tmd} whose normalization we have adopted here.\footnote{Note that, apart from the absence of $s_{01}$ in this work, the 
%most drastic
only difference to the open-string $2\times 2$ representations $r_{0,2}(\cdot)$ in~\cite{Broedel:2020tmd} occurs in
$R_{\eta}(\epsilon_0)= r_{0,2}(\epsilon_0)-2 \zeta_2 (s_{02}{+}s_{12}) - 2\pi i \bar \eta \partial_\eta$.}
By inspection of the explicit operators in~\eqref{gen.23} and comparison with~\eqref{eq:xops}, we see that,
\begin{align}
R_\eta(b_k) = R_\eta (x_{k-1}) \hspace{10mm} \text{for $k\geq 2$}
\end{align}
which holds in identical form for their open-string analogues in~\cite{Broedel:2020tmd}
by the integrability of the elliptic KZB system.

\sm

Similar operators $R_\eta(\epsilon_k)$ in the $\tau$-derivatives of generating functions
without the extra puncture~\cite{Mafra:2019ddf, Mafra:2019xms, Gerken:2019cxz} 
were conjectured to furnish a matrix operator representation of Tsunogai's derivation algebra~\cite{Tsunogai,Pollack}.
As we will see in sections \ref{sec:3.6} and \ref{sec:4.comm}, the operators $R_\eta(\epsilon_k), R_\eta(b_k)$ 
in (\ref{gen.23}) and their generalizations to $n\geq 3$ points obey modified commutation relations studied
in the context of an elliptic KZB system \cite{Broedel:2020tmd}.

\subsubsection{Implications for dihedral eMGFs as single-valued elliptic polylogarithms}
\label{imp2pt}

Since all the matrix entries $Y_{ij}$ generate dihedral eMGFs upon expansion in $\eta,\bar \eta,s_{02}$ and $s_{12}$, 
the form of the $\tau$-derivative in (\ref{eq:dtauY}) implies that they generalize Zagier's single-valued elliptic polylogarithms to arbitrary depth:
We have seen examples of eMGFs at depth two and three in sections \ref{sec:dp2} and \ref{sec:dp3}
whose $\tau$-derivatives introduce $f^{(k)}_{01} \rightarrow f^{(k)}(z)$ multiplying lower-depth objects
and thereby generalize the differential equations (\ref{diff.6}) at depth one.
From the $f^{(k)}_{01}$ that multiply the entire generating series in (\ref{eq:dtauY}), there cannot be 
an upper limit to the depth of the eMGFs in the $\alpha'$-expansions (\ref{eq:Yabij}) and (\ref{alt:Yabij}).
Conversely, a solution of (\ref{gen.9}) via Picard iteration will be explored in \cite{toappsoon}
where iterated integrals over $\tau$ with an unbounded number of integration kernels 
$f^{(k)}(u\tau{+}v|\tau)$ are introduced.

\subsubsection{Implications for component integrals}

The Cauchy--Riemann equation~\eqref{eq:dtauY} can be considered for the component integrals $Y_{ij}^{(a|b)}$ introduced in~\eqref{eq:Yab}. For these we obtain (for $j=1,2$ and any $a,b\geq 0$),
\begin{align}
-4\pi \nabla_\tau Y_{j1}^{(a|b)} &= aY_{j1}^{(a+1|b-1)} - s_{02} (\tau{-}\bar \tau)^2 f^{(2)}_{01} Y_{j1}^{(a|b)} - s_{02} Y_{j2}^{(a+2|b)} \notag \\
& \ \ \ \ + \frac{1}{2} s_{12}(a{-}1)(a{+}2) Y_{j1}^{(a+2|b)} 
+ \frac{1}{2} s_{02} a (a{+}1) Y_{j1}^{(a+2|b)}  \notag \\
&\ \ \ \ - s_{12} \sum_{k=4}^{a+2} (k{-}1) (\tau{-}\bar \tau)^k G_k Y_{j1}^{(a+2-k|b)} \notag \\
&\ \ \ \ + s_{02} \sum_{k=2}^{a+2} (-1)^k (k{-}1) (\tau{-}\bar \tau)^k f^{(k)}_{01} Y_{j2}^{(a+2-k|b)} \notag \\
%%%
%%%
-4\pi \nabla_\tau  Y_{j2}^{(a|b)} &= aY_{j2}^{(a+1|b-1)} -s_{12} (\tau{-}\bar \tau)^2 f^{(2)}_{01} Y_{j2}^{(a|b)}
- s_{12} Y_{j1}^{(a+2|b)} \label{gen.57}  \\
&\ \ \ \ + \frac{1}{2} s_{02}(a{-}1)(a{+}2) Y_{j2}^{(a+2|b)} 
+ \frac{1}{2} s_{12} a (a{+}1) Y_{j2}^{(a+2|b)}  \notag \\
&\ \ \ \ - s_{02} \sum_{k=4}^{a+2} (k{-}1) (\tau{-}\bar \tau)^k G_k Y_{j2}^{(a+2-k|b)} \notag \\
&\ \ \ \ + s_{12} \sum_{k=2}^{a+2}  (k{-}1) (\tau{-}\bar \tau)^k f^{(k)}_{01} Y_{j1}^{(a+2-k|b)} \notag
\end{align}
Similar to (\ref{gen.56}), these differential equations generate the $\tau$-derivatives of individual eMGFs in (\ref{gen.66Ctau}) upon $\alpha'$-expansion. Again, the factors of $ (\tau{-}\bar \tau)^k f^{(k)}_{01} $ and $(\tau{-}\bar \tau)^k G_k$ signal that dihedral HSRs (\ref{diff.15}) and their coincident limits (\ref{diff.17}) are automatically performed.
Furthermore, by the absence of $\dplus{1  \\ 0  }\! (z_{01})$ 
and $\dplus{2  \\ 0  }\! (0)$ from (\ref{gen.57}), the HSR incorporated in these
all-order differential equations are all from the subclass (\ref{diff.18}).

\sm

Let us consider an example of a $\tau$-derivative of the generating series of eMGFs, focussing again on the simplest case $Y_{11}^{(0|0)}$
\begin{align}
-4\pi \nabla_\tau Y_{11}^{(0| 0)} &=
 - s_{12} Y_{11}^{(2| 0)} - s_{02} Y_{12}^{(2| 0)}
 \label{simp:tau}
\end{align}
Referring back to the component expansion in~\eqref{eq:Yabex1} and \eqref{eq:Yabex2}, we can consider the order $s_{02}s_{12}^2$ of this equation which becomes,
\begin{align}
 \pi \nabla_{\tau} \cplus{1&1&1\\1&1&1\\z&0&0} = 2 \cplus{2&1&1\\0&1&1\\0&z&0} + \cplus{2&1&1\\0&1&1\\z&0&0}
\end{align}
which is consistent with~\eqref{gen.66Ctau}.
A variety of further examples of $\nabla_\tau Y_{11}^{(0| 0)}$ can be found in appendix \ref{app:compdiff.b}.

%%%%%%%%%%%%%%%%%%%%%%%%%%%%%%%%%%%%%%%%%%%%%%%%%%%%%%%%%%%
\subsection{Commutation relations from integrability}
\label{sec:3.6}
%%%%%%%%%%%%%%%%%%%%%%%%%%%%%%%%%%%%%%%%%%%%%%%%%%%%%%%%%%%

Similar to the discussion of the elliptic KZB system obeyed by the open-string analogue of
the generating series (\ref{eq:Y2pt}) \cite{Broedel:2020tmd}, the commutativity of mixed
derivatives\footnote{For the differential operators $\nabla_{\tau}$ and $\nabla_{z}$ defined
in (\ref{crdrv}) and (\ref{diff.1}), the following identity is useful for the comparison of mixed derivatives,
\[
-4\pi \nabla_\tau \big[ (\tau{-}\bar \tau)^k f^{(k)}(z|\tau) \big] = k \nabla_{z} 
(\tau{-}\bar \tau)^{k+1} f^{(k+1)}(z|\tau) 
\]
We recall that only the holomorphic versions $\nabla_{z}$ and $\nabla_{\tau}$ commute,
whereas $\overline{\nabla}_{z}$ does not commute with $\nabla_{\tau}$ at fixed $(u,v)$.} $\nabla_{z_0}\nabla_{\tau}$ implies commutation relations of the operators  $R_\eta(x_k),R_\eta(\epsilon_k),R_\eta(b_k)$. From their appearance in the $z_0$- and
$\tau$-derivatives (\ref{closedana}) and (\ref{eq:dtauY}), we obtain the direct analogues,
\begin{align}
R_\eta([x_0 ,\epsilon_k] ) &= \sum_{a=1}^{k/2-1} (-1)^a R_\eta([x_a , x_{k-1-a}]) \, , &&k\geq 4\label{likeopen}
\\
R_\eta([x_w,\epsilon_k] ) &= - \sum_{a=0}^{w-1} {w{-}1\choose a} R_\eta( [ x_{a+1}, x_{k+w-a-2}] ) \, , 
&&k\geq 4, \ w\geq 1
\notag
\end{align}
of the relations among the $r_{0,2}(\cdot)$ in the open-string setup \cite{Broedel:2020tmd}, see
appendix \ref{app:comm} for further details. Here
and below, we are using the shorthand $R_\eta([a,b])= R_\eta(a) R_\eta(b) - R_\eta(b) R_\eta(a)$.

\sm

However, the first term $2\pi i \bar \eta(j{-}i)Y_{ij}$ in the $z_0$-derivative (\ref{closedana})
cannot be absorbed into a redefinition of $R_\eta(x_0)$ acting solely on the second index $j$.
This leads to deformations of some of the commutation relations compared to the reference, i.e.\ the
terms $\sim 2\pi i \bar \eta$ in,
\begin{align}
R_\eta([x_0,\epsilon_0])_{j\ell} &=  2\pi i \bar \eta (\ell {-}j) R_\eta(\epsilon_0)_{j \ell} 
\notag
\\
R_\eta([x_w,\epsilon_0])_{j\ell} &=   \sum_{a=0}^{\lfloor w/2 \rfloor - 1}    {w \choose a} 
\frac{(w{-}1{-}2a) }{(a{+}1)} R_\eta( [x_a , x_{w-1-a}] )_{j\ell} \label{defbrack} \\
&\ \ \ \
  + 2\pi i(w{-}1) \bar \eta (j{-}\ell ) R_\eta(x_{w-1})_{j \ell} \, , \ \ \ \  \ \ \ \ w \geq 1 \notag
\end{align}
due to the last line of (\ref{high.16}), for instance,
\begin{align}
R_\eta([x_1,\epsilon_0])_{j\ell} &=  0 \label{exdefbrack}
\\
R_\eta([x_2,\epsilon_0])_{j\ell} &=   R_\eta([x_0,x_1])_{j\ell}
+  2\pi i  \bar \eta (j{-}\ell ) R_\eta(x_{1})_{j \ell} \notag
\end{align}
We observe the following relations for the adjoint action\footnote{The adjoint action is defined as ${\rm ad}_x y = [x,y]$.} of $\ep_0$ on the operators $\ep_k$ and $x_k$ (with $k \geq 1$) under $R_\eta(\cdot)$
\begin{align}
\label{gennilp}
R_\eta( {\rm ad}_{\ep_0}^k \epsilon_{k+1} )&=0 
\\
R_\eta( {\rm ad}_{\ep_0}^k x_k )&=0
\notag
\end{align}
which we have verified up to $k=6$. In the open-string case, when the $\bar\eta$-terms are absent, it was shown in~\cite{Broedel:2020tmd} that the relations~\eqref{defbrack} imply the second line of~\eqref{gennilp}. The first line is reminiscent of Tsunogai's derivation algebra~\cite{Tsunogai,Pollack} whose abstract generators
$\epsilon_k$ obey ${\rm ad}_{\ep_0}^k \epsilon_{k+1} =0$, see section \ref{sec:4.comm} for further details.

\sm

The $n$-point analogues of these relations and others will be discussed in section \ref{sec:4.comm}.
Given that the above commutators apply to the operators $R_\eta(b_k) = R_\eta(x_{k-1})$ in
the $\tau$-derivatives (\ref{eq:dtauY}), they constrain the possible differential equations of
eMGFs. As will be explored in future work \cite{toappsoon}, the above commutation
relations reduce the number of independent iterated integrals in a perturbative solution to the
differential equations of the series $Y_{ij}$. Hence, the commutators of $R_\eta(\epsilon_k)$
and $R_\eta(x_w)$ crucially enter the counting of independent eMGFs
at given modular and transcendental weights, see \cite{Gerken:2020yii} for the analogous counting
of MGFs.

%%%%%%%%%%%%%%%%%%%%%%%%%%%%%%%%%%%%%%%%%%%%%%%%%%%%%%%%%%%
\subsection{Extracting differential equations of eMGFs}
\label{sec:3.5}
%%%%%%%%%%%%%%%%%%%%%%%%%%%%%%%%%%%%%%%%%%%%%%%%%%%%%%%%%%%

We shall now revisit the $\tau$-derivatives (\ref{crexpl.3}), (\ref{finaldd41}) and (\ref{finaldd42})
of specific eMGFs and pinpoint the equivalent of the underlying HSRs at the level of the generating series.
The three examples of this section occur in the $\ap$-expansions (\ref{eq:Yabex1}) and (\ref{eq:Yabex2})
of component integrals via
\begin{align}
i\cplus{1&1&1\\0&1&1\\z&0&0} &= Y_{11}^{(1|0)} \, \big|_{s_{02}^2} \notag \\
\frac{1}{6} \cplus{1&1&1&1\\1&1&1&1\\z&0&0&0}&= Y_{11}^{(0|0)} \, \big|_{s_{02}^3 s_{12}}
\label{fromgen.1}
\\
\frac{1}{4} \cplus{1&1&1&1\\1&1&1&1\\z&z&0&0}&= Y_{11}^{(0|0)} \, \big|_{s_{02}^2 s_{12}^2}
\notag
\end{align}
and generalize the earlier ones in~\eqref{simp:z0} and~\eqref{simp:tau}.
The $\tau$-derivative of the odd eMGF in the first line can be read off from
the coefficient of $s_{02}^2$ on the right-hand side of
\begin{align}
-4\pi \nabchi_\tau Y_{11}^{(1|0)} &=   s_{02} (Y_{11}^{(3|0)} -Y_{12}^{(3|0)} )-2 s_{02}(\tau{-}\bar \tau)^3 f^{(3)}(z)Y_{11}^{(0|0)} \notag \\
&\ \ \ \
+ s_{02}(\tau{-}\bar \tau)^2 f^{(2)}(z)(Y_{12}^{(1|0)}-Y_{11}^{(1|0)})
\label{fromgen.2}
\end{align}
In combination with the $\alpha'$-expansions (\ref{eq:Yabex1}) and (\ref{eq:Yabex2}) of the component
integrals on the right-hand side, this leads to the following equivalent of (\ref{crexpl.3})
\begin{align}
(\pi \nabchi_\tau)\cplus{1 & 1 & 1 \\ 0 & 1 & 1 \\ z & 0 & 0} &=
 2 \dplus{4  \\ 1 }\!(z) 
+ \frac{1}{2}(\tau{-}\bar \tau)^2 f^{(2)}(z) \dplus{2  \\ 1}\!(z) 
\label{fromgen.3}
\end{align}
Similarly, the second $\tau$-derivatives of the weight-four examples in (\ref{fromgen.1}) follow
from the differential equation
\begin{align}
(-4\pi \nabchi_\tau)^2 Y_{11}^{(0|0)} &= -2s_{12}(s_{02}{+}s_{12}) Y_{11}^{(4|0)} -2s_{02}(s_{02}{+}s_{12}) Y_{12}^{(4|0)}  \notag \\
& \ \ \ \
+ 3 (s_{02}^2{+}s_{12}^2) (\tau{-}\bar \tau)^4G_4 Y_{11}^{(0|0)}- 6 s_{02}s_{12} (\tau{-}\bar \tau)^4  f^{(4)}(z)  Y_{11}^{(0|0)} \notag \\
&\ \ \ \ 
+ 2 s_{02}s_{12} (\tau{-}\bar \tau)^3  f^{(3)}(z)  ( Y_{12}^{(1|0)}  - Y_{11}^{(1|0)}  )
\label{fromgen.4}
\end{align}
$\alpha'$-expanding the integrals on the right-hand side via (\ref{eq:Yabex1}), (\ref{eq:Yabex2}), (\ref{eq:Yabex3})
and isolating the coefficients of $s_{02}^3 s_{12}$ and $s_{02}^2 s_{12}^2$ yields the equivalents
\begin{align}
(\pi \nabchi_\tau)^2 D_4^{(1)}(z)  & =
  -\frac{9}{8}(\tau{-}\bar \tau)^4 f^{(4)}(z)\dplus{2 \\ 2 }\!(0) 
+ \frac{9}{8}(\tau{-}\bar \tau)^4 G_4 \dplus{2 \\ 2 }\!(z)  
  \notag \\
&\ \ - \frac{3i}{4} (\tau{-}\bar \tau)^3 f^{(3)}(z)\cplus{1&1&1 \\ 0&1&1 \\ z&0&0} +36 \dplus{6 \\ 2}\!(z) 
+ 36 \dplus{6 \\ 2}\!(0)   - 6 \dplus{3 \\ 1 }\!(z)  ^2 
\notag \\
&\ \ 
- 6 \dplus{3 \\ 1}\!(0) ^2
+ 12 \dplus{4 \\ 1}\!(z)  \dplus{2 \\ 1}\!(z)  
- 12 \dplus{3 \\ 1}\!(z)  \dplus{3 \\ 1}\!(0) 
\label{fromgen.5} \\
%%%%%%%
(\pi \nabchi_\tau)^2 D_4^{(2)}(z)  & = 
\frac{3}{4} (\tau{-}\bar \tau)^4 G_4 \dplus{2 \\ 2 }\!(0)   
- \frac{3}{2} (\tau{-}\bar \tau)^4  f^{(4)}(z) \dplus{2\\ 2}\!(z)
\notag \\
& \ \ 
- i  (\tau{-}\bar \tau)^3 f^{(3)}(z) \cplus{1&1&1 \\ 0&1&1 \\ z&0&0} 
+48 \dplus{6 \\ 2}\!(z) 
+ 24 \dplus{6 \\ 2} \!(0)
\notag \\
&\ \ 
+ 16 \dplus{4 \\ 1}\!(z) \dplus{2 \\ 1} \!(z)
- 8 \dplus{3 \\ 1}\!(z) ^2 
- 16 \dplus{3 \\ 1}\!(z) \dplus{3 \\ 1} \!(0)
\label{fromgen.6}
\end{align}
of (\ref{finaldd41}) and (\ref{finaldd42}). We have again used the
identities (\ref{crexpl.5}) in intermediate steps which can be derived
either from the sieve algorithm as in section \ref{sec:sieveex} or from the 
last line of the all-order differential equations (\ref{gen.65a}).

\sm

The procedure of this section to extract the differential equations of eMGFs 
from those of Koba--Nielsen integrals in (\ref{gen.57}) is particularly efficient 
to systematically derive a large number of $\tau$-derivatives: Instead of 
performing HSR on a case-by-case basis in the $\alpha'$-expansion, the
exposure of $f^{(k)}(z)$ and $G_k$ in (\ref{fromgen.2}) or (\ref{fromgen.4})
applies to all orders in $\alpha'$ and therefore incorporates an infinity
of HSRs.

\newpage

%%%%%%%%%%%%%%%%%%%%%%%%%%%%%%%%%%%%%%%%%%%%%%%%%%%%%%%%%%%
%%%%%%%%%%%%%%%%%%%%%%%%%%%%%%%%%%%%%%%%%%%%%%%%%%%%%%%%%%%
\section{Higher-point eMGFs from Koba--Nielsen integrals}
\label{sec:4}
%%%%%%%%%%%%%%%%%%%%%%%%%%%%%%%%%%%%%%%%%%%%%%%%%%%%%%%%%%%
%%%%%%%%%%%%%%%%%%%%%%%%%%%%%%%%%%%%%%%%%%%%%%%%%%%%%%%%%%%

In this section, the two-point generating functions $Y_{ij}$ in (\ref{eq:Y2pt}) along with their differential
equations in $z_0$ and $\tau$ will be generalized to higher multiplicity. While the two-point
integrals $Y_{ij}$ only generate dihedral eMGFs, the $n$-point
Koba--Nielsen integrals to be introduced below generate eMGFs of arbitrary graph topologies.
The differential equations of all eMGF topologies will be shown to line up with higher-depth 
versions of Zagier's single-valued elliptic polylogarithms. The expansion of $(n\geq 3)$-point
generating series below for instance features the trihedral eMGFs in appendix \ref{app:tri}
and additional two-loop eMGFs that were absent in $Y_{ij}$. Hence, the higher-point
generating functions of this section furnish a variety of single-valued elliptic polylogarithms 
at higher depth that drop out from the two-point setup.

%%%%%%%%%%%%%%%%%%%%%%%%%%%%%%%%%%%%%%%%%%%%%%%
%%%%%%%%%%%%%%%%%%%%%%%%%%%%%%%%%%%%%%%%%%%%%%%
 \subsection{Higher-point generating series}
 \label{sechigh.0}
%%%%%%%%%%%%%%%%%%%%%%%%%%%%%%%%%%%%%%%%%%%%%%%
%%%%%%%%%%%%%%%%%%%%%%%%%%%%%%%%%%%%%%%%%%%%%%%

The $n$-point generating series of this section generalize the two-point integrands
in (\ref{eq:Y2pt}) to multiple factors of left- and right-moving Kronecker--Eisenstein 
series $\Omega$ and $\bar \Omega$.
Following the integrands of earlier generating series of genus-one integrals 
\cite{Mafra:2019ddf, Mafra:2019xms, Gerken:2019cxz, Broedel:2020tmd}, both left- and right-movers will be composed of building blocks
\beq
\varphi_{\vec{\eta}}(k_1,k_2,\ldots,k_r|\tau) = \Omega(z_{k_1 k_2},\eta_{k_2 k_3\ldots k_r}|\tau)
\Omega(z_{k_2 k_3},\eta_{k_3 k_4\ldots k_r}|\tau) \ldots 
\Omega(z_{k_{r-1} k_r},\eta_{k_r}|\tau)
\label{high.1}
\eeq
with $\varphi_{\emptyset}(k|\tau) =1$ and $\varphi_{\eta_\ell}(k,\ell|\tau) = \Omega(z_{k\ell},\eta_\ell|\tau)$ in case of one or two entries. The sums $\eta_{k_i \ldots k_j}=
\eta_{k_i}+\ldots+\eta_{k_j}$ of consecutive expansion variables $\eta_k$ in the second arguments
are engineered to obtain the simple differential equation for the complex conjugate
\beq
\partial_{z_j} \overline{ \varphi_{\vec{\eta}}(1,\rho(2,3,\ldots,r)|\tau) } 
= \frac{ 2\pi i \eta_j}{\tau{-}\bar \tau}  \overline{ \varphi_{\vec{\eta}}(1,\rho(2,3,\ldots,r)|\tau) }  \, , \ \ \ \ \ \ 
j=2,3,\ldots,r
\label{high.2}
\eeq
that does not depend on the permutation $\rho \in S_{r-1}$ of its entries. 

\sm

Following the open-string integrands of \cite{Broedel:2020tmd} that also accommodate an
extra unintegrated puncture $z_0$, 
we will consider products of $\varphi_{\vec{\eta}}(1,\ldots)$ and $\varphi_{\vec{\eta}}(0,\ldots)$ for both chiral halves of their closed-string versions below. The remaining legs $2,3,\ldots,n$ will be distributed over the ellipses of $\varphi_{\vec{\eta}}(1,\ldots)$ and $\varphi_{\vec{\eta}}(0,\ldots)$, and there are $n!$ admissible ways of doing so (including cases with a single entry $\varphi_{\emptyset}(0|\tau) =\varphi_\emptyset(1|\tau)=1$). In an open-string setting with a fixed integration domain $\gamma$ on the $A$-cycle of a torus,  these integrands give rise to the following $n!$-component vectors of integrals,
\beq
Z( \gamma |  \begin{smallmatrix} K \\ L \end{smallmatrix})
= \int_\gamma \bigg( \prod_{j=2}^n  \dd z_j \bigg) 
\varphi_{\vec{\eta}}(1,K)\varphi_{\vec{\eta}}(0,L)
\prod_{0\leq i < j}^n e^{s_{ij} g_{\rm open}(z_{ij}|\tau)}
\label{high.3}
\eeq
that close under derivatives w.r.t.\ $z_0$ and $\tau$ \cite{Broedel:2020tmd}.
The entries of the ordered sequences $K=(k_1k_2\ldots k_i)$ and $L=(\ell_1\ell_2\ldots \ell_j)$
are disjoint and yield $\{k_1,k_2,\ldots,k_i\} \cup \{\ell_1,\ell_2,\ldots,\ell_j\} = \{2,3,\ldots,n\}$ 
as their union. The details of the open-string Green function\footnote{Similar
to the differential equations (\ref{looprev.7}) and (\ref{f2}) of the closed-string
Green function, the open-string Green function in genus-one amplitudes can
be taken to obey
\[
\partial_z g_{\rm open}(z|\tau) = - f^{(1)}(z|\tau) \, , \ \ \ \ \ \
2\pi i \partial_\tau g_{\rm open}(z|\tau) = - f^{(2)}(z|\tau) - 2 \zeta_2
\]} $g_{\rm open}(z_{ij}|\tau)$, the integration cycles $\gamma$ and the association of  
$\varphi_{\vec{\eta}}(1,K)\varphi_{\vec{\eta}}(0,L)$ with two Cayley graphs 
rooted at the unintegrated punctures 
$z_1$ and $z_0$ can be found in the reference.

\sm

The corresponding closed-string generating functions generalizing
the $Y_{ij}$ in (\ref{eq:Y2pt}) are constructed from a double copy 
of the integrand of (\ref{high.3})
\begin{align}
\label{high.4}
Y(\begin{smallmatrix} M \\ N \end{smallmatrix} |  \begin{smallmatrix} K \\ L \end{smallmatrix})
&= (2i\Im\tau)^{n-1} \int\limits_{\Sigma^{n-1}} \bigg( \prod_{j=2}^n \frac{\dd^2 z_j}{\Im\tau}  \bigg) {\rm KN}_n(z_0|\tau)\\
&\hspace{10mm}\times \varphi_{(\tau-\bar \tau) \vec{\eta}}(1,K)\varphi_{(\tau-\bar \tau) \vec{\eta}}(0,L)
\overline{\varphi_{ \vec{\eta}}(1,M)\varphi_{\vec{\eta}}(0,N)}
\nonumber
\end{align}
up to a rescaling of all the formal variables $\eta_2,\eta_3,\ldots,\eta_n$ in the left-moving factors
by  $(\tau{-}\bar \tau)$. This rescaling was already present in the factors of
$\Omega(z_{j2}, (\tau{-}\bar\tau)\eta|\tau) $ in the two-point integrand (\ref{eq:Y2pt}) and ensures that the
holomorphic modular weight of the component integrals in (\ref{high.4}) vanishes. The Koba--Nielsen factor is 
again taken to exclude a contribution of $s_{01}g(z_{01}|\tau)$ that can be pulled out of the integral
\beq
{\rm KN}_n(z_0|\tau) = \prod_{0\leq i<j \atop{(i,j) \neq (0,1)}}^n e^{s_{ij} g(z_{ij}|\tau) }
\label{high.71}
\eeq
and we fix $z_1=0$ by translation invariance as in the previous section. 
We suppress the explicit dependence on $\eta_i,\bar\eta_i,z_0$ and $\tau$
in the definitions~\eqref{high.3} and \eqref{high.4}.

\sm

The entries of the $n! \times n!$ matrix of closed-string integrals in (\ref{high.4}) 
are labelled by $2+2$ ordered sequences for 
the left- and right movers, i.e.\ there are $n!$ choices of 
$\begin{smallmatrix} M \\N \end{smallmatrix}$ and 
$ \begin{smallmatrix} K \\ L \end{smallmatrix}$ each.
For instance, the $2\times 2$ matrix at two points
in (\ref{eq:Y2pt}) arises from the choices 
$\begin{smallmatrix} M \\ N \end{smallmatrix}, \begin{smallmatrix} K \\ L \end{smallmatrix}
\in \{   \begin{smallmatrix} 2 \\ \emptyset \end{smallmatrix},
 \begin{smallmatrix} \emptyset \\ 2 \end{smallmatrix} \}$,
associated with double copies of $\{ \varphi_{ \eta_2}(1,2), \varphi_{\eta_2}(0,2)\}$.
Similarly, the closed-string integrals (\ref{high.4}) at $n=3$ points form a $6\times 6$
matrix indexed by $\begin{smallmatrix} M \\ N \end{smallmatrix}, \begin{smallmatrix} K \\ L \end{smallmatrix}
\in \{   \begin{smallmatrix} 23 \\ \emptyset \end{smallmatrix},
 \begin{smallmatrix} 32 \\ \emptyset \end{smallmatrix},
  \begin{smallmatrix} 2 \\ 3 \end{smallmatrix},
   \begin{smallmatrix} 3 \\ 2 \end{smallmatrix},
 \begin{smallmatrix} \emptyset \\ 32 \end{smallmatrix},
  \begin{smallmatrix} \emptyset \\ 23 \end{smallmatrix} \}$, referring to a double copy of
\beq
n=3 \ \ \Rightarrow \ \ \varphi_{\vec{\eta}}(1,K) \varphi_{\vec{\eta}}(0,L) \in
\left\{ \begin{array}{c}
 \varphi_{ \eta_2,\eta_3}(1,2,3),
 \varphi_{ \eta_2,\eta_3}(1,3,2),
  \varphi_{ \eta_2}(1,2) \varphi_{\eta_3}(0,3), \\
  \varphi_{ \eta_3}(1,3) \varphi_{\eta_2}(0,2),
  \varphi_{ \eta_2,\eta_3}(0,3,2),\varphi_{ \eta_2,\eta_3}(0,2,3)
  \end{array} \right\} 
  \label{high.72}
  \eeq
As will be spelt out below, also the closed-string generating functions (\ref{high.4}) are closed
under differentiation in $z_0$ and $\tau$. This supports the expectation that the $n!$ families of
$\varphi_{ \vec{\eta}}(1,M)\varphi_{\vec{\eta}}(0,N)$ form a basis of chiral integrands in 
Koba--Nielsen integrals at genus one under Fay identities and integration by parts. It would be
interesting to find a general proof and to find a precise formulation in the language of
twisted deRham theory.

\subsubsection{Expansion of component integrals} 
 
The lattice-sum representations of its constituents implies that
the simultaneous Laurent expansion of (\ref{high.4}) in $s_{ij}$, $\eta_j$ and $\bar \eta_j$
yields eMGFs at each order. At fixed order in $\eta_j$ and $\bar \eta_j$, one encounters
component integrals over Kronecker--Eisenstein coefficients as in (\ref{eq:Yab}).

\sm

At three points, for instance, the first entry in the $6\times 6$ matrix yields
\begin{align}
Y^{(a,b|c,d)}_{\begin{smallmatrix} 23 \\ \emptyset \end{smallmatrix} \big| \begin{smallmatrix} 23 \\ \emptyset \end{smallmatrix}} &=  \frac{1}{(2\pi i)^{c+d} } Y(\begin{smallmatrix} 23 \\ \emptyset \end{smallmatrix} |  \begin{smallmatrix} 23 \\ \emptyset \end{smallmatrix})  \, \Big|_{\eta_{23}^{a-1} \eta_3^{b-1} \bar \eta_{23}^{c-1} \bar \eta_3^{d-1}}
\label{high.73} \\
&= (2i)^{a+b-c-d} \frac{( \Im \tau)^{a+b} }{\pi^{c+d}}
 \int_{\Sigma} \frac{ \dd^2 z_2}{\Im \tau} \int_{\Sigma} \frac{ \dd^2 z_3}{\Im \tau}
 f_{12}^{(a)} f_{23}^{(b)} \overline{ f_{12}^{(c)} }\overline{ f_{23}^{(d)} }  {\rm KN}_3(z_0) 
 \notag
\end{align}
Each term in its $\alpha'$-expansion can be lined up with the definition 
(\ref{basic.16}) or integral representation (\ref{basic.44}) of trihedral eMGFs 
\begin{align}
Y^{(a,b|c,d)}_{\begin{smallmatrix} 23 \\ \emptyset \end{smallmatrix} \big| \begin{smallmatrix} 23 \\ \emptyset \end{smallmatrix}} &= 
 (2i)^{a+b-c-d} (-1)^{c+d} \sum_{k_1,\ldots,k_5=0}^{\infty} \frac{ s_{12}^{k_1} s_{13}^{k_2} s_{23}^{k_3} s_{02}^{k_4} s_{03}^{k_5}  }{k_1! k_2!k_3!k_4!k_5! } 
\notag\\
&\ \ \ \ \times \cplustri{a &0 &1_{k_1} &1_{k_4} \\ 0 &c &1_{k_1} &1_{k_4} \\ 0&0 &0_{k_1} &(z)_{k_4} }{b &0 &1_{k_3} \\ 0 &d &1_{k_3} \\ 0 &0 &0_{k_3}}{ 1_{k_2} &1_{k_5} \\ 1_{k_2} &1_{k_5} \\ 0_{k_2} &(-z)_{k_5}}  \label{high.74} 
\end{align}
where we have assumed that $a,b,c,d \neq 0$ to avoid trivialization of edges
and $(a,c),(b,d) \neq (1,1)$ to avoid kinematic poles as in (\ref{subpole.1}).
Moreover, we have not performed any topological simplifications (\ref{topsimp})
for low values of the $k_i$ and again renamed $z_0 \rightarrow z$ to avoid cluttering.
When defining the three-point component integrals of the remaining matrix entries,
the expansion variables $\eta_{23},\eta_3,\bar\eta_{23},\bar\eta_3$ in (\ref{high.73})
need to be adapted to the Kronecker--Eisenstein integrand in (\ref{high.72}), e.g.\
$Y^{(a,b|c,d)}_{\begin{smallmatrix} 2 \\ 3 \end{smallmatrix} \big| \begin{smallmatrix}   \emptyset \\ 23
 \end{smallmatrix}}$ would be the coefficient of $\eta_{23}^{a-1}\eta_{3}^{b-1} \bar\eta_{2}^{c-1} \bar \eta_3^{d-1}$.
 
 \sm

The same approach leads to eMGFs of box-, kite- and tetrahedral eMGFs in the $\alpha'$-expansion
of four-point component integrals, and there are no limitations to the eMGF topologies obtained
from generating functions (\ref{high.4}) at higher multiplicity.

\subsubsection{Two-loop eMGFs and beyond at leading orders in $\alpha'$} 

We shall now focus on the $\alpha' \rightarrow 0$ limit of various three-point
component integrals (\ref{high.73}) and compare the contributions from different matrix entries.
The simplest choice of $\begin{smallmatrix} M \\ N \end{smallmatrix}
= \begin{smallmatrix} 23 \\ \emptyset \end{smallmatrix}
= \begin{smallmatrix} K \\ L \end{smallmatrix} $ in (\ref{high.74}) yields products of one-loop eMGFs
\beq
Y^{(a,b|c,d)}_{\begin{smallmatrix} 23 \\ \emptyset \end{smallmatrix} \big| \begin{smallmatrix} 23 \\ \emptyset \end{smallmatrix}} = (2i)^{a+b-c-d} \cplus{a &0 \\ c &0 \\ 0&0} \cplus{b &0 \\ d &0 \\ 0&0} + {\cal O}(s_{ij})
\label{high.75} 
\eeq
at leading order in $\alpha'$, while different matrix entries turn out to  yield two-loop eMGFs. Here and below, we assume $a,b,c,d \neq 0$ and $(a,c),(b,d) \neq (1,1)$ to avoid trivialization of edges and kinematic poles.  With a different choice of $\overline{ f_{ij}^{(c)} }\overline{ f_{pq}^{(d)} }$ in the  integrand of (\ref{high.73}), we obtain,
\begin{align}
Y^{(a,b|c,d)}_{\begin{smallmatrix} 32 \\ \emptyset \end{smallmatrix} \big| \begin{smallmatrix} 23 \\ \emptyset \end{smallmatrix}} &=  \frac{Y(\begin{smallmatrix} 32 \\ \emptyset \end{smallmatrix} |  \begin{smallmatrix} 23 \\ \emptyset \end{smallmatrix})}{(2\pi i)^{c+d} }   \, \Big|_{\eta_{23}^{a-1} \eta_3^{b-1} \bar \eta_{23}^{c-1} \bar \eta_2^{d-1}}
= (2i)^{a+b-c-d} (-1)^{a+c} \cplus{a &b &0 \\ c &0 &d \\ 0 &0 &0}  + {\cal O}(s_{ij}) \notag \\
Y^{(a,b|c,d)}_{\begin{smallmatrix} 2 \\ 3 \end{smallmatrix} \big| \begin{smallmatrix} 23 \\ \emptyset \end{smallmatrix}} &=  \frac{Y(\begin{smallmatrix} 2 \\ 3\end{smallmatrix} |  \begin{smallmatrix} 23 \\ \emptyset \end{smallmatrix})}{(2\pi i)^{c+d} }   \, \Big|_{\eta_{23}^{a-1} \eta_3^{b-1} \bar \eta_{2}^{c-1} \bar \eta_3^{d-1}}
= (2i)^{a+b-c-d} (-1)^{b+c+d} \cplus{b &a &0 \\ d &0 &c \\ z &0 &0}  + {\cal O}(s_{ij}) \notag \\
Y^{(a,b|c,d)}_{\begin{smallmatrix} 3 \\ 2 \end{smallmatrix} \big| \begin{smallmatrix} 23 \\ \emptyset \end{smallmatrix}} &=  \frac{Y(\begin{smallmatrix} 3 \\ 2 \end{smallmatrix} |  \begin{smallmatrix} 23 \\ \emptyset \end{smallmatrix})}{(2\pi i)^{c+d} }   \, \Big|_{\eta_{23}^{a-1} \eta_3^{b-1} \bar \eta_{3}^{c-1} \bar \eta_2^{d-1}}
= (2i)^{a+b-c-d} (-1)^{b+c+d} \cplus{b &a &0 \\ c &0 &d \\ 0 &0 &z}  + {\cal O}(s_{ij})
\label{high.76} \\
Y^{(a,b|c,d)}_{\begin{smallmatrix}  \emptyset \\ 32 \end{smallmatrix} \big| \begin{smallmatrix} 23 \\ \emptyset \end{smallmatrix}} &=  \frac{Y(\begin{smallmatrix}  \emptyset \\ 32 \end{smallmatrix} |  \begin{smallmatrix} 23 \\ \emptyset \end{smallmatrix})}{(2\pi i)^{c+d} }   \, \Big|_{\eta_{23}^{a-1} \eta_3^{b-1} \bar \eta_{23}^{c-1} \bar \eta_2^{d-1}}
= (2i)^{a+b-c-d}(-1)^{a+c} \cplus{a &b &0 \\ c &0 &d \\ z &0 &0}  + {\cal O}(s_{ij}) \notag \\
Y^{(a,b|c,d)}_{\begin{smallmatrix}   \emptyset  \\ 23 \end{smallmatrix} \big| \begin{smallmatrix} 23 \\ \emptyset \end{smallmatrix}} &=  \frac{Y(\begin{smallmatrix}   \emptyset \\ 23 \end{smallmatrix} |  \begin{smallmatrix} 23 \\ \emptyset \end{smallmatrix})}{(2\pi i)^{c+d} }   \, \Big|_{\eta_{23}^{a-1} \eta_3^{b-1} \bar \eta_{23}^{c-1} \bar \eta_3^{d-1}}
= (2i)^{a+b-c-d} (-1)^{a+c} \cplus{a &0 \\ c &0 \\ z &0}
\cplus{b &0 \\ d &0 \\ 0 &0}  + {\cal O}(s_{ij}) \notag 
\end{align}
When also varying the $ f_{ij}^{(a)}  f_{pq}^{(b)} $ integrands 
in (\ref{high.75}) and (\ref{high.76}),
the corresponding $\alpha'\rightarrow 0$ limits include for instance
\begin{align}
Y^{(a,b|c,d)}_{\begin{smallmatrix} 32 \\ \emptyset \end{smallmatrix} \big| \begin{smallmatrix} 2 \\ 3 \end{smallmatrix}} &=  \frac{Y(\begin{smallmatrix} 32 \\ \emptyset \end{smallmatrix} |  \begin{smallmatrix} 2 \\ 3\end{smallmatrix})}{(2\pi i)^{c+d} }   \, \Big|_{\eta_{2}^{a-1} \eta_3^{b-1} \bar \eta_{23}^{c-1} \bar \eta_2^{d-1}}
= (2i)^{a+b-c-d} (-1)^{c} \cplus{a &b &0 \\ d &0 &c \\ 0 &z &0}  + {\cal O}(s_{ij})  \label{high.77}\\
Y^{(a,b|c,d)}_{\begin{smallmatrix} 3 \\ 2 \end{smallmatrix} \big| \begin{smallmatrix} 2 \\ 3\end{smallmatrix}} &=  \frac{Y(\begin{smallmatrix} 3 \\ 2 \end{smallmatrix} |  \begin{smallmatrix} 2 \\ 3 \end{smallmatrix})}{(2\pi i)^{c+d} }   \, \Big|_{\eta_{2}^{a-1} \eta_3^{b-1} \bar \eta_{3}^{c-1} \bar \eta_2^{d-1}}
= (2i)^{a+b-c-d} (-1)^{a+d} \cplus{b &0 \\ c &0 \\z &0}  \cplus{a &0 \\ d &0 \\ z &0 }  + {\cal O}(s_{ij})
\notag
\end{align}
Note that the expansions (\ref{eq:Yabij}) and (\ref{alt:Yabij}) of the
two-point component integrals only generate the restricted class
$\cplus{a &0 &1 \\ 0 &b &1 \\ z_1 &z_2 &z_3}$ of two-loop eMGFs
with $z_i \in \{0,z\}$ and unit entries in the third column. Hence, 
already the $\alpha' \rightarrow 0$ limits
in (\ref{high.75}) to (\ref{high.77}) exemplify the extra value of three-point 
generating series to more flexibly access dihedral eMGFs. In fact, it
was shown in section \ref{sec:2.2loop} that any $\cplus{a_1 &a_2 &a_3 \\ b_1 &b_2 &b_3 
\\ z_1 &z_2 &z_3}$ with $z_i \in \{0,z\}$ can be expanded in terms
of the two-loop MGFs in (\ref{high.76}) and (\ref{high.77}) with one of the
$a_i$ and $b_j$ vanishing each.
Similarly, ($n\geq 4$)-point generating series are needed
to obtain trihedral eMGFs beyond the restricted class in (\ref{high.74}) and
the remaining three-point component integrals.

\sm

With no upper limit on the multiplicity $n$ of the generating series, all
convergent eMGFs (\ref{allemgfs}) will eventually be generated by the
$\alpha' \rightarrow 0$ terms of component integrals, regardless of their
topology. Convergence of the lattice sums is ensured by excluding those
combinations of integrands $f^{(1)}_{ij} \overline{ f^{(1)}_{ij}} \rightarrow |z_{ij}|^{-2}$ that
introduce kinematic poles (e.g.\ by imposing $(a,c),(b,d) \neq (1,1)$ in 
the above three-point examples). Still, one can use subtraction 
schemes generalizing the ones in appendix \ref{app:kinsub.1} to capture the (possibly nested)
kinematic poles of integrals over pairs of $f^{(1)}_{ij} \overline{ f^{(1)}_{ij}}$.

%%%%%%%%%%%%%%%%%%%%%%%%%%%%%%%%%%%%%%%%%%%%%%%
%%%%%%%%%%%%%%%%%%%%%%%%%%%%%%%%%%%%%%%%%%%%%%%
 \subsection{Differential equations}
 \label{sechigh.1}
%%%%%%%%%%%%%%%%%%%%%%%%%%%%%%%%%%%%%%%%%%%%%%%
%%%%%%%%%%%%%%%%%%%%%%%%%%%%%%%%%%%%%%%%%%%%%%%

The open-string prototypes (\ref{high.3}) of the generating functions of eMGFs
\cite{Broedel:2020tmd} obey KZB-type differential equations 
\begin{align}
\partial_{z_0} Z( \gamma |  \begin{smallmatrix} K \\ L \end{smallmatrix})
&= \sum_{P,Q}  X_{\rm open}(\begin{smallmatrix} K \\ L \end{smallmatrix} |  \begin{smallmatrix} P \\ Q \end{smallmatrix}) Z( \gamma |  \begin{smallmatrix} P \\ Q \end{smallmatrix})
\label{high.5} \\
2\pi i \partial_{\tau} Z( \gamma |  \begin{smallmatrix} K \\ L \end{smallmatrix})
&= \sum_{P,Q}  D_{\rm open}(\begin{smallmatrix} K \\ L \end{smallmatrix} |  \begin{smallmatrix} P \\ Q \end{smallmatrix}) Z( \gamma |  \begin{smallmatrix} P \\ Q \end{smallmatrix}) \notag
\end{align}
that do not depend on the choice of the integration cycle $\gamma$ as long as
the Koba--Nielsen factor vanishes on its boundary components. The $(n!\times n!)$-matrix valued differential operators $X_{\rm open}$ and $D_{\rm open}$ solely depend on $z_0$ and $\tau$ via
$f^{(k)}_{01}$ and $G_k$,
\begin{align}
X_{\rm open} &= \sum_{k=0}^\infty f^{(k)}_{01} r_{\vec{\eta}}(x_k)
\label{high.6} \\
D_{\rm open}&= -  r_{\vec{\eta}}(\epsilon_0) 
+ \sum_{k=4}^\infty (1{-}k) G_k r_{\vec{\eta}}(\epsilon_k)
+ \sum_{k=2}^\infty (k{-}1) f^{(k)}_{01} r_{\vec{\eta}}(b_k)
\notag
\end{align}
The notation $r_{\vec{\eta}}(\cdot)$ refers to ``open-string type'' matrix representations
of abstract operators $x_j,\epsilon_j,b_j$ in an elliptic KZB system. Their all-multiplicity form is
determined in section 4 of \cite{Broedel:2020tmd}, and each entry of the $n!\times n!$
matrices is linear in $s_{ij}$. The sums over $P,Q$ in (\ref{high.5}) and later 
equations are understood to run over the $n!$ distributions of $\{2,3,\ldots,n\}$
into the integrands $\varphi_{\vec{\eta}}(1,P)\varphi_{\vec{\eta}}(0,Q)$ as described
around (\ref{high.3}).

\sm

The closed-string analogues of the differential equations (\ref{high.5}) in $z_0$ and $\tau$ take 
the almost identical form: As will be justified below, the $n$-point generating series 
(\ref{high.4}) of eMGFs obeys the KZB-type differential equations
\begin{align}
\nabla_{z_0} Y( \begin{smallmatrix} M \\ N \end{smallmatrix} |  \begin{smallmatrix} K \\ L \end{smallmatrix})
&= \sum_{P,Q}  \bigg( \sum_{k=0}^{\infty} (\tau{-}\bar \tau)^k f^{(k)}_{01} R_{\vec{\eta}}(x_k)  \bigg)_{ \begin{smallmatrix} K \\ L \end{smallmatrix} \big|  \begin{smallmatrix} P \\ Q \end{smallmatrix}} 
Y( \begin{smallmatrix} M \\ N \end{smallmatrix} |  \begin{smallmatrix} P \\ Q \end{smallmatrix})
+ 2 \pi i  ( \bar \eta_{L}{-}\bar \eta_N )
Y( \begin{smallmatrix} M \\ N \end{smallmatrix} |  \begin{smallmatrix} K \\ L \end{smallmatrix})
\notag \\
-4\pi \nabchi_\tau Y( \begin{smallmatrix} M \\ N \end{smallmatrix} |  \begin{smallmatrix} K \\ L \end{smallmatrix})
&=
\sum_{P,Q}  \bigg( {-}R_{\vec{\eta}}(\epsilon_0) 
+ \sum_{k=2}^{\infty}(k{-}1) (\tau{-}\bar \tau)^k f^{(k)}_{01} R_{\vec{\eta}}(b_k) \label{high.12} \\
& \ \ \ \ \ \ \ \ \ \ \ \ 
+ \sum_{k=4}^{\infty}(1{-}k) (\tau{-}\bar \tau)^k G_k R_{\vec{\eta}}(\epsilon_k)
  \bigg)_{ \begin{smallmatrix} K \\ L \end{smallmatrix} \big|  \begin{smallmatrix} P \\ Q \end{smallmatrix}} 
Y( \begin{smallmatrix} M \\ N \end{smallmatrix} |  \begin{smallmatrix} P \\ Q \end{smallmatrix}) \notag
\end{align}
where we recall that $\nabla_{z_0}=(\tau{-}\bar \tau) \partial_{z_0}$ and use the shorthand
\beq
\bar\eta_L = \sum_{j \in L} \bar \eta_j  
\label{high.10}
\eeq
Most of the $(n!\times n!)$-matrix valued operators $R_{\vec{\eta}}(\cdot)$ are identical to their 
open-string counterparts $r_{\vec{\eta}}(\cdot)$ in (\ref{high.6}) and \cite{Broedel:2020tmd},
\begin{align}
R_{\vec{\eta}}(x_k)&=r_{\vec{\eta}}(x_k)  \, \big|_{s_{01} \rightarrow\, 0} \ \forall \ k \geq 0 \notag \\
R_{\vec{\eta}}(b_k)&=r_{\vec{\eta}}(b_k) \, \big|_{s_{01} \rightarrow\, 0} \ \forall \ k \geq 2   \label{high.13} \\
R_{\vec{\eta}}(\epsilon_k)&=r_{\vec{\eta}}(\epsilon_k) \ \forall \ k \geq 4  
\notag
\end{align}
and we therefore inherit the following relations
\beq
R_{\vec{\eta}}(b_k) = R_{\vec{\eta}}(x_{k-1}) \ \forall \ k \geq 2
\label{high.14} 
\eeq
The need to set $s_{01}\rightarrow 0$ on the open-string side in (\ref{high.13})
only affects $r_{\vec{\eta}}(x_1), r_{\vec{\eta}}(b_2)$ and
is just an artifact of $s_{01}g_{\rm open}(z_{01}|\tau)$ entering the Koba--Nielsen 
factor of (\ref{high.3}) while its closed-string counterpart (\ref{high.71}) is defined
without $s_{01}g(z_{01}|\tau)$. Apart from this matter of convention,
the only difference between open- and closed-string operators occurs
for $\epsilon_0$
\beq
R_{\vec{\eta}}(\epsilon_0) = r_{\vec{\eta}}(\epsilon_0) - 2 \zeta_2 \sum_{0\leq i <j}^n s_{ij} - 2\pi i \sum_{j=2}^n \bar \eta_j \partial_{\eta_j}
\label{high.15} 
\eeq
which closely follows the analogous relation from the $z_0$-independent
generating series \cite{Gerken:2019cxz, Gerken:2020yii}.
The combination $r_{\vec{\eta}}(\epsilon_0) - 2 \zeta_2 s_{012\ldots n}$ causes all
the $\zeta_2$ to vanish as expected for an operator relating closed-string quantities.
At two points, for instance, we have \cite{Broedel:2020tmd}
\beq
r_{\eta}(\epsilon_0) = \frac{1}{\eta^2} \begin{pmatrix} s_{12} &s_{02} \\ s_{12} &s_{02} \end{pmatrix} - \frac{1}{2}(s_{02}{+}s_{12}) \partial_\eta^2 + 2 \zeta_2(s_{01}{+}s_{02}{+}s_{12})
\eeq
which reproduces the operator $R_\eta(\epsilon_0)$ in (\ref{gen.23}) via (\ref{high.15}).
The remaining $2\times 2$-matrices $r_{\eta}(\cdot)$ in the open-string differential equation
(\ref{high.5}) at two points are identical to their closed-string counterparts $R_\eta(\cdot)$
in (\ref{eq:xops}) and (\ref{gen.23}).

\sm

At three points, explicit $6\times 6$ results for the three-point instances of 
$r_{\eta_2,\eta_3}(x_k),r_{\eta_2,\eta_3}(\epsilon_k)$
and $r_{\eta_2,\eta_3}(b_k)$
can be found in sections 4.2.3, 4.2.4, 4.3.2 and 4.3.3 of \cite{Broedel:2020tmd}. The
conventions for ordering the rows and columns of the matrices in the reference follow
the enumeration of basis integrands in (\ref{high.72}).

\subsubsection{Origin of the closed-string differential equations}

The $\tau$-dependent constituents $(\tau{-}\bar \tau)^k f^{(k)}_{01}$ and $(\tau{-}\bar \tau)^k G_k$
in (\ref{high.12}) arise from the expansion of $\Omega(z_{01},(\tau{-}\bar \tau)\eta_L)$
and $\wp((\tau{-}\bar \tau)\eta_L)$ similar to those in (\ref{2ptzdif}), (\ref{new2ptdiff}) and the open-string differential
equations in section 4 of \cite{Broedel:2020tmd}.
Both the relation (\ref{high.13}) between the operators $r_{\vec{\eta}}(\cdot)$
and $R_{\vec{\eta}}(\cdot)$ in the open- and closed-string differential equations
and the appearance of the term $2\pi i \sum_{j=2}^n \bar \eta_j \partial_{\eta_j} $ in
(\ref{high.15}) from the $\tau$-derivative closely follow the results of \cite{Gerken:2019cxz} on $z_0$-independent generating series.
The term $  2\pi i ( \bar \eta_{L} {-} \bar \eta_N )
Y( \begin{smallmatrix} M \\ N \end{smallmatrix} |  \begin{smallmatrix} K \\ L \end{smallmatrix})$ in
(\ref{high.12}) from the $z_0$-derivative, however, does not have any analogue in earlier work, 
and it generalizes the term $ 2\pi i \bar \eta  (j{-}i) Y_{ij} $ in (\ref{closedana}) to $n$ points.

\sm

The first contribution $ 2\pi i   \bar \eta_{L}
Y( \begin{smallmatrix} M \\ N \end{smallmatrix} |  \begin{smallmatrix} K \\ L \end{smallmatrix})$
stems from the integrations by parts in the simplification of the $z_0$-derivative. As one can
see in (4.16) of \cite{Broedel:2020tmd}, the total derivatives $\partial_{z_j}$ (with $j \in \{2,3,\ldots,n\}$) discarded in the open-string integrands are determined by the factor of $\varphi_{\vec{\eta}}(0,L)$ in the integrand.
The integrations by parts of these $\partial_{z_j}$ yield terms of the form
$ 2\pi i    \bar \eta_j$ via (\ref{high.2}) that do not depend
on the choices of $M,N$ in $\overline{ \varphi_{\vec{\eta}}(1,M) } \overline{ \varphi_{\vec{\eta}}(0,N) } $
and add up to $\bar \eta_L$ defined in (\ref{high.10}).

\sm

The second contribution $-  2\pi i   \bar\eta_{N}
Y( \begin{smallmatrix} M \\ N \end{smallmatrix} |  \begin{smallmatrix} K \\ L \end{smallmatrix})$ 
in turn is a direct consequence of (\ref{high.2}) and translation invariance which
implies $\partial_{z_0} \overline{ \varphi_{\vec{\eta}}(0,N) } = - \frac{2\pi i  }{\tau{-}\bar \tau} \bar \eta_{N} \overline{ \varphi_{\vec{\eta}}(0,N) }$,
regardless of the choice of $K,L$ in $\varphi_{\vec{\eta}}(1,K)\varphi_{\vec{\eta}}(0,L)$.

\subsubsection{Implications for eMGFs as single-valued elliptic polylogarithms}

By the form of the differential equation (\ref{high.12}), we arrive at the same type of conclusion 
as in section \ref{imp2pt}: The eMGFs in the expansion of the generating series $Y(\begin{smallmatrix} M \\ N \end{smallmatrix} |  \begin{smallmatrix} K \\ L \end{smallmatrix})$ in $\eta_j,\bar \eta_j$
and $s_{ij}$ generalize Zagier's single-valued elliptic polylogarithms $\dplus{a \\ b }$ to arbitrary depth.
Since the higher-point generating series of this section probe eMGFs beyond the
dihedral topology, (\ref{high.12}) implies that the general eMGF (\ref{allemgfs}) regardless
of the graph topology obeys a higher-depth analogue of the differential equations (\ref{diff.6})
of the $\dplus{a \\ b }$. 
In a perturbative solution of (\ref{high.12}) via Picard iteration \cite{toappsoon}, 
this will lead to a characterization of arbitrary eMGFs via iterated $\tau$-integrals 
with various numbers of integration kernels $f^{(k)}$ and $G_k$.

%%%%%%%%%%%%%%%%%%%%%%%%%%%%%%%%%%%%%%%%%%%%%%%%%%%%%%%%%%%
\subsection{Commutation relations from integrability and beyond}
\label{sec:4.comm}
%%%%%%%%%%%%%%%%%%%%%%%%%%%%%%%%%%%%%%%%%%%%%%%%%%%%%%%%%%%

We shall now discuss the commutation relations of the operators
$R_{\vec{\eta}}(x_w) = R_{\vec{\eta}}(b_{w+1})$ and $R_{\vec{\eta}}(\epsilon_k)$ 
that govern the differential equations (\ref{high.12}) of the generating series of eMGFs.
This will extend the discussion in section \ref{sec:3.6} not only by the generalization
to $n$ points but also by classes of commutation relations that do not follow
from the integrability of the KZB system and connect with Tsunogai's derivation algebra.

\sm

As in the two-point case, the commutativity of $\nabla_{z_0}$ and
$\nabla_\tau$ imposes commutation relations among the operators in (\ref{high.12}),
see appendix \ref{app:comm} for intermediate steps in deriving the subsequent results.
The commutation relations (\ref{likeopen}) of the two-point $2\times 2$ operators
$R_{\eta}(\cdot)$ can be uplifted in identical form to their $n$-point $n!\times n!$ counterparts $R_{\vec{\eta}}(\cdot)$
\begin{align}
R_{\vec{\eta}}([x_0 ,\epsilon_k] ) &= \sum_{a=1}^{k/2-1} (-1)^a R_{\vec{\eta}}([x_a , x_{k-1-a}]) \, , &&k\geq 4\label{likenopen}
\\
R_{\vec{\eta}}([x_w,\epsilon_k] ) &= - \sum_{a=0}^{w-1} {w{-}1\choose a} R_{\vec{\eta}}( [ x_{a+1}, x_{k+w-a-2}] )  
&&k\geq 4, \ w\geq 1
\notag
\end{align}
where we use the shorthand 
$R_{\vec{\eta}}([a,b])= R_{\vec{\eta}}(a) R_{\vec{\eta}}(b) - R_{\vec{\eta}}(b) R_{\vec{\eta}}(a)$.
In the generalization of (\ref{defbrack}) to higher multiplicity in turn,
the terms $\sim 2\pi i \bar \eta(j{-}\ell)$ in the two-point relations are promoted
to differences of $\bar \eta_L$ and $\bar \eta_Q$ in (\ref{high.10}):
\begin{align}
R_{\vec{\eta}}([x_0,\epsilon_0])_{ \begin{smallmatrix} K \\ L \end{smallmatrix} \big|  \begin{smallmatrix} P \\ Q \end{smallmatrix}}  &=  2\pi i (\bar \eta_Q {-} \bar \eta_L) R_{\vec{\eta}}(\epsilon_{0})_{ \begin{smallmatrix} K \\ L \end{smallmatrix} \big|  \begin{smallmatrix} P \\ Q \end{smallmatrix}} 
\notag \\
R_{\vec{\eta}}([x_w,\epsilon_0])_{ \begin{smallmatrix} K \\ L \end{smallmatrix} \big|  \begin{smallmatrix} P \\ Q \end{smallmatrix}}  &= \sum_{a=0}^{ \lfloor w/2 \rfloor  -1} {w \choose a} 
\frac{ (w{-}1{-}2a)}{(a{+}1)} R_{\vec{\eta}}([x_a,x_{w-1-a}])_{ \begin{smallmatrix} K \\ L \end{smallmatrix} \big|  \begin{smallmatrix} P \\ Q \end{smallmatrix}}  \label{high.19} \\
&\ \ \ \
+ 2\pi i(w{-}1) (\bar \eta_L {-} \bar \eta_Q) R_{\vec{\eta}}(x_{w-1})_{ \begin{smallmatrix} K \\ L \end{smallmatrix} \big|  \begin{smallmatrix} P \\ Q \end{smallmatrix}} 
\notag
\end{align}
While (\ref{likenopen}) agrees literally with the commutation relations
of the operators $r_{\vec{\eta}}(x_w) =r_{\vec{\eta}}(b_{w+1}) $ and $ r_{\vec{\eta}}( \epsilon_k ) $ in
the open-string differential equations (\ref{high.5}) \cite{Broedel:2020tmd},
the extra terms $\sim 2\pi i (\bar \eta_L {-} \bar \eta_Q)$ in (\ref{high.19}) are specific
to closed strings. They balance the discrepancy $\sim 2\pi i \bar \eta_j \partial_{\eta_j}$
between $r_{\vec{\eta}}(\epsilon_{0})$ and $R_{\vec{\eta}}(\epsilon_{0})$ in (\ref{high.15}).
In view of the analogous commutation relations among the $r_{\vec{\eta}}(\cdot)$,
(\ref{high.19}) is equivalent to
\begin{align}
\bigg[ \sum_{j=2}^n \bar \eta_j \partial_{\eta_j} , R_{\vec{\eta}}(x_0) \bigg]_{ \begin{smallmatrix} K \\ L \end{smallmatrix} \big|  \begin{smallmatrix} P \\ Q \end{smallmatrix}}  &=  (\bar \eta_Q {-} \bar \eta_L) R_{\vec{\eta}}(\epsilon_0)_{ \begin{smallmatrix} K \\ L \end{smallmatrix} \big|  \begin{smallmatrix} P \\ Q \end{smallmatrix}}  \\
\bigg[ \sum_{j=2}^n \bar \eta_j \partial_{\eta_j} , R_{\vec{\eta}}(x_k) \bigg]_{ \begin{smallmatrix} K \\ L \end{smallmatrix} \big|  \begin{smallmatrix} P \\ Q \end{smallmatrix}}  &=  (k{-}1)(\bar \eta_L {-} \bar \eta_Q) R_{\vec{\eta}}(x_{k-1})_{ \begin{smallmatrix} K \\ L \end{smallmatrix} \big|  \begin{smallmatrix} P \\ Q \end{smallmatrix}} 
 \, , &&k\geq 1 \notag
\end{align}

\subsubsection{Beyond integrability}

On top of the commutation relations (\ref{likenopen}) and (\ref{high.19}) derived from integrability,
the operators $R_{\vec{\eta}}(x_w) = R_{\vec{\eta}}(b_{w+1})$ and $R_{\vec{\eta}}(\epsilon_k)$ 
at $n$ points are expected to obey variants of the relations 
of Tsunogai's derivation algebra~\cite{Tsunogai,Pollack}.
We henceforth suppress the matrix indices $ \begin{smallmatrix} K \\ L \end{smallmatrix} \big|  \begin{smallmatrix} P \\ Q \end{smallmatrix}$ to avoid cluttering and use the notation 
\beq
\ad_x y= [x,y]\, , \ \ \ \ \ \ R_{\vec{\eta}}(\ad_{\epsilon_0}^{N} y) = \ad_{R_{\vec{\eta}}(\epsilon_0)}^{N} R_{\vec{\eta}}(y) \, , \ \ \ \ \ \ N \geq 0
\eeq
for the adjoint Lie action. The $(n!\times n!)$-matrix
valued operators in (\ref{high.13}) and (\ref{high.15}) are conjectured to enjoy the following
nilpotency property, generalizing~\eqref{gennilp} at two points,
\begin{align}
R_{\vec{\eta}}(\ad_{\epsilon_0}^{k-1} \epsilon_k) &= 0  \, , \ \ \ \ \ \ k\geq 4  \notag
\\
R_{\vec{\eta}}(\ad_{\epsilon_0}^{k-1} b_k) &= 0  \, , \ \ \ \ \ \ k\geq 2 
\label{nilpot}
\end{align}
The first relation would imply that the property $\ad_{\epsilon_0}^{k-1} \epsilon_k=0$ of the derivations
are preserved, though we will give examples below around~\eqref{promot} that not all their commutation
relations are preserved by $R_{\vec{\eta}}(\cdot)$. This is in contrast to the $\epsilon_k$-type 
operators in the differential equations of generating functions of elliptic multiple 
zeta values \cite{Mafra:2019ddf, Mafra:2019xms} and MGFs \cite{Gerken:2019cxz, Gerken:2020yii}: In the
simplified setting without the elliptic variable $z_0$, the operators of the reference are believed
to furnish matrix representations of Tsunogai's derivation algebra which has been supported
by evaluating a large number of nested commutators at multiplicities $n\leq 5$.

\sm

Even though the $R_{\vec{\eta}}(\epsilon_k), R_{\vec{\eta}}(b_k)$ in this work are not
expected to be matrix representations of the derivations $\epsilon_k$, the
nilpotency property (\ref{nilpot}) is believed to hold at any multiplicity and supported
by a broad range of explicit checks. On the basis of (\ref{nilpot}), the differential equations
(\ref{high.12}) of generating series of eMGFs will be solved by Picard iteration in follow-up
work \cite{toappsoon} and related to iterated $\tau$-integrals. 

\sm

Apart from the nilpotency property $\ad_{\epsilon_0}^{k-1} \epsilon_k=0$, Tsunogai's derivations
obey a variety of further commutation relations \cite{LNT, Pollack, Broedel:2015hia} starting with
\begin{align}
0 &=[\ep_{10},\ep_4]-3[\ep_{8},\ep_6] \label{eq2.43a}  \\
0&=2 [\ep_{14},\ep_4] - 7[\ep_{12},\ep_6] + 11 [\ep_{10},\ep_8]   
\notag
\end{align}
at depth two and
\begin{align}
0&=80[\ep_{12},[\ep_4,\ep_{0}]] + 16 [\ep_4,[\ep_{12},\ep_0]] - 250 [\ep_{10},[\ep_6,\ep_0]]\label{eq2.43c}  \\
& \ \ \ \  - 125 [\ep_6,[\ep_{10},\ep_0]] + 280 [\ep_8,[\ep_8,\ep_0]]- 462 [\ep_4,[\ep_4,\ep_8]] - 1725 [\ep_6,[\ep_6,\ep_4]] 
\notag
\end{align}
at depth three. Relations at higher weight and depth can also be downloaded from
\cite{WWWe}. 

\sm

The $n!\times n!$ matrices $R_{\vec{\eta}}(\epsilon_k)$ preserve (\ref{eq2.43a}) in the sense that
for instance
\beq
0 = R_{\vec{\eta}}([\ep_{10},\ep_4])-3 R_{\vec{\eta}}([\ep_{8},\ep_6])
\label{eq2.43z}
\eeq
since their open-string counterparts obey $r_{\vec{\eta}}([\ep_{10},\ep_4])-3 r_{\vec{\eta}}([\ep_{8},\ep_6])$ \cite{Broedel:2020tmd} and agree
with the $R_{\vec{\eta}}(\epsilon_k)$ for the cases $k\geq 4$ relevant to depth-two relations.

\sm

The depth-three relation (\ref{eq2.43c}), by contrast, no longer holds under the naive
replacement $\epsilon_k \rightarrow R_{\vec{\eta}}(\epsilon_k)$. Still, both
(\ref{eq2.43a}) and (\ref{eq2.43c}) are preserved when promoting
\beq
\epsilon_k \rightarrow \left\{ \begin{array}{cl}
R_{\vec{\eta}}(\epsilon_0) &: \ k=0 
\\
R_{\vec{\eta}}(\epsilon_k) + R_{\vec{\eta}}(b_k) &: \ k\geq 4
\end{array} \right.
\label{promot}
\eeq
and the same has been observed for the open-string operators $r_{\vec{\eta}}(\epsilon_k)$ 
in the place of $R_{\vec{\eta}}(\epsilon_k)$ in (\ref{promot}) \cite{Broedel:2020tmd}.

\sm

In summary, the derivation algebra~\cite{Tsunogai,Pollack} is a rich source of 
candidate commutation relations of $R_{\vec{\eta}}(\epsilon_k)$ and
$R_{\vec{\eta}}(b_k)$ that cannot be derived from integrability. In some
cases such as (\ref{nilpot}) and (\ref{eq2.43z}), these candidate relations have
been verified to hold on a case-by-case basis for low values of $k$ or $n$
and are then conjectured to generalize to all $k$ and $n$. As exemplified by
the depth-three relation (\ref{eq2.43c}), however, some of the $\epsilon_k$
relations \cite{LNT, Pollack, Broedel:2015hia, WWWe} need to be modified
to find an echo at the level of $R_{\vec{\eta}}(\epsilon_k)$ and $R_{\vec{\eta}}(b_k)$.
The systematics is likely to match the analysis of the open-string operators
$r_{\vec{\eta}}(\cdot)$ \cite{Broedel:2020tmd} and will be left as an open problem.

\subsubsection{Implication for the counting of independent eMGFs}

While the above commutation relations among $R_{\vec{\eta}}(\epsilon_k)$
and $R_{\vec{\eta}}(b_k)$ are believed to be multiplicity agnostic, the
two-point examples in (\ref{gen.23}) obey additional relations that
are absent at $n\geq 3$. For instance, the diagonal form of
$R_{\eta}(\epsilon_{k\geq 4}) = \eta^{k-2} {\rm diag}(s_{12},s_{02})$
implies that all two-point commutators $R_{\eta}([\epsilon_{k_1}, \epsilon_{k_2}]) $
vanish for $k_1,k_2\geq 4$. 
Hence, only their $(n\geq 3)$-point analogues 
can serve as meaningful
testing grounds for relations like (\ref{eq2.43z}), where the two-point example
would be insensitive to the relative factor of $-3$.

\sm

The relations among $R_\eta(\epsilon_k)$ and $R_\eta(b_k)$ lead to a
constrained class of differential equations for the dihedral eMGFs generated
by $Y_{ij}$ in (\ref{eq:Y2pt}). This resonates with the finding of section
\ref{sechigh.0} that $(n\geq 3)$-point generating functions feature additional
eMGFs, say those of trihedral topologies or more general
two-loop eMGFs in case of $n=3$.

\sm

This situation is similar to that of generating series of MGFs that was investigated in \cite{Gerken:2019cxz, Gerken:2020yii}. The $\epsilon_k$-type operators governing their $\tau$-derivatives
also exhibit degenerate commutators
at two points. As a consequence, the simplest imaginary MGF cusp forms \cite{DHoker:2019txf}
were found to drop out from the two-point generating series of MGFs \cite{Gerken:2020yii}.
Moreover, the absence of higher-depth MZVs in the degeneration
of the two-point MGFs at the cusp \cite{DHoker:2019xef, Zagier:2019eus, Gerken:2020yii, Vanhove:2020qtt} 
obstructs the appearance of single-valued iterated Eisenstein integrals 
that require irreducible MZVs beyond depth one in their completion to be a modular form.
The connection between iterated Eisenstein integrals and MZVs as their multiple
modular values is actively discussed in the recent mathematics literature 
\cite{Brown:mmv, Brown:2017qwo, Brown:2017qwo2, Saad:2020mzv}.

\sm

In summary, the differential equations of the $(n\geq 3)$-point eMGF
generating series (\ref{high.4}) are crucial for obtaining a reliable
picture of the multiplicity-agnostic commutation relations of $R_{\vec{\eta}}(\epsilon_k)$
and $R_{\vec{\eta}}(b_k)$. Those in turn will serve as a starting point to 
count the number of independent eMGFs (under $\mathbb Q$-relations
over MZVs and MGFs) at given transcendental and modular 
weight \cite{toappsoon}. This strategy has been used for the counting of
independent MGFs of arbitrary graph topology \cite{Gerken:2020yii}, 
where each relation in the derivation algebra beyond $\ad_{\epsilon_0}^{k-1} 
\epsilon_k=0$ was found to yield additional dropouts.

\newpage

%%%%%%%%%%%%%%%%%%%%%%%%%%%%%%%%%%%%%%%%%%%%%%%%%%%%%%%%%%%
%%%%%%%%%%%%%%%%%%%%%%%%%%%%%%%%%%%%%%%%%%%%%%%%%%%%%%%%%%%
\section{Conclusion}
\label{sec:concl}
%%%%%%%%%%%%%%%%%%%%%%%%%%%%%%%%%%%%%%%%%%%%%%%%%%%%%%%%%%%
%%%%%%%%%%%%%%%%%%%%%%%%%%%%%%%%%%%%%%%%%%%%%%%%%%%%%%%%%%%

In this work, we have defined eMGFs as non-holomorphic single-valued elliptic functions associated to decorated Feynman graphs of a conformal scalar on a torus. They generalize
the MGFs in the configuration-space integrals of closed-string amplitudes at genus one by additionally depending
on an arbitrary number of torus punctures $z_r$ through the graph decorations. In string perturbation
theory, eMGFs are naturally introduced in the non-separating degeneration limit of higher-genus MGFs.

\sm

We have derived infinite families of algebraic and differential identities among eMGFs, in many cases
by following the key ideas in the derivation of identities among MGFs. A key feature of eMGFs which was
absent for MGFs are the meromorphic Kronecker--Eisenstein coefficients $f^{(k)}$ in their differential equations
w.r.t.\ $z_r$ and $\tau$ which we have exposed from two complementary perspectives: an elliptic analogue of the holomorphic subgraph reduction formulas for MGFs and generating functions of Koba--Nielsen integrals.

\sm

The differential properties of eMGFs generalize those of Zagier's single-valued elliptic polylogarithms
to higher depth. This motivates a variety of follow-up studies of relevance to both physicists and mathematicians:
\begin{itemize}
\item In the same way as MGFs reduce to iterated Eisenstein integrals and their complex 
conjugates, eMGFs admit representations in terms of iterated integrals of meromorphic Kronecker--Eisenstein 
coefficients over the modular parameters. On the one hand, the iterated-integral perspective will
expose all algebraic relations among eMGFs that are obscured by their lattice-sum representations
and result in practical procedures to analytically and numerically control their functional dependence on $z_r$, and $\tau$. 
On the other hand, this may guide first steps in 
generalizing Brown's construction \cite{Brown:2017qwo, Brown:2017qwo2} of 
non-holomorphic modular forms to include elliptic variables.
\item 
It would be interesting to express eMGFs in terms of finite combinations of
Brown--Levin elliptic polylogarithms \cite{BrownLev} and their complex conjugates
as done at depth one in \cite{Broedel:2019tlz}. 
This may lead to an explicit construction of a single-valued map for the Brown--Levin elliptic polylogarithms.
Upon evaluation at $z=0$, such a single-valued map for elliptic polylogarithms
should yield another explicit realization of MGFs as single-valued elliptic MZVs,
complementing the construction in \cite{Gerken:2020xfv}. This amounts to
identifying the meromorphic counterparts in the realization of MGFs via
$z\rightarrow 0$ limits of eMGFs \cite{DHoker:2015wxz}.
\end{itemize}
At the level of computational practicalities, an important follow-up step is to systematically analyze and generate the relations among eMGFs, similar to the {\tt Mathematica} package for MGF relations \cite{Gerken:2020aju}.
Moreover, the Koba--Nielsen integrals and generating-function techniques in sections \ref{sec:3} and \ref{sec:4} are likely to streamline
the non-separating degenerations of genus-two 
four- and five-point integrals initiated in \cite{DHoker:2018mys, DHoker:2020tcq}.

\newpage

\appendix

%%%%%%%%%%%%%%%%%%%%%%%%%%%%%%%%%%%%%%%%%%%%%%%%%%%%%%%%%%%
%%%%%%%%%%%%%%%%%%%%%%%%%%%%%%%%%%%%%%%%%%%%%%%%%%%%%%%%%%%
\section{Trihedral eMGFs}
\label{app:tri}
%%%%%%%%%%%%%%%%%%%%%%%%%%%%%%%%%%%%%%%%%%%%%%%%%%%%%%%%%%%
%%%%%%%%%%%%%%%%%%%%%%%%%%%%%%%%%%%%%%%%%%%%%%%%%%%%%%%%%%%

In this appendix, we define trihedral eMGFs and obtain some of their most important properties in explicit form.  A trihedral graph is connected; may have arbitrary depth and weight; and has  three vertices of valence three or higher in addition to an arbitrary number of bivalent vertices. The first definition is in terms of Kronecker--Eisenstein sums, generalizing (\ref{gen.66}) of the dihedral case,
\bea
\label{basic.16} 
\cplustri{A \\ B \\ Z}{C \\ D \\ Y}{E \\ F \\ X}\!(\tau)  & = & 
\frac{(\Im \tau)^{|A| + |C| + |E| } }{ \pi^{ |B|+|D|+|F|} }  
 \sum_{p_r, k_r,\ell_r \in \Lambda' }
\! \!\delta\bigg( \sum_{r=1}^{R_1} p_r - \sum_{r=1}^{R_2} k_r  \bigg)
\delta\bigg(  \sum_{r=1}^{R_2} k_r - \sum_{r=1}^{R_3} \ell_r \bigg)  %
\no \\ &&  \times  \! 
\bigg( \prod_{r=1}^{R_1} \frac{ \chi_{p_r}(z_r) }{p_r^{a_r} \bar p_r^{b_r} }\bigg)
\bigg( \prod_{r=1}^{R_2} \frac{ \chi_{k_r}(y_r) }{k_r^{c_r} \bar k_r^{d_r} }\bigg)
\bigg( \prod_{r=1}^{R_3} \frac{ \chi_{\ell_r}(x_r) }{\ell_r^{e_r} \bar \ell_r^{f_r} }\bigg)
\eea
The arrays $A,B,Z$ have entries $a_r,b_r,z_r$ for $r=1,\cdots, R_1$; the arrays $C,D,Y$ have entries $c_r, d_r, y_r$ for $r=1,\cdots, R_2$ and the arrays $E,F,X$ have entries $e_r, f_r, x_r$ for $r=1,\cdots, R_3$. The trihedral eMGFs defined this way have two different permutation symmetries. First, the eMGF is invariant under the $R_1!$ permutations of the triplets $(a_r,b_r,z_r)$ within the column of arrays $A,B,Z$. Second, it is also invariant under the $3!$ permutations of the columns of arrays. The trihedral eMGF of (\ref{basic.16}) transforms under $SL(2,\ZZ)$ as a modular form of weight $(0,|B|{+}|D|{+}|F|{-}|A|{-}|C|{-}|E|)$, thereby generalizing (\ref{basic.27}) for the dihedral case. The trihedral analogues of (\ref{basic.23}) may be  found in appendix \ref{app:tri.1}. 

\sm

The notation of (\ref{basic.16}) to arrange the arrays $A,B,C,D,E,F$ and $X,Y,Z$ into three columns
reflects the connectivity of the vertices and edges. A similar notation can be found in \cite{Gerken:2020aju} for 
the more complicated graph topologies in four-point MGFs, namely the box-, kite- and tetrahedral topology.
The notation for additional topologies of MGFs in the reference can be similarly adapted to eMGFs by
adjoining an array of characteristics $Z$ to each pair of arrays $A,B$ that encode the holomorphic and antiholomorphic decorations in a given part of the graph.

\sm

Alternatively, the trihedral eMGFs may also be obtained via an integral representation in terms of the functions 
$\cD^+$ of (\ref{basic.13}), thereby generalizing the dihedral formula (\ref{basic.41}) to the trihedral case, 
and we have, 
\begin{align}
\cplustri{A \\ B \\ Z}{C \\ D \\ Y}{E \\ F \\ X}\!(\tau)  &=
 \int_\Sigma \frac{ \dd^2 z }{\Im \tau} \int_{\Sigma} \frac{ \dd^2 w}{\Im \tau}
\bigg( \prod_{r=1}^{R_1}  \dplus{a_r\\b_r} \!(z_r{-}z|\tau) \bigg)  
\label{basic.44}
\\
&\ \ \ \ \ \ \  \times 
\bigg( \prod_{r=1}^{R_2} \dplus{c_r \\ d_r} \!(y_r{+}z{-}w|\tau)\bigg)
\bigg( \prod_{r=1}^{R_3} \dplus{e_r \\ f_r} \!(x_r{+}w|\tau) \bigg) 
\notag
\end{align}
The trihedral analog to (\ref{basic.43}) when a single column vanishes is given by,\footnote{Throughout, the dependence on $\tau$ will be understood but not exhibited explicitly.}
\begin{align}
\cplustri{0 &A \\ 0& B \\ 0&Z}{C \\ D \\ Y}{E \\ F \\ X}  &=
\bigg( \prod_{r=1}^{R_1} 
\dplus{a_r \\ b_r}\!(z_r)\bigg) 
(-1)^{|E|+|F|}\cplus{C &E \\ D&F \\ Y &-X}
- \cplustri{ A \\ B \\ Z}{C \\ D \\ Y}{E \\ F \\ X}
 \label{basic.45}
\end{align} 
and there are topological simplifications to dihedral eMGFs in case of empty arrays
or pairs of single-entry arrays such as $A=a$ and $C=c$, e.g.
\begin{align}
\cplustri{A \\ B \\ Z}{C \\ D \\ Y}{ \\  \\ }  &=  \cplus{A \\ B \\ Z} \cplus{C \\ D \\ Y} \label{topsimp}
\\
\cplustri{a \\ b \\ z}{c \\ d \\ y}{E \\ F \\ X}  &= (-1)^{a+b+c+d} \cplus{a+c &E \\ b+d &F \\
-z-y &X}  \notag
\end{align}
In the remainder of this appendix, we gather the generalizations of various identities among 
dihedral eMGFs (\ref{gen.66}) to the trihedral case (\ref{basic.16}). 

\subsection{Trihedral momentum conservation}
\label{app:tri.1}

The trihedral eMGFs in (\ref{basic.16}) satisfy the following
analogue of the dihedral momentum-conservation identities (\ref{basic.23})
\bea
\sum_{j=1}^{R_1} \cplustri{A-S_j \\B \\Z  }{ C \\D \\Y }{ E \\F \\X}
= \sum_{j=1}^{R_2} \cplustri{A\\B \\Z  }{ C-S_j  \\D \\Y }{ E \\F \\X}
= \sum_{j=1}^{R_3} \cplustri{A \\B \\Z  }{ C \\D \\Y }{ E-S_j \\F \\X}
\nn \\
\sum_{j=1}^{R_1} \cplustri{A \\B-S_j \\Z  }{ C \\D \\Y }{ E \\F \\X}
= \sum_{j=1}^{R_2} \cplustri{A\\B \\Z  }{ C  \\D-S_j \\Y }{ E \\F \\X}
= \sum_{j=1}^{R_3} \cplustri{A \\B \\Z  }{ C \\D \\Y }{ E \\F-S_j \\X}
\label{basic.23tri} 
\eea
where we again use the notation $S_j= [0, \cdots 0, \, 1, \, 0, \cdots ,0]$
for the array with non-vanishing entry $j$. Similarly, the trihedral generalization
of translation invariance in (\ref{basic.23}) is given by
\beq
 \cplustri{a_1 &\ldots &a_{R_1} \\b_1 &\ldots &b_{R_1} \\z_1 &\ldots &z_{R_1} }{ c_1 &\ldots &c_{R_2} \\d_1 &\ldots &d_{R_2} \\y_1 &\ldots &y_{R_2} }{ E \\F \\X}
=  \cplustri{a_1 &\ldots &a_{R_1} \\b_1 &\ldots &b_{R_1} \\z_1-w &\ldots &z_{R_1}-w }{ c_1 &\ldots &c_{R_2} \\d_1 &\ldots &d_{R_2} \\y_1+w &\ldots &y_{R_2}+w }{ E \\F \\X}
\label{basic.24tri} 
\eeq

\subsection{Two-point HSR in trihedral eMGFs}
\label{app:tri.2}

By inserting the Fay identity (\ref{diff.12}) among the Kronecker--Eisenstein coefficients
$f^{(a)}$ into the integral representation (\ref{basic.44}) of trihedral eMGFs, we obtain the
following generalization of (\ref{diff.15}) for two-point holomorphic subgraphs
\begin{align}
 &\cplustri{a_1 & a_2 & A \\ 0 & 0 & B \\ z_1 & z_2 & Z}{C \\ D \\ Y}{E \\ F \\ X}  = 
(-1)^{a_2} \dplus{a_0  \\ 0  }\! (z_{12}) \, \cplustri{A  \\ B \\ Z}{C \\ D \\ Y}{E \\ F \\ X}
\label{trih2pt}
\\ 
 &\hskip 0.5in  - \tbinom{a_0-1}{a_1} \, \cplustri{a_0  & A \\ 0 & B \\ z_2 & Z}{C \\ D \\ Y}{E \\ F \\ X}
 + \sum _{k=1}^{a_1} \tbinom{a_0-1-k}{a_1-k} \,
\dplus{k   \\ 0    }\! (z_{12})
\cplustri{a_0-k  & A \\ 0 & B \\ z_2 & Z}{C \\ D \\ Y}{E \\ F \\ X}
\notag  \\ 
&\hskip 0.5in - \tbinom{a_0-1}{a_2} \, \cplustri{a_0  & A \\ 0 & B \\ z_1 & Z}{C \\ D \\ Y}{E \\ F \\ X}+ 
\sum _{k=1}^{a_2} \tbinom{a_0-1-k}{a_2-k} \,
\dplus{k   \\ 0   }\! (z_{21})
\cplustri{a_0-k  & A \\ 0 & B \\ z_1 & Z}{C \\ D \\ Y}{E \\ F \\ X}
\notag
\end{align}
where $a_0=a_1+a_2 \geq 3$ and $z_{12}=z_1{-}z_2\neq 0$. The coincident limit $z_1 \rightarrow z_2$
can again be taken by isolating the singular behavior $\dplus{1 \\ 0 }\! (z_{12}|\tau)
 \rightarrow - \frac{ \Im \tau}{ z_{12}}$ of the $k=1$ terms in (\ref{trih2pt}) and Taylor
 expanding the coefficients of the simple pole around $z_1=z_2$. Accordingly, the
relation in (\ref{diff.17}) straightforwardly generalizes to the trihedral case by carrying out the following substitution
 \bea
 \cplus{\ldots  & A \\ \ldots & B \\ \ldots & Z}
\rightarrow
 \cplustri{\ldots  & A \\ \ldots & B \\ \ldots & Z}{C \\ D \\ Y}{E \\ F \\ X}
 \eea 
 in each term.

\subsection{Three-point HSR in trihedral eMGFs}
\label{app:tri.3}

While (\ref{trih2pt}) resolves two-point holomorphic subgraphs within trihedral eMGFs, 
an independent HSR is needed for three-point holomorphic subgraphs that correspond
to cycles $\dplus{a\\0}(z_1{-}z) \, \dplus{c\\0}(y_1{+}z{-}w) \, \dplus{e\\0}(x_1+w)$
in the integrand of (\ref{basic.44}). By applying the Fay identity (\ref{diff.12})
to the first two factors (assuming that $a,c\neq 0$)
\begin{align}
\dplus{a\\0}&(z_1{-}z) \dplus{c\\0}(y_1{+}z{-}w) = (-1)^{a+c} \dplus{a+c\\0}(w{-}z_1{-}y_1) \notag \\
&-(-1)^a \tbinom{a+c-1}{a} \dplus{a+c\\0}(y_1{+}z{-}w) - (-1)^c \tbinom{a{+}c{-}1}{c} \dplus{a+c\\0}(z_1{-}z) \notag \\
&+(-1)^a \sum_{k=1}^a \tbinom{a{+}c{-}k{-}1}{a{-}k} \dplus{k\\0}(w{-}z_1{-}y_1) \dplus{a+c-k\\0}(y_1{+}z{-}w) \label{trifayid}\\
&+(-1)^c \sum_{k=1}^c \tbinom{a{+}c{-}k{-}1}{c{-}k} \dplus{k\\0}(w{-}z_1{-}y_1) \dplus{a+c-k\\0}(z_1{-}z) \notag
\end{align}
one arrives at one out of three possible resolutions of the three-point holomorphic subgraph
\begin{align}
&\cplustri{a_1 &A \\ 0& B \\ z_1&Z}{c_1 &C \\ 0 &D \\ y_1 &Y}{e_1 &E \\ 0 &F \\x_1 &X}
 =(-1)^{a+c} \cplustri{ A \\  B \\  Z}{ C \\  D \\  Y}{a_1+c_1 &e_1 &E \\ 0 &0 &F \\-y_1-z_1 &x_1 &X}  \notag \\
&\hskip 1in  -(-1)^{a_1} \tbinom{ a_1{+}c_1{-}1}{a_1} \cplustri{ A \\  B \\  Z}{a_1+c_1 &C \\ 0 &D \\ y_1 &Y}{e_1 &E \\ 0 &F \\x_1 &X}
 \notag \\
&\hskip 1in -(-1)^{c_1}  \tbinom{a_1{+}c_1{-}1}{c_1}\cplustri{a_1+c_1 &A \\ 0& B \\ z_1&Z}{ C \\  D \\  Y}{e_1 &E \\ 0 &F \\x_1 &X} \label{trihsr} \\
&\hskip 1in + (-1)^{a_1} \sum_{k=1}^{a_1} \tbinom{a_1{+}c_1{-}k{-}1}{a_1{-}k } \cplustri{ A \\  B \\  Z}{a_1+c_1-k &C \\ 0 &D \\ y_1 &Y}{k &e_1 &E \\ 0 &0 &F \\-y_1-z_1 &x_1 &X} \notag\\
&\hskip 1in + (-1)^{c_1} \sum_{k=1}^{c_1} \tbinom{a_1{+}c_1{-}k{-}1}{c_1{-}k } \cplustri{ a_1+c_1-k &A \\  0 &B \\ z_1 &Z}{C \\ D \\ Y}{k &e_1 &E \\ 0 &0 &F \\-y_1-z_1 &x_1 &X} \notag
\end{align}
Every term on the right-hand side is free of three-point holomorphic subgraphs,
and the terms in the first, fourth and fifth line feature two-point holomorphic subgraphs
which can be resolved via (\ref{trih2pt}). One can repeat this procedure in different 
ways of performing the Fay identity (\ref{trifayid}) on two out of three factors 
$\dplus{a\\0}(z_1{-}z) \dplus{c\\0}(y_1{+}z{-}w) \dplus{e\\0}(x_1+w)$ 
in the integrand of (\ref{basic.44}): Equating (\ref{trihsr}) with the two alternative expressions
(obtained from permutations of the three groups of labels) is a convenient source of identities 
among eMGFs. For MGFs, this method of generating identities has been used in 
setting up a database of identities \cite{Gerken:2020aju}, see \cite{Gerken:2018zcy} for 
the analogous trihedral HSR.

\subsection{Fay identities beyond three-point HSR}
\label{app:tri.4}

Even in absence of holomorphic subgraphs, the Fay identity (\ref{trifayid}) 
can be imported to the integrand of (\ref{basic.44}) to yield identities among
trihedral eMGFs with two vanishing entries in different parts of the graph
(see \cite{Gerken:2018zcy} for the analogous identity among MGFs)
\begin{align}
&\cplustri{a_1 &A \\ 0& B \\ z_1&Z}{c_1 &C \\ 0 &D \\ y_1 &Y}{E \\ F \\ X}
 =(-1)^{a+c} \cplustri{ A \\  B \\  Z}{ C \\  D \\  Y}{a_1+c_1 &E \\ 0 &F \\-y_1-z_1  &X}  
 \notag \\
&\hskip 1in -(-1)^{a_1} \tbinom{ a_1{+}c_1{-}1}{a_1} \cplustri{ A \\  B \\  Z}{a_1+c_1 &C \\ 0 &D \\ y_1 &Y}{E \\ F \\ X}
 \notag \\
&\hskip 1in  -(-1)^{c_1}  \tbinom{a_1{+}c_1{-}1}{c_1}\cplustri{a_1+c_1 &A \\ 0& B \\ z_1&Z}{ C \\  D \\  Y}{E \\ F \\ X}
 \label{alttrihsr} \\
&\hskip 1in + (-1)^{a_1} \sum_{k=1}^{a_1} \tbinom{a_1{+}c_1{-}k{-}1}{a_1{-}k } \cplustri{ A \\  B \\  Z}{a_1+c_1-k &C \\ 0 &D \\ y_1 &Y}{k  &E \\ 0  &F \\-y_1-z_1  &X}
 \notag\\
&\hskip 1in + (-1)^{c_1} \sum_{k=1}^{c_1} \tbinom{a_1{+}c_1{-}k{-}1}{c_1{-}k } \cplustri{ a_1+c_1-k &A \\  0 &B \\ z_1 &Z}{C \\ D \\ Y}{k &E \\ 0  &F \\-y_1-z_1  &X}
 \notag
\end{align}

\newpage

%%%%%%%%%%%%%%%%%%%%%%%%%%%%%%%%%%%
%%%%%%%%%%%%%%%%%%%%%%%%%%%%%%%%%%%
\section{Differential identities of two-point component integrals}
\label{app:compdiff}
%%%%%%%%%%%%%%%%%%%%%%%%%%%%%%%%%%%
%%%%%%%%%%%%%%%%%%%%%%%%%%%%%%%%%%%

In this appendix, we shall list a variety of further examples of the differential
equations (\ref{gen.56}) and (\ref{gen.57}) of two-point component integrals
$Y_{ij}^{(a|b)}$ defined in (\ref{eq:Yab}). In this appendix, we will use the shorthand notation 
\begin{align}
\newf_{ij}^{(a)} &= (\tau{-} \bar \tau)^a f^{(a)}_{ij} =  (\tau{-} \bar \tau)^a f^{(a)}(z_i{-} z_j|\tau) \label{shortapp} \\
\newg_k &= (\tau{-}\bar \tau)^k G_k(\tau) = -\newf^{(k)}(0|\tau) \, , \ \ \ \ \ \ k\geq 4
\notag
\end{align}

%%%%%%%%%%%%%%%%%%%%%%%%%%%%%%%%%%%
%%%%%%%%%%%%%%%%%%%%%%%%%%%%%%%%%%%

\subsection{$z_0$-derivatives}
\label{app:compdiff.a}

The simplest examples of the $z_0$-derivatives (\ref{gen.56}) include (\ref{simp:z0}) as well as
\begin{align}
 \nabla_{z_0}  Y_{11}^{(1| 0)} &=
 -s_{02} \newf_{01}^{(2)} Y_{11}^{(0| 0)} -  s_{02} \newf_{01}^{(1)} Y_{11}^{(1| 0)} - s_{02} Y_{11}^{(2| 0)} + s_{02} \newf_{01}^{(1)} Y_{12}^{(1| 0)} -  s_{02} Y_{12}^{(2| 0)}
\notag \\
\nabla_{z_0}  Y_{11}^{(0| 1)} &=
 - s_{02} Y_{12}^{(1| 1)}
\notag \\
 \nabla_{z_0} Y_{11}^{(2| 0)} &=
s_{02} \newf_{01}^{(3)} Y_{11}^{(0| 0)}  -  s_{02} \newf_{01}^{(1)} Y_{11}^{(2| 0)} - 2 s_{02} Y_{11}^{(3| 0)}  \notag \\
& \ \ \ \ - s_{02} \newf_{01}^{(2)} Y_{12}^{(1| 0)} + 
 s_{02} \newf_{01}^{(1)} Y_{12}^{(2| 0)} - s_{02} Y_{12}^{(3| 0)}
\notag \\
 \nabla_{z_0} Y_{11}^{(2| 2)} &=
s_{02} \newf_{01}^{(3)} Y_{11}^{(0| 2)}  -  s_{02} \newf_{01}^{(1)} Y_{11}^{(2| 2)} - 2 s_{02} Y_{11}^{(3| 2)}  \notag \\
& \ \ \ \ - s_{02} \newf_{01}^{(2)} Y_{12}^{(1| 2)}  +  s_{02} \newf_{01}^{(1)} Y_{12}^{(2| 2)} - s_{02} Y_{12}^{(3| 2)}
\label{gen.63a}
\end{align}
and
\begin{align}
\nabla_{z_0}  Y_{12}^{(1| 0)} &=
s_{12} \newf_{01}^{(2)} Y_{11}^{(0| 0)} + s_{12} \newf_{01}^{(1)} Y_{11}^{(1| 0)} + s_{12} Y_{11}^{(2| 0)}  - s_{12} \newf_{01}^{(1)} Y_{12}^{(1| 0)} + s_{12} Y_{12}^{(2| 0)}
\notag \\
  \nabla_{z_0}  Y_{12}^{(2| 0)}  &=
s_{12} \newf_{01}^{(3)} Y_{11}^{(0| 0)} + s_{12} \newf_{01}^{(2)} Y_{11}^{(1| 0)} + 
 s_{12} \newf_{01}^{(1)} Y_{11}^{(2| 0)}   \notag \\
& \ \ \ \ + s_{12} Y_{11}^{(3| 0)} -  s_{12} \newf_{01}^{(1)} Y_{12}^{(2| 0)} + 2 s_{12} Y_{12}^{(3| 0)}
\notag \\
\nabla_{z_0}  Y_{12}^{(2| 2)}  &=
s_{12} \newf_{01}^{(3)} Y_{11}^{(0| 2)} + s_{12} \newf_{01}^{(2)} Y_{11}^{(1| 2)} + 
 s_{12} \newf_{01}^{(1)} Y_{11}^{(2| 2)} + s_{12} Y_{11}^{(3| 2)} \notag \\
& \ \ \ \  - s_{12} \newf_{01}^{(1)} Y_{12}^{(2| 2)} + 2 s_{12} Y_{12}^{(3| 2)}+ Y_{12}^{(2| 1)}  
\label{gen.63}
\end{align}
%

%%%%%%%%%%%%%%%%%%%%%%%%%%%%%%%%%%%
%%%%%%%%%%%%%%%%%%%%%%%%%%%%%%%%%%%

\subsection{$\tau$-derivatives}
\label{app:compdiff.b}

The simplest examples of the $\tau$-derivatives (\ref{gen.57}) include (\ref{simp:tau}) as well as
\begin{align}
 {-}4\pi \nabla_\tau Y_{11}^{(1| 0)} &=
 -2 s_{02} \newf_{01}^{(3)} Y_{11}^{(0| 0)}  -  s_{02} \newf_{01}^{(2)} Y_{11}^{(1| 0)} + s_{02} Y_{11}^{(3| 0)} + s_{02} \newf_{01}^{(2)} Y_{12}^{(1| 0)} -  s_{02} Y_{12}^{(3| 0)}
\notag \\
 {-}4\pi \nabla_\tau Y_{11}^{(0| 1)} &=
 - s_{12} Y_{11}^{(2| 1)} - s_{02} Y_{12}^{(2| 1)}
\notag \\
 {-}4\pi \nabla_\tau Y_{11}^{(2| 0)} &=
3 s_{02} \newf_{01}^{(4)} Y_{11}^{(0| 0)} - 3 s_{12} \newg_4 Y_{11}^{(0| 0)}  - s_{02} \newf_{01}^{(2)} Y_{11}^{(2| 0)}   + 3 s_{02} Y_{11}^{(4| 0)} \notag \\
&\ \ \ \  +  2 s_{12} Y_{11}^{(4| 0)} - 2 s_{02} \newf_{01}^{(3)} Y_{12}^{(1| 0)} + s_{02} \newf_{01}^{(2)} Y_{12}^{(2| 0)} -  s_{02} Y_{12}^{(4| 0)}
\notag \\
{-}4\pi \nabla_\tau Y_{11}^{(2| 2)} &=
3 s_{02} \newf_{01}^{(4)} Y_{11}^{(0| 2)} - 3 s_{12} \newg_4 Y_{11}^{(0| 2)}  - s_{02} \newf_{01}^{(2)} Y_{11}^{(2| 2)}  \notag \\
&\ \ \ \   +  3 s_{02} Y_{11}^{(4| 2)} + 2 s_{12} Y_{11}^{(4| 2)} - 2 s_{02} \newf_{01}^{(3)} Y_{12}^{(1| 2)} \notag \\
&\ \ \ \   +  s_{02} \newf_{01}^{(2)} Y_{12}^{(2| 2)} - s_{02} Y_{12}^{(4| 2)} + 2 Y_{11}^{(3| 1)} \notag \\
%%%%%%%%%%
{-}4\pi \nabla_\tau Y_{11}^{(4| 0)} &=
5 s_{02} \newf_{01}^{(6)} Y_{11}^{(0| 0)} 
- 5 s_{12} \newg_6 Y_{11}^{(0| 0)} 
- 4 s_{02} \newf_{01}^{(5)} Y_{12}^{(1| 0)}
  - 2 s_{02} \newf_{01}^{(3)} Y_{12}^{(3| 0)} 
\notag \\
&\ \ \ \ 
+  3 s_{02} \newf_{01}^{(4)} Y_{12}^{(2| 0)}
-  3 s_{12} \newg_4 Y_{11}^{(2| 0)} 
- s_{02} \newf_{01}^{(2)} Y_{11}^{(4| 0)} 
+ s_{02} \newf_{01}^{(2)} Y_{12}^{(4| 0)}
\notag \\
&\ \ \ \ 
+ 10 s_{02} Y_{11}^{(6| 0)} 
+  9 s_{12} Y_{11}^{(6| 0)} 
  - s_{02} Y_{12}^{(6| 0)}
 \label{gen.65a}
\end{align}
and
\begin{align}
 {-}4\pi \nabla_\tau Y_{12}^{(1| 0)} &=
2 s_{12} \newf_{01}^{(3)} Y_{11}^{(0| 0)} + s_{12} \newf_{01}^{(2)} Y_{11}^{(1| 0)} - s_{12} Y_{11}^{(3| 0)}  - s_{12} \newf_{01}^{(2)} Y_{12}^{(1| 0)} + s_{12} Y_{12}^{(3| 0)}
\notag \\
{-}4\pi \nabla_\tau Y_{12}^{(2| 0)} &=
3 s_{12} \newf_{01}^{(4)} Y_{11}^{(0| 0)} - 3 s_{02} \newg_4 Y_{11}^{(0| 0)} +  2 s_{12} \newf_{01}^{(3)} Y_{11}^{(1| 0)}
+ s_{12} \newf_{01}^{(2)} Y_{11}^{(2| 0)}  \notag \\
&\ \ \ \  - s_{12} Y_{11}^{(4| 0)}  - s_{12} \newf_{01}^{(2)} Y_{12}^{(2| 0)} + 2 s_{02} Y_{12}^{(4| 0)} +  3 s_{12} Y_{12}^{(4| 0)}
\notag \\
 {-}4\pi \nabla_\tau Y_{12}^{(2| 2)} &=
3 s_{12} \newf_{01}^{(4)} Y_{11}^{(0| 2)} - 3 s_{02} \newg_4 Y_{11}^{(0| 2)} + 
 2 s_{12} \newf_{01}^{(3)} Y_{11}^{(1| 2)}  \notag \\
&\ \ \ \  + s_{12} \newf_{01}^{(2)} Y_{11}^{(2| 2)} - s_{12} Y_{11}^{(4| 2)}  - s_{12} \newf_{01}^{(2)} Y_{12}^{(2| 2)}  \notag \\
&\ \ \ \  
 + 2 s_{02} Y_{12}^{(4| 2)} + 3 s_{12} Y_{12}^{(4| 2)} + 2 Y_{12}^{(3| 1)}
 \label{gen.65}
\end{align}

\newpage

%%%%%%%%%%%%%%%%%%%%%%%%%%%%%%%%%%%
%%%%%%%%%%%%%%%%%%%%%%%%%%%%%%%%%%%
\section{Singular two-point component integrals}
\label{app:kinsub}
%%%%%%%%%%%%%%%%%%%%%%%%%%%%%%%%%%%
%%%%%%%%%%%%%%%%%%%%%%%%%%%%%%%%%%%

This appendix is dedicated to the two-point component integrals
$Y^{(1|1)}_{ij}$ defined by (\ref{eq:Yab}) which have kinematic 
poles or singular $z_0 \rightarrow 0$ limits.

\subsection{Treatment of kinematic poles}
\label{app:kinsub.1}

By the singularity of $f^{(1)}(z)\overline{ f^{(1)}(z)} \sim \frac{1}{|z|^2}$ when $z \to 0$,
the component integrals $Y^{(1|1)}_{11}$ and $Y^{(1|1)}_{22}$ have kinematic poles
in $s_{12}$ and $s_{02}$, respectively. The representation of $Y^{(1|1)}_{11}$
in (\ref{subpole.1}) which exposes its kinematic pole can be attained by splitting
the Koba--Nielsen factor in its definition via (\ref{eq:Yab})
\beq
Y^{(1|1)}_{11} = \frac{ \Im \tau }{\pi} \int_{\Sigma} \frac{ \dd^2 z_2 }{\Im \tau} e^{s_{12} g(z_{12}|\tau)} \big(
\underbrace{ e^{s_{02} g(z_{02}|\tau)} - e^{s_{02} g(z_{01}|\tau)} }_{(i)} + \underbrace{ e^{s_{02} g(z_{01}|\tau)} }_{(ii)}
\big) f^{(1)}_{12} \overline{ f^{(1)}_{12}} 
\label{appp.1}
\eeq
The integrand of the first part $(i)$ no longer has the $\frac{1}{|z_{12}|^2}$ singularity 
since the factor of $e^{s_{02} g(z_{02}|\tau)} - e^{s_{02} g(z_{01}|\tau)}$ vanishes as $z_2 \rightarrow z_1$.
Hence, the $\ap$-expansion can be performed at the level of the integrand, and we obtain the
first line of (\ref{subpole.1}),
\beq
Y^{(1|1)}_{11} \, \big|_{(i)} =  \frac{\Im \tau}{ \pi } 
\sum_{\ell\geq 0} \sum_{k\geq 1} \frac{s_{02}^k s_{12}^\ell}{k! \, \ell!}
\int_{\Sigma} \frac{ \dd^2 z_2 }{\Im \tau} 
\,(g(z_{12}|\tau))^\ell \big( g(z_{02}|\tau)^k - g(z_{01}|\tau)^k \big) f^{(1)}_{12} \overline{f^{(1)}_{12}}
\label{appp.2}
\eeq
In the second part of (\ref{appp.1}), the factor of $e^{s_{02} g(z_{01}|\tau)} $ does not depend
on the integration variable $z_2$, and one can integrate by parts after identifying
$f^{(1)}_{12} e^{s_{12} g(z_{12}|\tau)} = \frac{1}{s_{12}} \partial_{z_2} e^{s_{12} g(z_{12}|\tau)}$
\begin{align}
Y^{(1|1)}_{11} \, \big|_{(ii)} &=  \frac{\Im \tau}{ \pi } e^{s_{02} g(z_{01}|\tau)} \int_{\Sigma} \frac{ \dd^2 z_2 }{\Im \tau}
\frac{1}{s_{12}}  \big( \partial_{z_2} e^{s_{12} g(z_{12}|\tau)}   \big)  \overline{ f^{(1)}_{12}}  \notag \\
&= -  \frac{\Im \tau}{ \pi s_{12} } e^{s_{02} g(z_{01}|\tau)} \int_{\Sigma} \frac{ \dd^2 z_2 }{\Im \tau}
e^{s_{12} g(z_{12}|\tau)} \partial_{z_2}  \overline{ f^{(1)}_{12}} 
\label{appp.3}\\
&= -  \frac{1}{ s_{12} } e^{s_{02} g(z_{01}|\tau)} \int_{\Sigma} \frac{ \dd^2 z_2 }{\Im \tau}e^{s_{12} g(z_{12}|\tau)} 
\notag
\end{align}
In passing to the last line, we have used $\partial_{z_2}  \overline{ f^{(1)}_{12}} = \pi( \frac{1}{\Im \tau} - \delta^2(z_{12}))$, where the delta function does not contribute in presence of $e^{s_{12} g(z_{12}|\tau)} $.
The last line of (\ref{appp.3}) is amenable to $\alpha'$-expansion at the level of the integrand and reproduces the
second line of (\ref{subpole.1}).

\subsection{$\alpha'$-expansion of $Y^{(1|1)}_{ij}$}
\label{app:kinsub.2}

Based on the splitting of $Y^{(1|1)}_{ij}$ into (\ref{appp.2}) and (\ref{appp.3}), we shall
now assemble the $\alpha'$-expansion to the subsubleading order. Based on standard
MGF techniques, the polar part gives rise to
\beq
Y^{(1|1)}_{11} \, \big|_{(ii)} = - \frac{1}{s_{12}} \Big\{ 1 + s_{02} g(z_{01}|\tau) + \frac{1}{2} s_{02}^2 g(z_{01}|\tau)^2 + \frac{1}{2} s_{12}^2 E_2(\tau) + {\cal O}(s_{ij}^3) \Big\}
\label{appp.4}
\eeq
The regular part (\ref{appp.2}) in turn only receives a single contribution at these orders
in $\alpha'$,
\begin{align}
Y^{(1|1)}_{11} \, \big|_{(i)} &= s_{02}  I_{11}  
+ {\cal O}(s_{ij}^2) 
\label{appp.5} \\
I_{11}  &= \frac{\Im \tau}{\pi} \int_{\Sigma} \frac{\dd^2 z_2}{\Im \tau}  \big[
g(z_{02}|\tau) - g(z_{01}|\tau)
\big] f^{(1)}_{12} \overline{f^{(1)}_{12}} \notag
\end{align}
and the integral $I_{11}$ can be determined by similar methods as in (\ref{appp.3}): Identifying
$f^{(1)}_{12} = \partial_{z_2}  g(z_{12}|\tau)$ and integrating by parts yields
\beq
I_{11} = \frac{\Im \tau}{\pi} \int_{\Sigma} \frac{\dd^2 z_2}{\Im \tau}  g(z_{12}|\tau)
\Big\{ \Big( \frac{ \pi }{\Im \tau} - \pi \delta^2(z_{12}) \Big) \big[
g(z_{01}|\tau) - g(z_{02}|\tau) \big]
- f^{(1)}_{02} \overline{f^{(1)}_{12}} \Big\}
\label{appp.6}
\eeq
The delta function drops out by the vanishing of
$ g(z_{12}|\tau)\big[g(z_{01}|\tau) - g(z_{02}|\tau)\big]$ as $z_2\rightarrow z_1$,
and the contributions from $g(z_{01}|\tau)$ and $- g(z_{02}|\tau)$ integrate to $0$ and
$-  g_2(z_{01}|\tau)$, respectively. The last term can be further simplified by
identifying  $\overline{f^{(1)}_{12}} g(z_{12}|\tau) = \frac{1}{2} \partial_{\bar z_2} g(z_{12}|\tau)^2$
and integrating by parts. After using $ \partial_{\bar z_2} f^{(1)}_{02} =  \pi( \frac{1}{\Im \tau} - \delta^2(z_{02}))$ and
taking the delta function into account, we finally arrive at
\beq
I_{11} = 
-  g_2(z_{01}|\tau)+ \frac{1}{2} E_2(\tau)  - \frac{1}{2}  g(z_{01}|\tau)^2
\label{appp.7}
\eeq
In combination with (\ref{appp.4}), the
leading orders in the $\alpha'$-expansion of (\ref{subpole.1}) are found to be
\begin{align}
Y_{11}^{(1|1)} &= - \frac{1}{s_{12}} - \frac{s_{02}}{s_{12}}  g(z_{01}) - \frac{1}{2} \Big( \frac{ s_{02}^2 }{s_{12}} + s_{02} \Big)  g(z_{01})^2 \notag \\
&\ \ \ \ - s_{02} g_2(z_{01}) + \frac{1}{2}(s_{02}-s_{12}) E_2 + {\cal O}(s_{ij}^2)
\label{subpole.2}
\end{align}
The analogous $\alpha'$-expansion of $Y_{12}^{(1|1)} $ can be obtained along
the same lines, or by exploiting (\ref{subpole.2}) and the integration-by-parts identity (\ref{subpole.3}):
\begin{align}
Y_{12}^{(1|1)} &=  g(z_{01}) + \frac{1}{2} (s_{02}{+}s_{12}) \big[ g(z_{01})^2 -   E_2 \big] + {\cal O}(s_{ij}^2)
\label{subpole.4}
\end{align}
However, we note that this $\alpha'$-expansion does not commute with the $z_0 \rightarrow z_1$ limit:
On the one hand, already the leading term in (\ref{subpole.4}) exhibits a logarithmic divergence
as $z_0 \rightarrow z_1$. On the other hand, the integrand $f^{(1)}_{02} \overline{f^{(1)}_{12}}$
of $Y_{12}^{(1|1)}$ coincides with the one of $Y_{11}^{(1|1)}$ as $z_0 \rightarrow z_1$, so performing the limit
before integration and $\ap$-expansion would yield the kinematic
pole in (\ref{subpole.2}). We will rely on the $\ap$-expansion (\ref{subpole.4}) obtained from the 
integration-by-parts relation (\ref{subpole.3}) since the latter is essential for the derivation 
of the differential equations of the generating series $Y_{ij}$.

\newpage

\section{Derivation of the commutation relations among $R_{\vec{\eta}}$}
\label{app:comm}

This appendix provides intermediate steps in deriving the commutation
relations of the operators $R_{\vec{\eta}}(\cdot)$ in the differential equations
(\ref{closedana}), (\ref{eq:dtauY}) of two-point and (\ref{high.12}) of $n$-point
generating series of eMGFs. In an $n$-point context, commutativity of the 
$z_0$- and $\tau$-derivatives in (\ref{high.12}) implies that
\begin{align}
0 &= [ -4 \pi \nabchi_\tau ,  \nabla_{z_0}  ] Y( \begin{smallmatrix} M \\ N \end{smallmatrix} |  \begin{smallmatrix} K \\ L \end{smallmatrix}) \notag \\
&= \sum_{P,Q} \bigg( 
\sum_{k=2}^\infty (k{-}1) (\tau{-}\bar \tau)^k f^{(k)}_{01} R_{\vec{\eta}}([x_0,x_{k-1}])
-\sum_{k=0}^\infty   (\tau{-}\bar \tau)^k f^{(k)}_{01} R_{\vec{\eta}}([x_k,\epsilon_{0}]) \notag \\
&\ \ \ \ \ \ + \sum_{\ell=0}^\infty  (\tau{-}\bar \tau)^\ell f^{(\ell)}_{01} 
\sum_{k=4}^\infty (1{-}k) (\tau{-}\bar \tau)^k G_k R_{\vec{\eta}}([x_\ell,\epsilon_k]) \label{high.16} \\
&\ \ \ \ \ \ + \sum_{1\leq a<b}^\infty  (\tau{-}\bar \tau)^{a+b+1} (b f_{01}^{(a)} f_{01}^{(b+1)} - a f_{01}^{(a+1)} f_{01}^{(b)} ) R_{\vec{\eta}}([x_a,x_b]) \notag \\
&\ \ \ \ \ \ + 2\pi i (\bar \eta_L {-} \bar \eta_Q) \Big[
{-}R_{\vec{\eta}}(\epsilon_0) + \sum_{k=2}^\infty (k{-}1) (\tau{-}\bar \tau)^k f^{(k)}_{01} R_{\vec{\eta}}(x_{k-1})
\Big]
\bigg)_{ \begin{smallmatrix} K \\ L \end{smallmatrix} \big|  \begin{smallmatrix} P \\ Q \end{smallmatrix}} 
Y( \begin{smallmatrix} M \\ N \end{smallmatrix} |  \begin{smallmatrix} P \\ Q \end{smallmatrix}) 
\notag
\end{align}
In obtaining the last line of (\ref{high.16}), we have used that $(\bar \eta_L {-} \bar \eta_Q) R_{\vec{\eta}}(\epsilon_{k\geq 4})_{ \begin{smallmatrix} K \\ L \end{smallmatrix} \big|  \begin{smallmatrix} P \\ Q \end{smallmatrix}} =0$:
It can be seen from (4.42) of \cite{Broedel:2020tmd} that all the nonzero matrix entries of the 
derivations $R_{\vec{\eta}}(\epsilon_{k\geq 4})$ preserve $\{L\}=\{Q\}$ as
sets (i.e.\ $R_{\vec{\eta}}(\epsilon_{k\geq 4})_{ \begin{smallmatrix} K \\ L \end{smallmatrix} \big|  \begin{smallmatrix} P \\ Q \end{smallmatrix}}\neq 0$ only if $L$ is a permutation of $Q$).
Once the factor of $b f_{01}^{(a)} f_{01}^{(b+1)} - a f_{01}^{(a+1)} f_{01}^{(b)}$ in (\ref{high.16})
is simplified via
\begin{align}
&b f_{01}^{(a)} f_{01}^{(b+1)} - a f_{01}^{(a+1)} f_{01}^{(b)} \nonumber\\
&= \frac{ (b{-}a) (a{+}b{+}1)! }{(a{+}1)!(b{+}1)!} f_{01}^{(a+b+1)}  - (-1)^b (a{+}b) G_{a+b+1} 
+ a G_{a+1}
\theta(a{-}3) f^{(b)}_{01}
- b G_{b+1} 
\theta(b{-}3) f^{(a)}_{01}  \label{gen.27}  \nonumber\\
&\phantom{=}
+ \sum_{k=4}^a { a{+}b{-}k \choose b{-}1 } (k{-}1) G_{k} f_{01}^{(a+b+1-k)}
- \sum_{k=4}^b { a{+}b{-}k \choose a{-}1 } (k{-}1) G_{k} f_{01}^{(a+b+1-k)}
\end{align}
one can read off the commutation relations of sections \ref{sec:3.6} and \ref{sec:4.comm}
from the coefficients of $f^{(k)}_{01}$ and $f^{(\ell)}_{01} G_k$:
\begin{itemize}
\item Imposing the coefficients of $(\tau{-}\bar\tau)^{k+w}G_k f^{(w)}_{01}$ in (\ref{high.16}) with $k\geq4$ 
to vanish implies the relations (\ref{likenopen}) that hold in identical form for the open-string operators $r_{\vec{\eta}}(\cdot)$.
\item The coefficients of $(\tau{-}\bar\tau)^{k} f^{(k)}_{01}$ in (\ref{high.16}) vanish by virtue
of (\ref{high.19}) which necessitates the terms $\sim 2\pi i(w{-}1) (\bar \eta_L {-} \bar \eta_Q) $
specific to the closed-string generating series.
\end{itemize}
Note that the step functions $\theta(c)$ in the second line of (\ref{gen.27})
are taken to be $1$ for $c\geq 0$ and zero for $c<0$, e.g.\ $a G_{a+1}
\theta(a{-}3)= 3G_4$ if $a=3$.

\providecommand{\href}[2]{#2}\begingroup\raggedright\endgroup

\end{document}